\documentclass[11pt]{article}
\usepackage[a4paper, total={6in, 8in}]{geometry}
\usepackage{graphicx}
\usepackage{xcolor}
\usepackage{amsmath}
\usepackage{amssymb, bm}
\usepackage[utf8]{inputenc} 
\def \eff{{\text{eff}}}
\newcommand{\matel}[3]{\langle #1|#2|#3\rangle}
\newcommand{\mi}{\!-\!}
\newcommand{\rGamma}{\mathrm\Gamma}
\begin{document}

\title{Recent progress in decays of $b$ and $c$ hadrons}



\author{Aoife Bharucha\footnote{aoife.bharucha@.cpt.univ-mrs.fr} \\ Aix Marseille Univ, Universit\'{e} de Toulon, CNRS, CPT, Marseille, France
                        %  \\
}

\date{February 2, 2024}

\maketitle

\begin{abstract}
In the last ten years there has been great progress in calculations of decays of $B$ and $D$ mesons, and baryons containing a heavy $b$ or $c$ quark. 
One propelling factor has been the measurement of several anomalies in $b\to s$ and $b\to c$ transitions, these are one of the only signs of physics beyond the Standard Model. The deviations included measurements of branching ratios, angular observables and lepton universality ratios. Another factor is the exclusive-inclusive discrepancy in the determination of the CKM elements $V_{ub}$ and $V_{cb}$. We will first review recent calculations involving  $b\to s$ and $c\to u$ transitions that could shed light on the neutral current anomalies. We will then summarise the progress the determination of the CKM elements, $V_{ub}$ and $V_{cb}$. Finally we will discuss the current theoretical status and experimental prospects for the lepton universality ratios in $b\to s$ and $b\to c$ semileptonic decays.

\end{abstract}

\section{Introduction}
\label{sec:intro}

The last decade has witnessed impressive changes in the way we approach heavy flavour physics.
A central role in orchestrating these changes has been played by the so-called $B$ anomalies,
and these must be mentioned by any review of heavy flavour physics.
The $B$ anomalies are a set of deviations observed by experiment in semileptonic $b\to s$ and $b\to c$ transitions, including branching ratio measurements, angular observables and lepton universality ratios. For the $b\to s$ case, this involves branching ratios for $B_s\to\mu^+\mu^-$\footnote{While there has been a longstanding disagreement in this channel, recent results from CMS~\cite{CMS:2022dbz} diminish the global significance of the discrepancy to below $2~\sigma$.}, $B\to K^{(\ast)}\mu^+\mu^-$ and $B_s\to \phi\mu^+\mu^-$ the latter two being consistently low across several $q^2$ bins (around a 2-3\,$\sigma$ effect)~\cite{LHCb:2021vsc,LHCb:2014cxe,CMS:2015bcy,LHCb:2015wdu,LHCb:2021zwz}, the angular observable $P_5'$ which deviates by 2-3\,$\sigma$ from the SM prediction~\cite{CMS:2015bcy,LHCb:2020lmf}, and the ratios $R_{K_{s}^{(\ast)}}$~\cite{LHCb:2017avl,Belle:2019oag,BELLE:2019xld,LHCb:2021trn,LHCb:2021lvy} and $R_{pK}$~\cite{LHCb:2019efc}.
For $b\to c$, the deviations involve lepton universality ratios, this time for ratios of branching fractions for $B\to D^{(\ast)}\ell\nu$ decays for $\ell=\tau$ and $\ell=\mu$, and have been measured by Babar, Belle and LHCb \cite{BaBar:2012obs,BaBar:2013mob,Belle:2015qfa,Belle:2016ure,Belle:2016dyj,LHCb:2015gmp,LHCb:2017smo}.

These deviations are some of the most intriguing signs of physics beyond the SM (BSM) that we have seen at colliders, and it is imperative that they are thoroughly explored.
This has lead to research in the theory community in the following directions: to gain an improved understanding of the long-distance QCD contributions to $b\to s \ell^+\ell^-$ exclusive decays, 
to study all possible channels which could provide complementary information about the $b\to s$ and $b\to c$ transition, particularly the lepton flavour structure and finally to improve the form factors describing the matrix element for the hadronic transition which are essential in the calculation of exclusive semileptonic decays. 
The exclusive-inclusive determination of $|V_{cb}|$ and $|V_{ub}|$ (note that in this review we will denote $|V_{qb}|$ by $V_{qb}$) has also played a major role in heavy flavour physics, and the last decade has seen a great deal of progress in both the exclusive determination (relying on improvements in the form factors) and the inclusive determination.  We will dedicate this review to these issues.
Note that the subject of heavy flavour physics, is very broad, and rapidly advancing, and it is difficult to do justice to all that has happened in the last ten years.
One of the most important topics which has been omitted is CP violation in $B$ and $D$ decays, for which excellent recent reviews can be found in~\cite{Gershon:2016fda,Nir:2020mgy}.

The rest of this review will be structured as follows, in the following Sect.~\ref{sec:theo} we will introduce the theoretical framework: the form factors and the Wilson coefficients.
In Sect.~\ref{sec:rare-lu}, we will discuss decays involving flavour changing neutral currents (FCNCs), focussing on $b\to s$ decays but also mentioning $b\to d$ and $c\to u$ decays. We will then come to the subject of charged current transitions in Sect.~\ref{sec:semilep}, with an emphasis on the tensions between the exclusive and inclusive determinations of $V_{cb}$ and $V_{ub}$, but also mentioning the determination of $V_{cs}$ and $V_{cd}$. In Sect.~\ref{sec:lu} we will finally discuss the lepton universality ratios for both $b\to s$ and $b\to c$ transitions, before summarising in Sect.~\ref{sec:summary}.

\section{Theoretical framework}
\label{sec:theo}
\subsection{Form factors}
\label{sec:FFs}
For exclusive processes, on which much of this review will focus, the calculation of the relevant decay rates relies on knowledge of the decay constants and form factors parametrising the appropriate matrix elements. The calculation on the Lattice of the decay constants is extremely precise, details and latest averages can be found in~\cite{Aoki:2021kgd}. The form factors for semileptonic decays can be calculated in Lattice QCD (LQCD) at low recoil energy of the final state meson (high $q^2$ the invariant mass of the leptons), and in light-cone sum rules (LCSR) at large recoil energy or low $q^2$. In order to obtain the form factor over the full kinematic region, a parametrisation is used, this is in general a series-type parametrisation in a parameter $z(q^2)$, which allows a unitarity bound to be placed on the series coefficients (see e.g.~\cite{Boyd:1994tt,Caprini:1997mu,Bourrely:2008za}). For completeness, here we will briefly review the definitions of the form factors and the recent calculations of these in the literature via LQCD or LCSR. We first discuss the form factors for $B$ and $D$ meson decays to pseudoscalars and then to vectors before coming to $b$-baryon decays.
\paragraph{Decays to pseudoscalar mesons:}
Only three independent form factors are needed to describe decays to pseudoscalar mesons, the scalar $f_0$, the vector $f_+$ and the tensor $f_T$ form-factors. These are defined via
\begin{eqnarray} 
	 \nonumber \langle P(k)|\bar{u}\gamma^\mu (1\pm\gamma_5) c| M (p)\rangle &=&\, f_+(q^2) \left( (p+k)^\mu -q^\mu \frac{m_M^2 -m_P^2}{q^2} \right) \\
	 \nonumber &&\,+ f_0(q^2) \frac{m_M^2 -m_P^2}{q^2}q^\mu ,\quad\mbox{and}\\
			\langle P(k)|\bar{u}\sigma^{\mu\nu}(1\pm\gamma_5) c| M(p)\rangle &=&\, i\frac{f_T(q^2)}{m_M+m_P} \biggr( (p+k)^\mu q^\nu - (p+k)^\nu q^\mu \nonumber \\
			&&\quad
			\pm i \epsilon^{\mu\nu\alpha\beta}(p+k)_{\alpha} q_\beta \biggr).
	   \label{eq:ffbasisP}
		\end{eqnarray}
			Note that, at zero momentum transfer, the relation $f_+(0) = f_0(0)$ reduces the number of independent form factors to two.

Starting with heavy-to-light $B$ meson decays, the form factors for $B\to K$ ($B\to\pi$ and $B_{s}\to K$) involving a $b\to s$ ($b \to d$) quark transition are needed in the search for new physics via rare decays as discussed in Sect.~\ref{sec:rare-lu}. The form factors for $B\to\pi$ and $B_{s}\to K$ transitions are also required for measurements of the CKM element $V_{ub}$, these will be discussed in detail in Sect.~\ref{sec:ExVub}. 

Calculations  of form factors for heavy-to-heavy $B$ meson decays ($B_{(s)}\to D_{(s)}$) are crucial for measurements of $V_{cb}$ and lepton universality ratios, these will be summarised in Sect.~\ref{sec:ExVcb}. 
For heavy-to-heavy form factors, the series parametrisations that are typically used by experiment are the Caprini-Neubert-Lellouch (CLN)~\cite{Caprini:1997mu} or Boyd-Lebed-Grinstein (BGL) parametrisation~\cite{Boyd:1994tt,Boyd:1995sq,Boyd:1997kz}. This will be discussed further in Sect.~\ref{sec:ExVcb}.

For $D$ decays, a great number of LQCD calculations exist for the form factors for $D\to\pi$ and $D\to K$ transitions~\cite{Aoki:2021kgd}, providing possibilities to search for BSM physics via rare decays, see Sect.~\ref{sec:rare-lu}, and high accuracy determinations of $V_{cd}$ and  $V_{cs}$, see Sect.~\ref{sec:Vcdcs}.

\paragraph{Decays to vector mesons:}
For decays to vector mesons, seven independent form factors are required. These can be expressed via~\cite{Straub:2015ica}\footnote{The factors $c_V^{b \to q}$ are attached to the matrix elements on the left-hand side, cf.~\cite{Ball:2004rg} as a result of form of the wave functions  $\rho^0 \sim 1/\sqrt{2} (\bar u u - \bar d d  )  $ and $\omega \sim 1/\sqrt{2} (\bar u u - \bar d d  )$, 
$c^{ u}_{\rho^0} = -c^{ d}_{\rho^0}  =c^{ u}_{\omega} = c^{ d}_{\omega}= 
  \sqrt{2}$ and $c_V=1$ in all other cases.}
 \begin{eqnarray}
\matel{V(k)}{\bar s \gamma^\mu(1 \mp \gamma_5) b}{\bar M(p)}  
&=& P_1^\mu \, \mathcal{V}_1(q^2) \pm \sum_{i=2,3,P} P_i^\mu \, \mathcal{V}_i(q^2) 
   \; ,\nonumber  \\[0.1cm] 
 \matel{V(k)}{\bar s iq_\nu \sigma^{\mu\nu} (1 \pm \gamma_5) b}{\bar M(p)} 
&=& P_1^\mu  T_1(q^2)   \pm  P_2^\mu  T_2(q^2) \pm  P_3^\mu  T_3(q^2) 
   \; , 
   \label{eq:ffbasisV}
\end{eqnarray}
where the Lorentz structures $P_i^\mu$ are defined as in \cite{Lyon:2013gba} as
\begin{eqnarray}
\label{eq:Vprojectors}
\nonumber P_1^\mu  &=&  2 \epsilon^{\mu}_{\phantom{x} \alpha \beta \gamma} \eta^{*\alpha} k^{\beta}q^\gamma  \; , \quad
 P_2^\mu = i \{(m_M^2\mi m_V^2) \eta^{*\mu} \mi 
(\eta^*\!\cdot\! q)(p+k)^\mu\} \; , \\
P_3^\mu &=&  i(\eta^*\!\cdot\! q)\{q^\mu \mi  \frac{q^2 }{m_M^2\mi m_{V}^2} (p+k)^\mu \}   \;, \quad P_P^\mu = i (\eta^* \cdot q) q^\mu \; ,
\end{eqnarray} 
with the $\epsilon_{0123}=+1$    convention for the Levi-Civita tensor and with $\eta$ the polarization of the vector meson.
Note that $T_1(0)= T_2(0)$ holds algebraically. This parameterisation is chosen as it simplifies results for decay rates, and makes the relation between the tensor and vector form factors clear.
The ${\cal V}_{P,1,2,3}$  can be related to the more traditional $A_{0,1,2,3}$ and $V$ via
 \begin{eqnarray}
 \mathcal{V}_P(q^2) &=&  \frac{- 2 m_{V}}{q^2} A_0(q^2) \;,  \quad \mathcal{V}_1(q^2) =  \frac{-V(q^2)}{m_M+m_{V}} \;, \quad     \mathcal{V}_2(q^2) =    \frac{-A_1(q^2)}{m_M-m_{V}} 
\;, \nonumber  \\[0.1cm]
   \mathcal{V}_3(q^2) &=&  \big( \frac{m_M+m_{V}}{q^2}    A_1(q^2) -   \frac{m_M-m_{V}}{q^2}    A_2(q^2) \big) \equiv \frac{2 m_{V}}{q^2} A_3(q^2) \; .
 \label{eq:VAs}
\end{eqnarray}
where $A_3(0) = A_0(0)$ insures that the matrix elements are finite at $q^2 =0$.

Heavy to light form factors for decays to vector mesons are particularly crucial given the $B$ anomalies in $B\to K^*\ell+\ell^-$ decays, the latest calculations will be discussed in Sect.~\ref{sec:rare-lu}. 
Heavy to heavy form factors include $B_{(s)}\to D^*_{(s)}$ transitions, required for determining $V_{cb}$, see Sect.~\ref{sec:ExVcb} for details about the latest results.
    	
\paragraph{Decays of baryons:}    	
The large number of $b$ and $c$ baryons produced at the LHC has motivated calculations of baryonic form factors, with particularly impressive results from the Lattice~\cite{Detmold:2015aaa} and in LCSR~\cite{Mannel:2011xg,Khodjamirian:2011jp,Mannel:2015osa}, allowing the precise measurement of $V_{ub}$ and $V_{cb}$, as discussed in Sect.~\ref{sec:VubVcb}.
In light of the anomalies in rare $B$ decays, the form factors for $\Lambda_b\to \Lambda$ (and to excited states), have also been calculated in LQCD~\cite{Detmold:2016pkz} (see also \cite{Meinel:2020owd,Meinel:2021mdj}) and LCSR~\cite{Mannel:2011xg}.
Further, results have been obtained for charmed baryon decays, i.e.~for $\Lambda_c\to\Lambda$~\cite{Meinel:2016dqj} and $\Lambda_c\to N$~\cite{Meinel:2017ggx} and to excited states~\cite{Meinel:2021mdj}.

\subsection{Wilson coefficients}

Since a substantial part of this review will focus on $b\to s$, $c\to u$ (in Sect.~\ref{sec:rare-lu}) and $b\to c$  (in Sect.~\ref{sec:ckm}) transitions, here we include the definition of the Wilson coefficients required for completeness. The relevant effective Hamiltonian for decays involving $b\to s\gamma$ and $b\to s\ell^+\ell^-$ can be expressed by \cite{Bobeth:1999mk,Bobeth:2001jm}
\begin{equation} \label{eq:Heff}
    {\cal H}_{\eff} = - \frac{4\,G_F}{\sqrt{2}}\left(
\lambda_t {\cal H}_{\eff}^{(t)} + \lambda_u {\cal
  H}_{\eff}^{(u)}\right)
\end{equation}
where $G_F$ is the Fermi constant,  $\lambda_q=V_{qb}V_{qs}^*$ and
\begin{eqnarray*}
{\cal H}_{\eff}^{(t)} 
& = & 
C_1 \mathcal O_1^c + C_2 \mathcal O_2^c + \sum_{i=3}^{6} C_i 
\mathcal O_i + \sum_{i=7,8,9,10,P,S} (C_i \mathcal O_i + C'_i \mathcal
O'_i)\,,
\\
{\cal H}_{\eff}^{(u)} 
& = & 
C_1 (\mathcal O_1^c-\mathcal O_1^u)  + C_2(\mathcal O_2^c-\mathcal
O_2^u)\,.
\end{eqnarray*}
The operators $\mathcal O_{i\leq 6}$ are given by the $P_i$ given
in \cite{Bobeth:1999mk}, while the remaining ones are given by
\begin{align}
\label{eq:O7}
{\mathcal{O}}_{7} &= \frac{e}{g^2} m_b
(\bar{s} \sigma_{\mu \nu} P_R b) F^{\mu \nu} ,&
{\mathcal{O}}_{7}^\prime &= \frac{e}{g^2} m_b
(\bar{s} \sigma_{\mu \nu} P_L b) F^{\mu \nu} ,\\
\label{eq:O8}
{\mathcal{O}}_{8} &= \frac{1}{g_s} m_b
(\bar{s} \sigma_{\mu \nu} T^a P_R b) G^{\mu \nu \, a} ,&
{\mathcal{O}}_{8}^\prime &= \frac{1}{g_s} m_b
(\bar{s} \sigma_{\mu \nu} T^a P_L b) G^{\mu \nu \, a} ,\\
\label{eq:O9}
{\mathcal{O}}_{9} &= \frac{e^2}{g^2} 
(\bar{s} \gamma_{\mu} P_L b)(\bar{\mu} \gamma^\mu \mu) ,&
{\mathcal{O}}_{9}^\prime &= \frac{e^2}{g^2} 
(\bar{s} \gamma_{\mu} P_R b)(\bar{\mu} \gamma^\mu \mu) ,\\
\label{eq:O10}
{\mathcal{O}}_{10} &=\frac{e^2}{g^2}
(\bar{s}  \gamma_{\mu} P_L b)(  \bar{\mu} \gamma^\mu \gamma_5 \mu) ,&
{\mathcal{O}}_{10}^\prime &=\frac{e^2}{g^2}
(\bar{s}  \gamma_{\mu} P_R b)(  \bar{\mu} \gamma^\mu \gamma_5 \mu) ,\\
\label{eq:OS}
{\mathcal{O}}_{S} &=\frac{e^2}{16\pi^2}
m_b (\bar{s} P_R b)(  \bar{\mu} \mu) ,&
 {\mathcal{O}}_{S}^\prime &=\frac{e^2}{16\pi^2}
m_b (\bar{s} P_L b)(  \bar{\mu} \mu) ,\\
\label{eq:OP}
{\mathcal{O}}_{P} &=\frac{e^2}{16\pi^2}
m_b (\bar{s} P_R b)(  \bar{\mu} \gamma_5 \mu) ,&
 {\mathcal{O}}_{P}^\prime &=\frac{e^2}{16\pi^2}
m_b (\bar{s} P_L b)(  \bar{\mu} \gamma_5 \mu),
\end{align}
where $g_s$ is the strong coupling constant, the left, right projectors are defined via $P_{L,R}=(1\mp
\gamma_5)/2$ and $m_b$ is $b$ quark
mass in the $\overline{\rm MS}$ scheme. Note that the primed operators and the scalar/pseudoscalar operators either vanish or are highly suppressed in the SM and can be neglected. BSM contributions to the operators  $\mathcal O_{i\leq 6}$ are also neglected. The renormalisation group (RGE) evolution of the operators in the SM is numerically significant as a result of the four-quark operators, and is known to next-to-next-to-leading logarithmic (NNLL) accuracy~\cite{Gambino:2003zm,Bobeth:2003at,Gorbahn:2004my,Gorbahn:2005sa}.

For the related up-type FCNC $c\to u$ transitions, the effective Hamiltonian takes the form~\cite{FK2015}:
	\begin{equation}
		 {\cal H}_{\eff}^{\rm SM}(m_b>\mu>m_c) =  -\frac{4\,G_F} {\sqrt{2}} \sum_{q=d,b} \lambda_q  {\cal H}_{\eff}^{(q)},
		\label{eq:Heff_db}
	\end{equation}
		where now  $\lambda_{b/d}$ are given by $\lambda_q=V_{cq}^* V_{uq}$, and
	\begin{align}
\hspace{-.4cm}   {\cal H}_{\eff}^{(b)} = C_1 {\mathcal{O}}_1^s + C_2 {\mathcal{O}}_2^s + \sum_{i=3}^9 C_i {\mathcal{O}}_i,\quad
		 {\cal H}_{\eff} ^{(d)} = C_1 ({\mathcal{O}}_1^s -{\mathcal{O}}_1^d) + C_2 ({\mathcal{O}}_2^s -{\mathcal{O}}_2^d).
	\end{align}
Note that since $\lambda_b \ll \lambda_d$, all contributions entering ${\cal H}_{\eff}^{(b)}$ are heavily CKM suppressed. The Wilson coefficients are as defined in Eqs.~\eqref{eq:O7} to \eqref{eq:OP} for $b\to s$ decays, making the replacements $b$ by $c$ and $s$ by $u$, full definitions including for $\mathcal O_{i\leq 6}$, and a calculation of the RG evolution of the Wilson coefficients at NNLL can be found in \cite{Boer_WC}.

Finally the effective Hamiltonian for the charged current transitions $b\to c$ can be expressed by
\begin{equation}
	 {\cal H}_{\eff}^{b\to c\ell\nu} =  -\frac{4\,G_F} {\sqrt{2}} V_{cb}\sum_{i} C_i {\mathcal{O}}_i+h.c.,
\end{equation}
where the operators take the form
\begin{align}
\nonumber {\mathcal{O}}_{VL} &=(\bar{c}\gamma^\mu P_L b)(  \bar{\ell} \gamma_\mu P_L\nu) ,&
{\mathcal{O}}_{VR} &=(\bar{c}\gamma^\mu P_R b)(  \bar{\ell} \gamma_\mu P_L\nu) \\
\nonumber {\mathcal{O}}_{SL} &=(\bar{c} P_L b)(  \bar{\ell} P_L\nu) ,&
{\mathcal{O}}_{SR} &=(\bar{c} P_R b)(  \bar{\ell} P_L\nu) \\
{\mathcal{O}}_{T} &=(\bar{c} \sigma^{\mu\nu} b)(  \bar{\ell}\sigma_{\mu\nu} P_L\nu) ,&
\end{align}
and the Wilson coefficients vanish in the SM, with the exception of $C_{VL}=1+\mathcal{O}(\alpha).$ The evaluation of the Wilson coefficients is simpler than in the FCNC case due to the absence of strong interaction effects, the running at leading log suffices and there are no non-local contributions.

\section{Neutral current transitions}
\label{sec:rare-lu}
\subsection{Leptonic decays}
\label{sec:lep}
Fully leptonic rare decays are highly suppressed in the SM, not only by the loop suppression of the FCNC transition but also by the helicity suppression. These therefore provide a powerful constraint on possible BSM contributions, particularly possible scalar or pseudoscalar operators.
In order to illustrate how the branching ratios for leptonic decays of heavy mesons are calculated, let us insert the effective Hamiltonian between the heavy decaying meson $\mathcal{B}$ and the final state $\ell^+\ell^-$, 
\begin{equation}
\langle\ell^+\ell^-|\mathcal{H}_{\rm eff}|\mathcal{B}\rangle\sim \frac{4 G_F}{\sqrt{2}}\lambda\sum_i\langle\ell^+\ell^-|C_i\mathcal{O}_i|\mathcal{B}\rangle,
\end{equation}
where $\lambda$ represents the appropriate combination of CKM elements.
On writing the operators as a product of two fermion currents, $\mathcal{O}\sim j_\ell\,j_q$, this then allows us to factorize the matrix element into a leptonic and hadronic part, schematically
\begin{align}
\nonumber\langle\ell^+\ell^-|\mathcal{O}_i|\mathcal{B}\rangle\sim&\langle\ell^+\ell^-|j^i_\ell\,j_q|\mathcal{B}\rangle\\
\nonumber\sim&\langle\ell^+\ell^-|j^i_\ell |0\rangle\,\langle 0|j_q|\mathcal{B}\rangle\\
\sim& f_\mathcal{B} \langle\ell^+\ell^-|j^i_\ell |0\rangle
\end{align}
where we have introduced the decay constant of the heavy meson $f_\mathcal{B}$, such that the amplitude can be expressed as
\begin{equation}
\langle\ell^+\ell^-|\mathcal{H}_{\rm eff}|\mathcal{B}\rangle\sim \frac{4 G_F}{\sqrt{2}}\lambda f_\mathcal{B} \sum_i C_i  \langle\ell^+\ell^-|j^i_\ell |0\rangle.
\end{equation}
At leading order the amplitude depends only on the decay constant, the CKM elements and the Wilson coefficients.

For the $B_{(s)}$ decay the calculation of the branching ratio is at an advanced stage, and combining NLO electroweak corrections~\cite{Bobeth:2013tba} and NNLO QCD corrections~\cite{Hermann:2013kca} to the Wilson coefficients. 
Several calculations exist in LQCD with $N_f=2+1+1$, and following the 2017 update from FNAL/MILC, the uncertainty on the decay constants is now below the percent level~\cite{Bazavov:2017lyh}. More recently, the power-enhanced leading-logarithmic QED corrections were calculated in \cite{Beneke:2019slt}, where it was found that the overall effect on the branching fraction was $<\mathcal{O}( 1\%)$. This is negligible compared to the other uncertainties~\cite{Beneke:2019slt}:
\begin{align}
\mathcal{B}_{B_s\to\mu^+\mu^-}=3.660\big(1&+0.011|_{f_{B_s}}+0.031|_{\rm CKM}\\
&+0.011|_{m_t}+0.014|_{\rm rem.}\big)\, 10^{-9}
\end{align}
Note however that here the inclusive value of $V_{cb}$ is used, and  if one replaces the inclusive value of $V_{cb}$ by the exclusive value this would provoke an additional effect of 10\%; it is therefore clear that the uncertainties from the CKM elements by far dominate~\cite{Bobeth:2021cxm}.

 Experimentally, there has been exciting progress in the measurement of $B_q\to\mu^+\mu^-$ at LHCb, with the first observation in a single experiment for the $B_s$ decay and limits on the $B_d$ decay in \cite{LHCb:2017rmj}; for the latest update see \cite{CMS:2022mgd}. Note that the first observation of $B_s\to\mu^+\mu^-$ was in fact from the combined CMS and LHCb analysis, see Ref.~\cite{CMS:2014xfa}. With 300 fb$^{-1}$ LHCb will have measured the branching ratio for $B_s\to\mu^+\mu^-$ to high precision, with uncertainties close to those from theory. LHCb has also put limits on $B_{(s)}\to\tau^+\tau^-$~\cite{LHCb:2017myy}, and will improve this limit by an order of magnitude by the end of the Upgrade II~\cite{LHCb:2018roe}.
 
These measurements of the $B_{(s)}$ branching ratios in combination with the theory predictions allow us to place extremmely stringent constraints on BSM physics, particularly contributions of scalar and pseudoscalar operators, in $b\to d$ or $b\to s$ transitions~\cite{Altmannshofer:2021qrr}. For $b\to s$, a complementary observable is the effective lifetime $\tau_{\mu^+\mu^-}$~\cite{DeBruyn:2012wk},
\begin{equation}
\tau_{\mu^+\mu^-}=\frac{\int_0^\infty dt\, t \langle \rGamma (B_s(t)\to\mu^+\mu^-)\rangle }{\int_0^\infty dt \langle \rGamma (B_s(t)\to\mu^+\mu^-)\rangle }.
\end{equation}
which depends on the theoretically clean observable $\mathcal{A}_{\Delta\rGamma}$, which in turn is sensitive to BSM physics.
It turns out that the measurement of the effective lifetime can resolve the degeneracy allowed by the measurement of the branching ratio with respect to the BSM contribution to the scalar and pseudoscalar Wilson coefficients~\cite{Altmannshofer:2021qrr}.
 
For the $D$ meson decays, the theory prediction in the SM is less known due to the fact that long-distance contributions, for example from intermediate di-photon states, are thought to overpower the short-distance~\cite{Petrov:2017nwo}. Current limits from LHCb $\mathcal{B}(D\to\mu^+\mu^-)<6.2\cdot 10^{-9}$ at 90\% C.L.~\cite{LHCb:2013jyo} will be improved to $1.8\cdot 10^{-10}$ with 300 fb$^{-10}$~\cite{LHCb:2018roe}. While the SM result is not well known, it should be below $10^{-11}$~\cite{Petrov:2017nwo} and experimental bounds therefore provides important constraints on the BSM contribution to $D$ decays. Experimental results on lepton violating processes would also be very useful in constraining possible BSM physics.

\subsection{Semileptonic rare decays}
\label{sec:semilep}
\paragraph{Theoretical background:}
These decays are ideal probes of BSM physics as the three and four-body final states give rise to abundant possibilities for constructing angular observables.
For the generic decay from a heavy ($\mathcal{B}$) to light ($\mathcal{M}$)  meson, in order to calculate the amplitude one can make use of the effective Hamiltonian defined in Eqs.~\eqref{eq:Heff} or \eqref{eq:Heff_db} for the $b\to s/d$ or $c\to u$ transition respectively:
\begin{align}
\nonumber\mathcal{A}_{\mathcal{B}\to \mathcal{M}\ell^+\ell^-}=\,&\langle\ell^+\ell^- \mathcal{M}|\mathcal{H}_{\rm eff}|\mathcal{B}\rangle\\
=\,&\frac{4 G_F}{\sqrt{2}}\lambda\sum_i\langle\ell^+\ell^-\mathcal{M}|C_i\mathcal{O}_i|\mathcal{B}\rangle.
\end{align}
Contrary to the case of leptonic decays, we cannot simply write all operators as a product of two fermion currents ($\mathcal{O}\sim j_\ell\,j_q$), as there are also contributions from photonic and gluonic penguins and four quark operators which produce diagrams with a virtual photon (and in the latter case a quark loop). We therefore express the amplitude in this case schematically as 
\begin{align}
\hspace{-.5cm}\label{eq:schem_BMll}\mathcal{A}_{\mathcal{B}\to \mathcal{M}\ell^+\ell^-}\sim \mathcal{N}\Big(C\mathcal{F}^V(q^2)&+\frac{1}{q^2}\Big(C \mathcal{F}^T(q^2)- \mathcal{H} (q^2) \Big)\Big) \langle\ell^+\ell^-|j_\ell |0\rangle,
\end{align}
where $C$ represents the Wilson coefficients, $\mathcal{F}^V_i(q^2)$ and $\mathcal{F}^T_i(q^2)$ denote the vector and tensor form factors as defined either for the case of $\mathcal{M}$ being a pseudoscalar or vector meson in Eqs.~\eqref{eq:ffbasisP} and \eqref{eq:ffbasisV} respectively, and finally $\mathcal{\mathcal{H}} (q^2)$ denotes the non-local hadronic matrix element of four-quark operators, 
\begin{equation}
\mathcal{H}^\mu(q^2)\equiv i\int d^4x e^{iqx}\langle \mathcal{M}(k)|T\{j_\mu^{\rm em}(x),(C_1 \mathcal{O}_1+C_1 \mathcal{O}_2)(0)\}|\mathcal{B}(q+k)\rangle,
\end{equation}
where $j_\mu^{\rm em}=\sum_q Q_q \bar q \gamma_\mu q$ with $q$ being all accessible quarks, and $k$ the four-momentum of the final state meson $\mathcal{M}$. These non-local matrix elements are responsible for the dominant uncertainty for these decays. 

For the $B\to K$ transition, the most precise calculations of the form factors from LQCD are from the HPQCD~\cite{Bouchard:2013pna} and FNAL/MILC~\cite{Bailey:2015dka} collaborations, and from LCSR the state-of-the-art  2017 calculation~\cite{Khodjamirian:2017fxg} includes twist 5,6 corrections.
Form factors for decays to vector mesons being more complicated on the lattice, as they are unstable and therefore it is complicated to extract the relevant matrix element from correlation functions (see also  \cite{Horgan:2013hoa} for first LQCD results), one relies on the LCSR calculation presented in \cite{Straub:2015ica}, where results for $B\to\rho, B\to\omega$, $B\to K^*$ and $B_s\to\phi$ are also provided. Results using $B$ meson DAs are also available in \cite{Gubernari:2018wyi}. Recently there has also been interest in applying to LCSR technique to final states containing two mesons, and 
results for $B\to K\pi$ form factors can be found in \cite{Descotes-Genon:2019bud}, and for $B\to \pi\pi$ in  \cite{Cheng:2017sfk,Cheng:2017smj,Feldmann:2018kqr}. This helps in understanding the relation between the experimental measurement, and the contributions of  the $S$-wave and $P$-wave resonances.

Calculating the non-local matrix elements is only possible in certain limits. QCD factorization (QCDf) provides a framework to calculate these using a local operator product expansion (OPE) in the kinematic region where $q^2/m_\mathcal{B}^2$ is a good expansion parameter and where one is suitably far away from resonances.
In the high $q^2$ region, a local OPE can also be used, expanding in $1/q^2$, however due to the fact that one is above the open charm threshold, one relies on the principle of quark-hadron duality meaning that by choosing observables integrated over a sufficiently large region in $q^2$, the result from the local OPE should correspond to the physical result up to duality violating effects. We will mention other possibilities later.

Going to next-to-leading order in the perturbative expansion (and also for the weak annihilation diagrams at leading order), one finds non-factorizable contributions where the non-perturbative part cannot be absorbed into the form factors. These contributions can also be included via QCDf in the combined heavy-quark and large-energy (recoil) limit through convolutions of LCDAs of the initial and final meson with a perturbatively calculable hard kernel, schematically
	\begin{equation}
\langle \mathcal{M} \ell^+\ell^-|\mathcal{H}_{\rm eff}^{(q)}|\mathcal{B}\rangle \sim  \phi_\mathcal{B}^\pm \otimes T^{(q)} \otimes \phi_\mathcal{M} + \mathcal{O}(1/m_h),
		\label{eq:schem_QCDf}
	\end{equation}
where $\phi_{\mathcal{B}/\mathcal{M}}^\pm$ are the LCDAs of the heavy or light meson $\mathcal{B}$ or $\mathcal{M}$ and $m_h=m_b$ or $m_c$ as appropriate.
The QCDf corrections have been known for the $B\to K^{(*)}$ transition since 2001~\cite{Beneke:2001at} and for $B\to\rho/\omega$ since 2004~\cite{Beneke:2004dp}. The application to the $c\to u$ transition is more recent, the expressions were first calculated in \cite{Feldmann:2018kqr}.

One of the golden modes of LHCb is $B\to K^*\mu^+\mu^-$; a neutral current decay which is suppressed in the SM and boasts a four-body final state allowing the construction of a large number of angular observables.
It was generally accepted that QCDf could be applied in the region of $q^2$ between 1 and 6 GeV$^2$~\cite{Altmannshofer:2008dz,Bobeth:2008ij,Egede:2008uy}, though originally the authors of Ref.~\cite{Beneke:2001at} proposed the region between 1 and 7 GeV$^2$.
However, since the observation of the $B$ anomalies in certain observables of $B\to K^{(\ast)}\ell^+\ell^-$ discussed in Sect.~\ref{sec:intro}, there has been a heated debate in the community about the possible size of the non-perturbative effects in the low $q^2$ region. This has encouraged several analyses of the size of the non-perturbative contribution to the non-local matrix elements, and the effect on observables at low $q^2$. This has been done phenomenologically~\cite{Arbey:2018ics,Hurth:2020rzx,Jager:2012uw,Ciuchini:2015qxb},  
 in the case of $B\to K$, by including a sum over resonances modelled by Breit Wigners, which were fitted using $e^+e^-$ data~\cite{Lyon:2013gba}. The phase between the long and short distance components was also obtained from LHCb data~\cite{LHCb:2016due}.
An alternative approach, is to calculate the non-local matrix elements of four-quark operators using a light-cone OPE in the Euclidean~\cite{Khodjamirian:2010vf,Gubernari:2020eft}, and then relate these to the physical region using analyticity~\cite{Khodjamirian:2010vf,Bobeth:2017vxj,Gubernari:2020eft}. It is found that the uncertainty coming from these non-local matrix elements should not exceed $\mathcal{O}(10\%)$.

\paragraph{Branching ratios:} For the $b\to s$ transition, the differential branching ratios for $B^0\to K^{0\ast}$~\cite{Belle:2009zue,BaBar:2012mrf,CDF:2011tds,CMS:2015bcy,LHCb:2016eyu,Belle-II:2022fky}, $B\to K$~\cite{Belle:2009zue,BaBar:2012mrf,CDF:2011tds,LHCb:2016eyu}, $B_s\to\phi$~\cite{LHCb:2015wdu,LHCb:2021zwz} and $\Lambda_b\to \Lambda$~\cite{CDF:2011buy,LHCb:2015tgy} have all been measured. Note that the measurements from the $B$-factories are of the combined differential decay rates, combining both light lepton flavours $\ell=e$ and $\mu$, and isospin-related mesons $B^{+,0}\to K^{(\ast)+,0}$. LHCb (as well as CMS and CDF) on the other hand concentrates on final states to which the sensitivity is better in the more difficult hadronic environment, notably charged $K$ and $\pi$ mesons and muons in the final state. Note that the results from LHCb are now very precise, for the majority of decays the uncertainties are smaller than those on the theoretical predictions. In addition, for all channels except $\Lambda_b\to\Lambda$ the central values lie below the SM prediction, contributing to the list of the $b\to s$ ``anomalies".
While the $b\to d$ transitions are less studied, LHCb has also observed several of these for the first time: $B^+\to\pi^+\mu^+\mu^-$~\cite{LHCb:2015hsa}, $B^0\to\pi^+\pi^-\mu^+\mu^-$~\cite{LHCb:2014yov}, $B_s\to K^{\ast}\mu^+\mu^-$~\cite{LHCb:2018rym} and $\Lambda_b^0\to p\pi^-\mu^+\mu^-$~\cite{LHCb:2017lpt}, with measurements consistent with the predictions in the SM.

\paragraph{Isospin and CP asymmetries:} The isospin asymmetry is an asymmetry between the decay widths $\rGamma(\mathcal{B}^0\to \mathcal{M}^0\ell^+\ell^-)$ and $\rGamma(\mathcal{B}^+\to \mathcal{M}^+\ell^+\ell^-)$  and CP asymmetries between the decay width and the decay width of the CP conjugate process, are also measured by experiment~\cite{Belle:2009zue,BaBar:2012mrf,LHCb:2014cxe,LHCb:2014mit}. For decays of $B$ mesons to vector meson final states, predictions for the isospin asymmetry are available in a  QCDf/LCSR hybrid framework~\cite{Khodjamirian:2012rm,Lyon:2013gba}. As a consequence of the CKM structure, the direct CP asymmetry for $b\to d$ is expected to be larger than for $b\to s$, and LHCb has confirmed that this holds, and that the asymmetry is in good agreement with the prediction in the SM for the case of $B^+\to\pi^+\mu^+\mu^-$. While previous measurement of the $B\to K\ell\ell$ isospin asymmetry were in tension with the SM prediction, the most precise recent LHCb measurement agrees within 1.5\,$\sigma$. All other measurements of the isospin and CP asymmetries are in agreement with the theoretical predictions, the latter being extremely precise due to the cancellation of hadronic uncertainties.

\paragraph{Angular observables:} For decays to pseudoscalar final states the differential decay rate can be written as 
		\begin{equation}
		    \frac{d^2\rGamma(\mathcal{B}\to \mathcal{M})}{ds~ d\cos\theta} = N \lambda^{1/2} \beta\, \left[\,a_\ell +b_\ell\cos\theta +c_\ell\cos^2{\theta}\,\right],
		    \label{Eq:DecayWidthBtoP}
		\end{equation}
where $\theta$ is the angle between the directions of the $\ell^+$ and the $B$ in the dilepton rest frame, and $a_\ell$, $b_\ell$ and $c_\ell$ are angular coefficients. Therefore in full generality three independent angular observables can be constructed, the decay rate $\rGamma$, the flat term $F_{\rm H}$ and the forward backward asymmetry $A_{\rm FB}$~\cite{Bobeth:2008ij}:
\begin{equation}
\frac{d\rGamma}{ds} = 2 N \lambda^{1/2} \beta  \left[ a_\ell + \frac{c_\ell}{3} \right],  \,\, F_{\rm H}=\frac{a_\ell+c_\ell}{a_\ell+c_\ell/3},\,\, \mbox{and} \,\, A_{\rm FB}=\frac{ b_\ell}{2\left[ a_\ell+c_\ell/3 \right]}.
\end{equation}
However in the SM, and in the limit of massless leptons, we find $b_\ell= 0$ and
			$a_\ell = -c_\ell/\beta^2 $, and this set reduces to a single observable, the decay rate.
These angular observables have been measured by Belle~\cite{Belle:2009zue}, BaBar~\cite{BaBar:2006tnv,BaBar:2012mrf}, CDF~\cite{CDF:2011tds}, LHCb~\cite{LHCb:2014auh} and CMS~\cite{CMS:2018qih}, which are all in agreement with the SM; i.e.~the forward-backward asymmetry and the flat term (approximately) are compatible with zero.

For the decay to vector mesons $\mathcal{B}\to \mathcal{M}\ell^+\ell^-$, where $\mathcal{M}$ decays to $\mathcal{M}'\pi$, the four-body final state gives rise to many more observables~\cite{Kruger:1999xa,Kim:2000dq,Kruger:2005ep}:
\begin{align}
\nonumber\frac{d^4\rGamma(\mathcal{B}\to \mathcal{M}\ell^+\ell^-)}{d\cos\theta_\ell d\phi\,d\cos\theta_{\mathcal{M}'} dq^2}=&\frac{9}{32 \pi}\sum_i I_i(q^2)f_i(cos\theta_\ell,\phi,\cos\theta_{\mathcal{M}'})\\
\frac{d^4\rGamma(\overline{\mathcal{B}}\to\overline{ \mathcal{M}}\ell^+\ell^-)}{d\cos\theta_\ell d\phi\, d\cos\theta_{\mathcal{M}'} dq^2}=&\frac{9}{32 \pi}\sum_i \bar{I}_i(q^2)f_i(cos\theta_\ell,\phi,\cos\theta_{\mathcal{M}'}),
\end{align}
where $\theta_\ell$ is the angle between the $\mathcal{M}'$ and the $\mathcal{B}$ in the $\mathcal{M}$ rest frame; $\theta_\ell$ is the angle between the $\ell^+$ and the $B$ in the dilepton rest frame; and $\phi$ is the angle between the $\mathcal{M}'$, $\pi$ and the dilepton plane, where we follow \cite{Altmannshofer:2008dz,LHCb:2013zuf} for the angular definitions, see also \cite{Blake:2016olu} for definitions of the functions $f_i(cos\theta_\ell,\phi,\cos\theta_{\mathcal{M}'})$.These are further discussed and conversions between other conventions in the literature are given in \cite{Gratrex:2015hna}.
These observables may be CP averages or CP asymmetries~\cite{Altmannshofer:2008dz,Bobeth:2008ij},
\begin{equation}
S_i=\frac{I_i+\bar{I}_i}{d(\rGamma+\overline{\rGamma})/dq^2}\qquad\qquad A_i=\frac{I_i-\bar{I}_i}{d(\rGamma+\overline{\rGamma})/dq^2}.
\end{equation}
While for the case where  $\mathcal{M}=K*$, the decays are self-tagging, this is not the case for the $\rho$, $\phi$ meson, and a time-dependent flavour tagged CP analysis is required in order to differentiate between $S_i$ and $A_i$. These observables and their sensitivity to new physics have been in studied in detail in \cite{Altmannshofer:2008dz,Bharucha:2010bb}.
Since, several alternative observables have been proposed in the literature in order to optimise the experimental sensitivity and the theoretical uncertainty~\cite{Egede:2008uy,Descotes-Genon:2012isb,Descotes-Genon:2013vna}., for example the $P'$ observables~\cite{Descotes-Genon:2013vna},
\begin{equation}
P'_{4,5,6,8}=\frac{S_{4,5,7,8}}{2\sqrt{-S^s_2\,S^c_2}}.
\end{equation}

The angular observables have been measured  for $B^0\to K^{\ast 0}\mu^+\mu^-$ by the $B$-factories~\cite{BaBar:2006tnv,Belle:2016fev}, CDF~\cite{CDF:2011tds}, LHCb~\cite{LHCb:2013zuf,LHCb:2015svh,LHCb:2020lmf}, CMS~\cite{CMS:2015bcy,CMS:2017rzx,CMS:2020oqb} and ATLAS~\cite{ATLAS:2018gqc},
 and for $B^+\to K^{\ast +}\mu^+\mu^-$ in ~\cite{LHCb:2020gog}.
The great number of measurements performed is due to the fact that LHCb has observed a persistent anomaly in a particular bin in $q^2$ (4 to 6 GeV$^2$) of the observable $P'_5$ at low $q^2$.\footnote{A similar anomaly is seen in the 6 to 8 GeV$^2$ bin, however this bin is close to the charmonium region and there are conflicting ideas in the community concerning the theoretical uncertainties for this bin.} The latest and most precise measurements from LHCb~\cite{LHCb:2020lmf} find a deviation at the 2.5\,$\sigma$ level. The remaining observables are largely in agreement with the SM predictions~\cite{Straub:2015ica,Straub:2018kue,Horgan:2013hoa,Horgan:2013pva}. Note that the angular observables have also been measured for the $B^0\to K^{\ast 0} e^+ e^-$ case at low $q^2$~\cite{LHCb:2020dof}, but this was mostly with the aim to provide a measurement of the photon polarization.

In order to probe the $b\to s$ further, angular analyses have been performed by LHCb for $B_s\to\phi \mu^+\mu^-$ \cite{LHCb:2021xxq} and for $\Lambda_b\to\Lambda \mu^+\mu^-$ \cite{LHCb:2018jna}, but these channels do not provide access to the $P_5^{\prime}$ observable. However, for the case of $B^+\to K^{\ast +}\mu^+\mu^-$, $P_5'$ can be measured, and it displays behaviour similar to that of the $B^0\to K^{\ast 0}\mu^+\mu^-$~\cite{LHCb:2020gog}.

The prospects for the angular analyses of the rare semileptonic decays are promising, with the upgrade II, LHCb plan to measure the CP and isospin asymmetries at the \% level, and angular observables for $b\to s\mu^+\mu^-$ and $b\to d\mu^+\mu^-$ at sub-percent level~\cite{LHCb:2018roe}. 
While Belle II promise uncertainties on the angular observables and exclusive branching fractions of a few percent with 50 ab$^{-1}$, access to the electron final states and also angular resolution on the angle $\phi$ will be improved~\cite{Kou:2018nap}.
Belle II will further significantly improve the measurement of the branching fractions of the normalisation modes $B \to K^{(*)} J/\Psi$~\cite{Kou:2018nap}.

Note that here we have not discussed inclusive rare semileptonic decays, measured at the $B$-factories, as in the last ten years little theoretical progress has been made on this front, a review of the status of these modes can be found in \cite{Blake:2016olu}. As the Belle and BaBar measurements are close to being systematically limited, in order to make improvements at Belle II these uncertainties will need to be reduced.

\paragraph{Charm decays:} Rare semileptonic decays involving  the $c\to u\ell^+\ell^-$ transition could also provide valuable insight into the flavour-structure of BSM physics. LHCb has a rich program of searches, including both one and two mesons in the final state. While the theoretical framework for calculating these decays  is approximately as advanced as for the correponding $B$ decays discussed earlier, with NNLL WCs~\cite{deBoer:2017que}, precision LQCD form factors~\cite{Lubicz:2017syv,Chakraborty:2021qav} and QCD factorization corrections~\cite{Feldmann:2017izn},  in comparison to the $B$ decays, the $D$ decays are subject to larger long-distance contributions from intermediate $\rho$, $\omega$ and $\phi$ resonances. There has been a recent theory effort to improve the modelling of these long-distant effects, see~\cite{Bharucha:2020eup}. These decays can still therefore provide important constraints either from branching ratios (for the case where the BSM contribution exceeds that from the long-distance physics) or from angular observables or CP asymmetries insensitive to these long-distance effects, for a recent analyses for a single meson in the final state see \cite{deBoer:2015boa,Fajfer:2015mia,Feldmann:2017izn,Bharucha:2020eup}, and for two mesons in the final state see~\cite{Bigi:2011em,Cappiello:2012vg}. For the phenomenology of charmed baryon decays $\Lambda_c\to p\mu^+\mu^-$ see \cite{Meinel:2017ggx}.

\paragraph{Rare decays to neutrinos:}
The decays of a heavy meson into a light meson and a neutrino pair also provide important insight into BSM physics. These decays have been constrained at the $B$-factories~\cite{BaBar:2013npw,Belle:2013tnz,Belle:2017oht}, and while they are challenging at LHCb, they are now a priority at Belle II, incidentally the first physics paper published by Belle II was a limit on $B^+\to K^+\nu\bar\nu$~\cite{Belle-II:2021rof}.
Expressing the amplitude as
\begin{equation}
\langle\nu\bar\nu \mathcal{M}|\mathcal{H}_{\rm eff}|\mathcal{B}\rangle\sim \frac{4 G_F}{\sqrt{2}}\lambda\sum_i\langle\nu\bar\nu \mathcal{M}|C_i\mathcal{O}_i|\mathcal{B}\rangle,
\end{equation}
and again writing the operators as a product of two fermion currents, $\mathcal{O}\sim j_\ell\,j_q$, we can factorize the matrix element into a leptonic and hadronic part, schematically
\begin{align}
\nonumber\sim&\langle\nu\bar\nu |j^i_\ell |0\rangle\,\langle \mathcal{M}|j^i_q|\mathcal{B}\rangle\\
\sim&F_i \langle\nu\bar\nu |j^i_\ell |0\rangle,
\end{align}
where $F_i$ is the appropriate form factor. We then find
\begin{equation}
\langle\nu\bar\nu \mathcal{M}|\mathcal{H}_{\rm eff}|\mathcal{B}\rangle\sim \frac{4 G_F}{\sqrt{2}}\lambda F_i \sum_i C_i  \langle\nu\bar\nu |j^i_\ell |0\rangle,
\end{equation}
such that, since the factorization for decays to neutrinos is exact, the uncertainty is dominated by the form factors. The form factors are the same as those for semileptonic decays in Sect.~\ref{sec:FFs}.
The latest theoretical and experimental results for the branching ratios are collected in Tab.~\ref{tab:nunu}.

In light of the anomalies mentioned earlier, it is interesting to relate the $b\to s\nu\bar\nu$ decays to $b\to s\ell^+\ell^-$. This can be done either in specific models~\cite{Buras:2014fpa,Browder:2021hbl} or using effective field theories (EFTs)
~\cite{Calibbi:2015kma,Descotes-Genon:2020buf,Bause:2021cna}. Within EFTs one can use a symmetry to relate the charged lepton and neutrino couplings. In \cite{Bause:2021cna}, $SU(2)_L$ is used to relate the charged dilepton and dineutrino decays within Standard Model EFT (SMEFT), and on how within this framework one could use the neutrino modes to learn more about the anomalies.
Note that an angular analysis of $B\to K^*\nu\bar\nu$ would allow one to access observables such as the longitudinal fraction, sensitive to right-handed currents~\cite{Altmannshofer:2009ma}.
As for the $b\to d$ decays,  requiring the initial and final state mesons to be charged, the experimental bound is weaker due to the larger background from $B^+ \to\tau\nu$ decays. Predictions and the phenomenology were studied for $B\to\pi$  in \cite{Du:2015tda} and for $B\to\rho$ in \cite{Straub:2015ica}.
Note that the experimental bounds on decays to neutrinos also provide bounds on searches for long-lived particles or Dark Matter candidates, for a recent summary of constraints on dark particles see \cite{Blake:2016olu,Hostert:2020gou,Felkl:2021uxi}.

\begin{table}
\centering
\begin{tabular}{ccccc}
\hline
Mode & $\mathrm{BR_{SM}}$ &Ref.& $\mathrm{BR_{exp}}$ &Ref.\\
\hline
$B^+\to K^+\nu\bar\nu$ & $(4.23\pm 0.56)\,10^{-6}$ &\cite{Bause:2021cna}  & $<1.6\,10^{-5}$ & \cite{BaBar:2010oqg}\\
$B^0\to K^0\nu\bar\nu$ & $(4.23\pm 0.56)\,10^{-6}$ &\cite{Bause:2021cna}& $<4.9\,10^{-5}$ &\cite{BaBar:2010oqg}\\
$B^+\to K^{*+}\nu\bar\nu$ & $(8.24\pm 0.99)\,10^{-6}$ &\cite{Bause:2021cna} & $<4.0\,10^{-5}$ &\cite{Belle:2013tnz}\\
$B^0\to K^{*0}\nu\bar\nu$ & $(8.24\pm 0.99)\,10^{-6}$ &\cite{Bause:2021cna}  & $<5.5\,10^{-5}$ &\cite{Belle:2013tnz} \\
$B^0\to \pi^0 \nu\bar\nu$ & $(4.6\pm 0.5)\,10^{-6}$&\cite{Du:2015tda}  & $<6.9\,10^{-5}$ &\cite{Belle:2013tnz}\\
$B^0\to \rho^0 \nu\bar\nu$ & $(4.6\pm 0.5)\,10^{-6}$ &\cite{Straub:2015ica} & $<2.1\,10^{-4}$ &\cite{Belle:2013tnz} \\
\hline
\end{tabular}
\caption{Branching ratios of decays to neutrinos, theory predictions (where form factors are taken from Ref~ \cite{Bailey:2015dka} for $B\to\pi$, \cite{Straub:2015ica} for $B\to\rho$ and $B\to K^*$, and \cite{FermilabLattice:2015cdh} for $B\to K$) and measurements. \label{tab:nunu}}
\end{table}

\subsection{Radiative rare decays}
\label{sec:rad}

\paragraph{Inclusive decays:}
The decay $B\to X_s\gamma$ is one of the most precise constraints on models of BSM physics. It is measured at the $B$-factories, this is conventionally done for photon energies above a certain cut-off $E_0$, and then extrapolated to $E_0=1.6$ GeV where the theoretical prediction is most precise. The current experimental world average is $(3.32\pm 0.15)\,10^{-4}$~\cite{HFLAV:2019otj}, and 
the latest theory prediction for $E_0=1.6$ GeV is $(3.40 \pm 0.17) \,10^{-4}$~\cite{Misiak:2020vlo}, showing excellent agreement.
This relies on a perturbative calculation, known to NNLO up to an interpolation in $m_c$ for the contribution of the $O_{1,2}-O_7$ operators. This interpolation introduces a 3\% uncertainty, it is therefore a priority to perform the calculation at the physical value of $m_c$, and this work is ongoing, see \cite{Misiak:2017woa,Misiak:2020vlo}. Note that missing four-body contributions at NLO, small but required for complete NNLO result were calculated in  \cite{Huber:2014nna}.

The contribution of non-perturbative effects for the leading $O_7-O_7$ contribution can be expressed via an OPE, 
\begin{equation}
\rGamma=\sum_{n=0}^\infty\frac{1}{m_b^n}\sum_k C_{k,n}\langle O_{k,n}\rangle,
\end{equation}
where $\langle O_{k,n}\rangle$ are the matrix elements of heavy quark effective theory (HQET) operators. At $n=0$ this corresponds to the perturbative result, at $n=1$ the coefficient vanishes, and at $n=2$ the Wilson coefficient is known to $\mathcal{O}(\alpha_s)$ and matrix element corresponds to the leading shape function. For $n=3,4,5$ the Wilson coefficients are known at leading order in the perturbative expansion, but the matrix elements are less known, corresponding to the sub-leading shape functions. It is important to improve the knowledge on these using data from Belle II, see \cite{Gambino:2016jkc,Heinonen:2016cwm,Gunawardana:2017zix,Bernlochner:2020jlt}.

Note that there are also non-perturbative effects arising from other operators, e.g.~$O_{1,2}-O_7$, $O_7-O_8$, $O_8-O_8$ arising at $\mathcal{O}(\Lambda_{\rm QCD}/m_b)$, which are more complicated, and can give rise to resolved photons, for example from $b\to s g\to s \bar{q} q\gamma$~\cite{Benzke:2010js}. These result in a total uncertainty $\sim 5\%$ on the total rate, which was claimed to be irreducible. However, more recently it was shown that this uncertainty can in fact be reduced~\cite{Bernlochner:2020jlt}. This latest analysis from the SIMBA collaboration consisted of fitting the shape function to data, expanding it in a suitable basis such that the approach is model independent~\cite{Ligeti:2008ac}. This treatment of the shape function, as opposed to the more conventional model dependent approach~\cite{Benson:2004sg,Lange:2005yw,Andersen:2005mj,Gambino:2007rp,Aglietti:2007ik}, allows a simultaneous fit to it and $|C_7|$, exploiting all available experimental information about the $B\to X_s\gamma$ spectrum and the theoretical knowledge about the perturbative contributions. It is found that while the uncertainty from the extrapolation in $m_c$ is marginally relevant, that from the resolved contributions is not relevant at the current level of accuracy.

\paragraph{Exclusive decays:}
While the inclusive decay rate provides an excellent test of the SM for $b\to s\gamma$, the exclusive channels provide additional observables and information due to the fact that the final state is fully reconstructed. Here by exclusive rare radiative decays we refer to FCNC decays of the $b$ or $c$ hadron, for example in the case of $b\to s$ the channels of interest are $B\to K^*\gamma$, $B\to K_1\gamma$ $B_s\to\phi\gamma$ or $\Lambda_b\to\Lambda\gamma$, for $b\to d$ these include $B_s\to K^*\gamma$ and $B\to\omega\gamma$ and for $c\to u$  $D\to\rho\gamma$ or $D_s\to K^*\gamma$.
One of the major motivators in designing observables for these decays is to obtain information about the photon polarisation, for example via the up-down asymmetries exploiting the four-body final state in $B\to K_1(1400)(\to K\pi\pi)\gamma$~\cite{Gronau:2001ng,Gronau:2002rz,Kou:2010kn,Kou:2016iau} or the three-body final state in  $\Lambda_b\to\Lambda^{(\ast)} \gamma$\cite{Hiller:2001zj,Legger:2006cq}. Other important observables are the isospin asymmetry and the direct and indirect CP asymmetries.
In order to see how we can access the photon polarisation, let us consider the expression for the time-dependent decay rate:
\begin{align}
\label{eq:GammaBMgamma}
\nonumber\rGamma(\mathcal{B}\to \mathcal{M}\gamma)(t)\sim e^{-\rGamma t} \bigg(\cosh(\Delta\rGamma/2)-\mathcal{A}_{\Delta\rGamma}\sinh (\Delta\rGamma/2)&\\
\pm\mathcal{C}_{\mathcal{M}\gamma}\cos(\Delta m\, t)\mp \mathcal{S}_{\mathcal{M}\gamma}\sin (\Delta m\,t)&\bigg),
\end{align} 
where $m$ and $ \rGamma$ are the mass and total decay width of the decaying hadron $\mathcal{B}$, $\mathcal{M}$ is the hadron in the final state and $\mathcal{A}_{\Delta\rGamma}$, $\mathcal{C}_{\mathcal{M}\gamma}$ and $\mathcal{S}_{\mathcal{M}\gamma}$ depend on the photon polarisation~\cite{LHCb:2018roe}.  Access to these quantities is therefore the priority in selecting channels and designing observables for the rare radiative decays.

The prediction for the two-body decays is the same as that for the semileptonic rare decays in the kinematic limit $q^2=0$. In this limit the rate depends on a single tensor form factor (see Sect.~\ref{sec:FFs}), however contributions from the non-local matrix elements defined in Eq.~\eqref{eq:schem_BMll} must also be taken into account. These give rise to the strong phases and also to isospin violating effects, and therefore need to be well understood in order to provide reliable predictions for the CP and isospin asymmetries. As in the semileptonic case, QCD factorization provides a framework to calculate these contributions, but partial LCSR results also exist (see e.g.~ \cite{Ball:2006eu,Dimou:2012un}) which include leading $\mathcal{O}(\Lambda_{\rm QCD}/m_b)$ effects and avoid the endpoint divergences found in QCDf, permitting results to be obtained in a hybrid QCDf/LCSR approach~\cite{Lyon:2013gba}.

The latest predictions for the isospin asymmetries are $(4.9\pm 2.6)\%$ for $B\to K^*\gamma$ and $(5.2\pm 2.8)\%$ for $B\to\rho\gamma$, where the latter is in slight tension with the most recent HFLAV average of $-46^{+17}_{-16}\%$~\cite{HFLAV:2019otj}.
At the LHC, in order to access the photon polarisation via CP asymmetries, the most sensitive observables in $\mathcal{A}_{\Delta\rGamma}$ for $B_s\to K^{\ast 0}$ which was recently measured to be $-0.98^{+0.46}_{-0.52}\mathrm{(stat)}^{+0.23}_{-0.20}\mathrm{(syst)}$.
The  up-down asymmetry in $B\to K\pi\pi\gamma$ was measured by LHCb in \cite{LHCb:2014vnw} in four different mass regions. In these different regions, different combinations of resonances contribute, whereas predictions are provided for specific resonances, however a recent BaBar measurement ~\cite{BaBar:2015chw} could provide the necessary information to disentangle the photon polarization.

In the future, with the full 50 ab$^{-1}$ dataset Belle II will measure many branching ratios and asymmetries to high precision. For example, the time dependent CP asymmetries will be measured, with 3\% precision on $\mathcal{S}_{K*\gamma}$ and 6.4\% precision on $\mathcal{S}_{\rho\gamma}$~\cite{Belle-II:2018jsg}.
The isospin asymmetries for $K^*\gamma$ and $\rho\gamma$ will also be measured to 0.53\% and 1.9\% respectively (where the latter assumes the current central value to be confirmed)~\cite{Belle-II:2018jsg}. 

As for LHCb, it is forecast that the statistical uncertainty on $\mathcal{A}_{\Delta\rGamma}$ can be reduced to $\sim 0.07$ with 50 fb$^{-1}$ and $\sim 0.02$ with 300 fb$^{-1}$~\cite{LHCb:2018roe}. Further, the time-dependent decay rate of $B_s\to K_s^0\pi^+\pi^-\gamma$ should provide a competitive measurement of $\mathcal{S}_{K_s^0\pi^+\pi^-\gamma}$. Finally the photon polarisation parameter should be determined via $B^+\to K^+\pi^-\pi^+\gamma$ to 1\% with 300 fb$^{-1}$ of data~\cite{LHCb:2018roe}.

Exclusive radiative $D$ decays suffer from large long-distance QCD contributions, such that the cleanest observables are the CP asymmetry, for the state-of-the-art theory predictions see \cite{deBoer:2017que}. In 2016, Belle obtained a limit on $A_{CP}$ for $D\to V\gamma$ in \cite{Belle:2016mtj}. In the future at LHCb and Belle II, the most promising channels are $D_{(s)}\to V\gamma$, with $V=\rho, K*,\phi$ at Belle II to be measured at the percent level~\cite{Kou:2018nap} and $V=\phi$ at LHCb~\cite{LHCb:2018roe} to be measured at the few percent level.
With the Upgrade II at LHCb, efforts will further be made to further study the photon polarization via the more challenging channel $D\to K\pi\pi\gamma$, for details see Ref.~\cite{LHCb:2018roe}.

\section{Charged current transitions}
\label{sec:ckm}
In this section we will first discuss the inclusive and exclusive determinations of  $V_{cb}$ and  $V_{ub}$, followed by a note on radiative semileptonic decays.

\subsection{Inclusive $V_{cb}$}
\label{sec:InVcb}
The determination of $V_{cb}$ via inclusive $B\to X_c\ell\nu$ decays relies on the idea that by summing over all possible final states we can remove the dependence on long-distance physics of the final state. The OPE then provides a framework for describing the non-perturbative physics via matrix elements of local operators of dimensions $d\geq 5$, for which the Wilson coefficients can be calculated perturbatively, expanding up to a given order in $\alpha_s$. The decay width then takes the form (see e.g.~\cite{Gambino:2015ima})
\begin{align}
\nonumber\rGamma \propto\, &\, |V_{cb}|^2 m_b^5\Big\{\rGamma_0^{(0)}+\frac{\alpha_s}{\pi}\,\rGamma_0^{(1)}
+\left(\frac{\alpha_s}{\pi}\right)^2\rGamma_0^{(2)}+\left(\frac{\alpha_s}{\pi}\right)^3\rGamma_0^{(3)}\\
\nonumber&+\frac{\mu_\pi^2}{m_b^2}\left(\rGamma^{(\pi,0)}+\frac{\alpha_s}{\pi}\,\rGamma^{(\pi,1)}\right)
+\frac{\mu_G^2}{m_b^2}\left(\rGamma^{(G,0)}+\frac{\alpha_s}{\pi}\,\rGamma^{(G,1)}\right)\\
&+\frac{\rho_D^3}{m_b^3}\left(\rGamma^{(D,0)}+\frac{\alpha_s}{\pi}\,\rGamma_0^{(1)}\right)+\mathcal{O}\left(\frac{1}{m_b^4}\right)+\ldots\Big\},
\end{align}
where $\rGamma_0=A_{ew}|V_{cb}|^2G_F^2 m_b^5(1-8\rho+8\rho^3-\rho^4-12\rho^2\ln\rho)/(192\pi^3)$ for $\rho=m_c^2/m_b^2$ and $A_{ew}\simeq 1.1014$, for more details see \cite{Gambino:2015ima}.
Two non-perturbative matrix elements appear at $\mathcal{O}(1/m_b^2)$:
\begin{align}
\mu_\pi^2(\mu)=\frac{1}{2 M_B}\langle B|\bar{b}_v\overrightarrow{\pi}^2b_v|B\rangle_\mu
\\
\mu_G^2(\mu)=\frac{1}{2 M_B}\langle B|\bar{b}_v\frac{i}{2}\sigma_{]mu\nu}G^{\mu\nu}b_v|B\rangle_\mu
\end{align}
where $\overrightarrow{\pi}=-i\overrightarrow{D}$ and $b_v(x)=e^{-i m_bv\cdot x}b(x)$ is the $b$-quark field with the high-frequency modes removed.
These quantities can be extracted from a fit to the central moments of the lepton energy and the hadronic invariant mass distribution~\cite{Alberti:2014yda,Gambino:2016jkc}.

The leading term $\rGamma_0$ is simply given by the partonic result, independent of hadronic physics, and is known at third order in $\alpha_s$~\cite{Fael:2020tow,Bordone:2021oof} (see \cite{Gambino:2020jvv} for a full list of citations).
The next terms in the heavy quark expansion appear at order $1/m_b^2$, and the coefficients $\rGamma^{(\pi)}$ and $\rGamma^{(G)}$ of the relevant matrix elements $\mu_\pi^2(\mu)$ and $\mu_G^2(\mu)$ are known at first order in the expansion in $\alpha_s$~\cite{Becher:2007tk,Alberti:2012dn,Alberti:2013kxa,Mannel:2014xza,Mannel:2015jka}. 
Finally, at order $1/m_b^3$ the matrix elements $\rho_D$ appears, for which the coefficient $\rGamma^{(D)}$ is known at first order in $\alpha_s$~\cite{Gremm:1996df,Mannel:2019qel}.

Based on this, new fit results in \cite{Bordone:2021oof} obtain a result for $V_{cb}$ with small dependence on assumptions for theoretical correlations and the choice of inputs; the uncertainty is at the $1.2\%$ level. The kinetic mass scheme is used for $m_b$,  latest FLAG results for $\bar{m}_b(\bar{m}_b)$ and $\bar{m}_c(3~\mathrm{GeV})$ are used. From the fit one can obtain a determination of the OPE parameters with 15-20\% level uncertainties. Note that on excluding $\bar{m}_b(\bar{m}_b)$ from the fit, one obtains a competitive determination of this quantity which is compatible with the FLAG value.

Higher order corrections $\mathcal{O}(1/m_b^{4,5})$ were studied in \cite{Gremm:1996df,Dassinger:2006md,Mannel:2010wj,Bigi:2005bh} where the Wilson coefficients are calculated at tree level, however the knowledge about the large number of corresponding matrix elements is limited. The Lowest Lying State Approximation (LLSA)~\cite{Mannel:2010wj} provides a means to estimate the size of $\mathcal{O}(1/m_b^4,1/m_b^5)$ effects, and reparameterization invariance allows one to express these numerous matrix elements in terms of a reduced set. It was suggested that one could fit these matrix elements using the $q^2$ moments of the inclusive distribution. The $q^2$ moments are reparameterization invariant quantities and depend on this reduced set of matrix elements~\cite{Fael:2018vsp}. Recently both Belle~\cite{Belle:2021idw} and Belle II~\cite{Belle-II:2022fug} have measured these to high precision and a fit has been performed to the combined data~\cite{Bernlochner:2022ucr}.

On the theoretical side, the most pressing next step would be to calculate the $\mathcal{O}(\alpha_s/m_b^3)$ contribution. Another promising direction would be to use LQCD to provide information about differential distributions and moments~\cite{Gambino:2020crt,Gambino:2022dvu}. While it is difficult to calculation the semileptonic width at the physical $b$ mass, this could provide information about the non-perturbative quantities which cannot be precisely measured by experiment.
For the moment, the dominant uncertainty comes from the measurement of the moments and the semileptonic branching fraction~\cite{Bernlochner:2022ucr}, and these will be made more precise from Belle II~\cite{Kou:2018nap}.

\subsection{Exclusive $V_{cb}$}
\label{sec:ExVcb}
\begin{figure}
\begin{center}
\includegraphics[width=.48\textwidth]{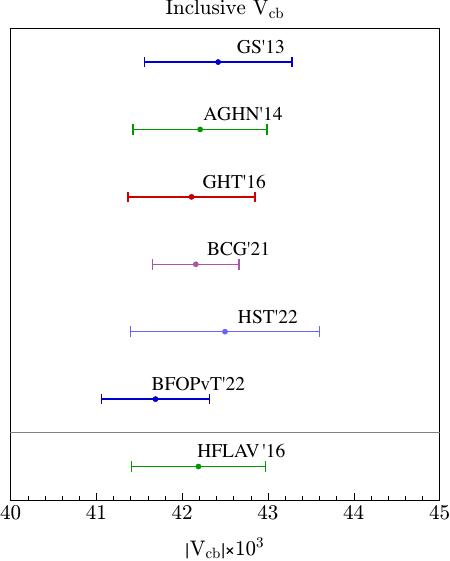}\hspace{.35cm}
\includegraphics[width=.46\textwidth]{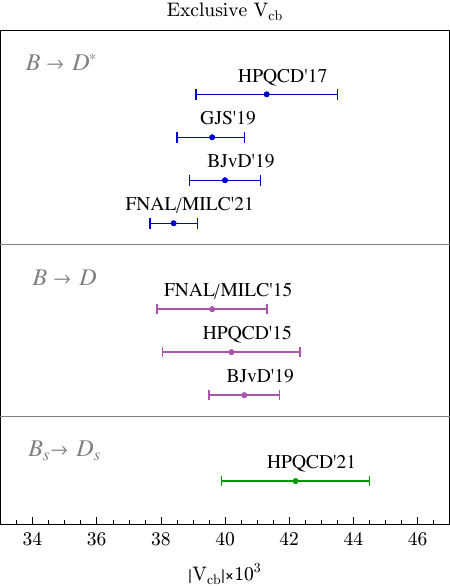}
\caption{We summarise the most recent determinations for inclusive $V_{cb}$ (left) (GS'13~\cite{Gambino:2013rza}, AGHN '14~\cite{Alberti:2016fba}, GHT'16~\cite{Gambino:2016fdy}, BCG'21~\cite{Bordone:2021oof}, HST'22~\cite{Hayashi:2022hjk} and BFOPvT'22~\cite{Bernlochner:2022ucr}) and compare this to the HFLAV experimental average from 2016~\cite{HFLAV:2016hnz} and for exclusive $V_{cb}$ (right), from $B\to D$ (HPQCD'17~\cite{Harrison:2017fmw}, GJS'19~\cite{Gambino:2019sif} and BJvD'19~\cite{Bordone:2019vic}), from $B\to D$ (FNAL/MILC'15~\cite{MILC:2015uhg}, HPQCD'15~\cite{Na:2015kha} and  BJvD'19~\cite{Bordone:2019vic}), and from $B_s\to D_s$ (HPQCD'21~\cite{Harrison:2021tol})\label{fig:Vcb}}
\end{center}
\end{figure}

The exclusive determination of $V_{cb}$ has been the source of much discussion, particularly over the past five years. The most accurate determination comes from the $B\to D^*$ channel, using LQCD results for the form factors from ~\cite{FermilabLattice:2021cdg} and a combination of Belle and BaBar data. This result is in tension with that from $B\to D$ (see e.g.~\cite{MILC:2015uhg,Na:2015kha}), and with the inclusive determination discussed in the following subsection, the latter being under good theoretical control, including third order corrections in $\alpha_s$ and $1/m_b$.  Alternative determinations could help clarify the situation, e.g.~from $B_s\to D^{(*)}_s\ell\nu$~\cite{Monahan:2018lzv,McLean:2019qcx,Harrison:2021tol}, as well as $\Lambda_b\to \Lambda_c\ell\nu$~\cite{Detmold:2015aaa}, both being actively pursued by LHCb and LQCD collaborations. The latest results for exclusive $V_{cb}$  are summarised in Fig.~\ref{fig:Vcb}.

The existing calculations for the form factors include results from LCSR~\cite{Gubernari:2018wyi} in the maximum recoil region and LQQCD for $B\to D$ ~\cite{MILC:2015uhg,Na:2015kha} and $B\to D^*$ at zero~\cite{FermilabLattice:2014ysv} and finite recoil~\cite{FermilabLattice:2021cdg} respectively. Latest experimental data includes results from Belle in 2017~\cite{Belle:2017rcc} and 2018~\cite{Belle:2018ezy}.
 In 2017, on analysing Belle data it was noticed that using a different parameterisation (BGL instead of CLN, see Sect.~\ref{sec:FFs}) of the $q^2$ dependence of the form factor improved the agreement with $V_{cb}^{\rm incl.}$~\cite{Bigi:2017njr,Bernlochner:2017jka,Grinstein:2017nlq,Jaiswal:2017rve}, however it was observed that in fact the slope of the form factors in the BGL fit were incompatible with heavy quark effective theory predictions~\cite{Bernlochner:2017jka,Bigi:2017jbd}. Therefore the consensus was that more experimental information about the spectrum in $q^2$, and/or a Lattice calculation of the form factor at non-zero recoil were needed before the results could be considered conclusive~\cite{Bordone:2019vic,Gambino:2019sif}. An alternative approach was therefore also advocated, involving an analysis of the full set of ten form factors in the heavy quark expansion (HQE) including corrections at order ($\alpha_s, 1/m_b,1/m_c^2)$~\cite{Bordone:2019vic}.

Since then there has been much progress both from experimental and from Lattice groups, in particular the calculation of the $B\to D^\ast$ form factors by two independent Lattice collaborations at non-zero recoil.  In addition, a recently updated LCSR calculation of the form factor using the $B$ meson distribution amplitude (DA) provides information on the full set of form factors, albeit for $q^2\lesssim 0$~\cite{Gubernari:2018wyi}.  However, the situation is not yet entirely resolved as the slopes of the form factors from Fermilab-MILC~\cite{FermilabLattice:2021cdg} and experiment are not in complete agreement with each other nor with HQE predictions. Here future results from the JLQCD collaboration will provide interesting complementary information.
On the other hand, for $B\to D$ the shape of the form factors from Lattice, LCSR, experiment and HQE show good compatibility. An upcoming BaBar result will further improve the accuracy of $V_{cb}$ in this channel. In addition, latest results on $V_{cb}$ from $B_s\to D_s^*$ form factors at non-zero recoil~\cite{Harrison:2021tol} and the LHCb measurement~\cite{LHCb:2020cyw} also reduce the tension. Note that the form factors for $B\to D_1(2420)$ and $B\to D_1'(2430)$ were also recently calculated in LCSR~\cite{Gubernari:2022hrq}, and for $B_c\to J/\Psi$  by the HPQCD collaborations, employing the NRQCD and HISQ actions for the $b$ quark and the HISQ action for the $c$ quark~\cite{Harrison:2020gvo}. These additional channels providing even more possibilities to probe $V_{cb}$ and lepton universality ratios.

The future prospects for the exclusive determination are good; a new HQE analysis is in order taking into account the $B_s\to D_s^\ast$ result from LHCb and the latest Lattice results at non-zero recoil for $B\to D^\ast$,  particularly following imminent updates from Belle II. Note that when 1 ab$^{-1}$ dataset is collected, the systematic uncertainties will dominate. Results of such a fit will hopefully help in resolving the difference amongst the exclusive $V_{cb}$ determinations, and perhaps also shed light on the inclusive-exclusive discrepancy.

\subsection{Inclusive $V_{ub}$}
\label{sec:InVub}
In the same way that one can obtain $V_{cb} $  from inclusive channels using a local OPE, one can also determine $V_{ub}$.
However, due to the large charm background one needs to impose experimental cuts, which reduces the degree of ``inclusiveness" of the prediction. 
In the region where the invariant mass of the hadronic system is less than the $D$ meson, the decay spectra are described using the non-local OPE, where perturbative coefficients are convoluted with non-perturbative ``Shape Functions"(SFs). This is known as the Shape Function region. For a recent review, see~\cite{Gambino:2020jvv}.
In the heavy quark limit, there is a single SF, but on adding higher order corrections in the $1/m_b$ expansion there are several SFs.
The perturbative coefficients can be factorized into hard and jet functions such that, schematically, the factorization form of the decay rate in the SF region is given by~\cite{Gambino:2020jvv}
\begin{equation}\label{eq:factoVubInc}
d\rGamma\sim H\cdot J\otimes S+\frac{1}{m_b}\sum_i h\cdot J_0\otimes s_i+\frac{1}{m_b}\sum_k h\cdot j_k\otimes S+\mathcal{O}(1/m_b^2)
\end{equation}
where $H$ is the leading power hard function, $J$ the leading power jet function, which are both known up to $\mathcal{O}(\alpha_s^2)$, $J_0$ is the $\mathcal{O}(\alpha_s^0)$ contribution to $J$, $h$ is the contribution to $H$ at $\mathcal{O}(\alpha_s)$ and $j_k$
There are several different approaches to the extraction of $V_{ub}$ from the measured inclusive spectrum, three of which are included by HFLAV in their averages:
\begin{itemize}
\item  BLNP: This is a precision extraction of $V_{ub}$ via simultaneous fits to $B\to X_u\ell\nu$ and $B\to X_s\gamma$.
The first two term in Eq.~\ref{eq:factoVubInc} are included, at $\mathcal{O}(\alpha_s)$ for the first term and $\mathcal{O}(\alpha_s^0)$ for the second term. Note that $H$ is evolved from the hard to jet scales in order to to resum Sudakov logarithms. The uncertainty is then estimated by comparing $\sim$ 700 models.
In order to model the sub-leading SFs, an exponential or Gaussian model is adopted, constrained by the first two moments of $S$ obtained from the global fit to HQE parameters in the kinetic scheme. Higher order corrections to $H$ and $J$ , the running from the hard to jet scale, and  as well as corrections to the description of $B\to X_S\gamma$ are not yet included.
\item GGOU: Perturbative coefficients are calculated to $\mathcal{O}(\alpha_s^2\beta_0)$ in the kinetic scheme.
There are three sub-leading SFs containing the effect of RGE evolution in the SF region, fitted from semileptonic fits. Here the uncertainty is estimated by comparing $\sim$ 100 models.
\item DGE: The leading  SF is obtained perturbatively via a resummation of running coupling corrections in the Sudakov exponent. 
\end{itemize}
The latest results for $V_{ub}$ obtained via these three methods using Belle data ~\cite{Belle:2021eni} show a large degree of model dependence, as seen in Fig.~\ref{fig:Vub}. To reduce the dependence on the sub-leading SFs, an alternative approach has been suggested by two independent groups, NNVub~\cite{Gambino:2016fdy} and SIMBA, whereby these are fitted using differential distributions in $B\to X_u\ell\nu$, e.g.~the lepton energy or various invariant mass distributions. This can additionally validate SF models or perturbative approaches (i.e.~DGE). While Belle has already measured the differential spectra~\cite{Belle:2021ymg}, upcoming results from Belle II will provide the necessary information to obtain precise results for inclusive $V_{ub}$. Quantitatively, the expected sensitivity of Belle II to inclusive $V_{ub}$ will be around 2\% with the full data sample of 5 ab$^{-1}$~\cite{Kou:2018nap}.
It is therefore important that theory errrors are reduced as far as possible to match this uncertainty, and in anticipation of this, in \cite{Capdevila:2021vkf} $\mathcal{O}(\alpha_s)$ corrections to the Wilson coefficients of dimension 5 operators were calculated.

\subsection{Exclusive $V_{ub}$}
\label{sec:ExVub}
\begin{figure}
\begin{center}
\includegraphics[width=.47\textwidth]{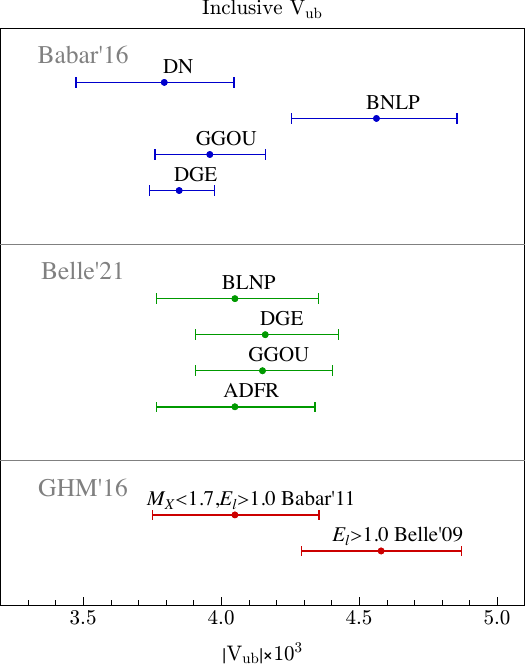}\hspace{.3cm}
\includegraphics[width=.47\textwidth]{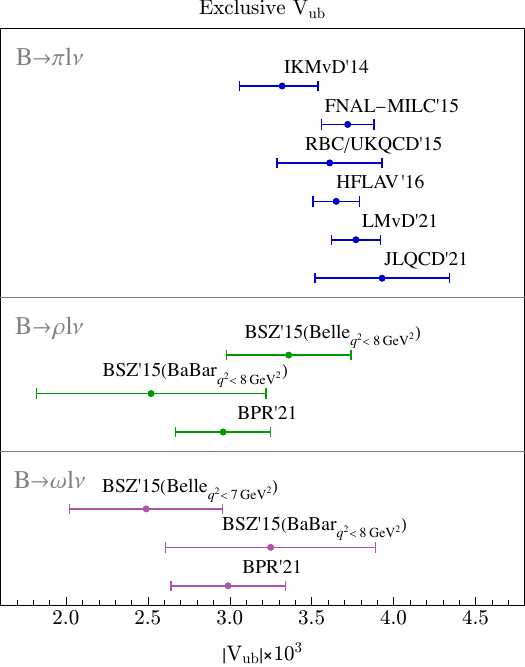}
\caption{We summarise the most recent determinations for inclusive $V_{ub}$ (left), from BaBar'16~\cite{BaBar:2016rxh}, Belle'21~\cite{Belle:2021ymg}, (where DN, BLNP, GGOU, DGE and ADFR are different theoretical methods described in the text) and GHM'16~\cite{Gambino:2016fdy} and for exclusive $V_{ub}$ (right), from $B\to\pi\ell\nu$ (IKMvD'14~\cite{Imsong:2014oqa}, FNAL/MILC'15~\cite{FermilabLattice:2015mwy}, RBC/UKQCD'15~\cite{Flynn:2015mha}, HFLAV'16~\cite{HFLAV:2016hnz}, LMvD'21~\cite{Leljak:2021vte} and JLQCD'21~\cite{Colquhoun:2022atw}), from $B\to\rho\ell\nu$ and $B\to\omega\ell\nu$ (BSZ'15~\cite{Straub:2015ica} and BPR'21~\cite{Bernlochner:2021rel}).  \label{fig:Vub}}
\end{center}
\end{figure}

The measurement of $V_{ub}$ requires the measurement of the differential branching ratio of the relevant semileptonic decay from experiment as a function of $q^2$, the lepton neutrino invariant mass, and the form factors from Lattice QCD at high $q^2$ and/or QCD sum rules on the light cone at low $q^2$.
In Fig.~\ref{fig:Vub} we compare various determinations of exclusive $V_{ub}$ since 2012.
The golden channel for the exclusive determination of $V_{ub}$ is $B\to\pi$, where the theoretical understanding is at an advanced stage and the experimental measurement was one of the major successes of the $B$ factories.
The LQCD results employing $N_f=2+1$ dynamical configurations include HPQCD in 2006~\cite{Dalgic:2006dt}, FNAL/MILC~\cite{FermilabLattice:2015mwy} and RBC/UKQCD in 2015~\cite{Flynn:2015mha}. These were averaged in the 2021 FLAG review~\cite{Aoki:2021kgd}.
The latest result for the $B\to\pi$ form factors on the Lattice is from the  JLQCD group, using M\"{o}bius  domain Wall fermions~\cite{Colquhoun:2022atw}, using a fully-relativistic lattice fermion action, and $N_f=2+1$.
Results from the Fermilab-MILC collaboration should appear soon, where the form factors are calculated with $N_f=2+1+1$, the first such calculation for the vector form factor. 

Since 2012, updates to the LCSR calculation include a NNLO ($\mathcal{O}(\alpha_s^2\beta_0)$) calculation of $f_+(0)$ was performed, with the result $f_+(0)= (0.262^{+0.020}_{-0.023})$ with uncertainties $\lesssim 9\%$~\cite{Bharucha:2012wy}. This calculation tested the argument that radiative corrections to $f_+f_B$ and $f_B$ should cancel when both calculated in sum rules (the 2-loop contribution to $f_B$ in QCDSR is sizeable). It was found that despite $\sim 9\%$ $\mathcal{O}(\alpha_s^2\beta_0)$ change to $f_B$, the effect on $f_+(0)$ was only $\sim 2\%$. Further twist 5 and 6 corrections were included in \cite{Khodjamirian:2017fxg}. The very small size of these corrections shows that the form factor converges very well in the 
expansion in twist, and higher twist corrections are not required..
More recently unitarity bounds and extrapolation were used to perform a Bayesian analysis of the form factor $f_+(q^2)$ for $B\to\pi$~\cite{Imsong:2014oqa}. This method was extended to a study including the $B$ meson distribution amplitude in \cite{Gubernari:2018wyi}.
A further analysis of $B\to\pi$ was performed in 2021 with a Bayesian approach to the uncertainties, where all the three form factors were considered and updated fits were performed with the most precise LQCD results~\cite{Leljak:2021vte}.
The prospects for measuring $V_{ub}$ at Belle II are exciting, with sensitivities at the 2-3\% level expected with 5 ab$^{-1}$~\cite{Kou:2018nap} (assuming LQCD projected uncertainties from~\cite{Kou:2018nap}).

Due to the lack of agreement between inclusive and exclusive results for $V_{ub}$, it is important to utilise additional channels. The $B_s\to K$ is an interesting possibility, particularly given the sensitivity of LHCb to this channel. Latest results can be found in \cite{Khodjamirian:2017fxg} from LCSR (see also \cite{Gubernari:2018wyi}) and \cite{Bouchard:2014ypa,Flynn:2015mha,FermilabLattice:2019ikx} from LQCD. A recent approach to $B \to \pi \ell \nu_\ell$ and $B_s \to K \ell \nu_\ell$ form factors using unitarity and LQCD  can be found in \cite{Martinelli:2022tte}.

However, the LHCb measurement is a ratio of the branching ratio of $B_s\to K\ell\nu$ to that of $B_s\to D\ell\nu$.
For $B$ decays to vector mesons, while LCSR results for $\rho$ and $\omega$ are available using the final state light cone distribution amplitudes (LCDAs)~\cite{Straub:2015ica} and the $B$ meson LCDA~\cite{Gubernari:2018wyi}, the situation is less advanced on the LQCD side.
In \cite{Bernlochner:2021rel} , the authors performed a fit to the combined measurements of the $B\to\rho\ell\nu$ and $B\to\omega\ell\nu$ spectra from Belle and BaBar, obtaining updated results for $V_{ub}$.
One manner to obtain an improved determination via the vector resonances is to obtain a better understanding of the full $B\to\pi\pi\ell\nu$ decay~\cite{Faller:2013dwa,Feldmann:2018kqr}. In fact the full $B\to\pi\pi\ell\nu$ spectrum was recently measured by Belle~\cite{Belle:2020xgu}, modelled in terms of four resonances the $\rho$, the $\rho'$, $f_0 (500)$ and the  $f_2 (1270)$. The form factors for this channel are not yet accessible via LQCD. However in LCSR, results have been obtained both using the $\pi\pi$~\cite{Hambrock:2015aor}  and the $B$-meson ~\cite{Cheng:2017sfk} distribution amplitudes.

Finally, one can measure $V_{ub}$ via the baryonic channel $\Lambda_b\to p\ell\nu$, again to be measured at LHCb via the ratio of branching fractions $\Lambda_b\to p\ell\nu$ to $\Lambda_b\to \Lambda_c\ell\nu$. Here Lattice QCD has presented precise predictions for the form factors~\cite{Detmold:2015aaa}. Independent results on these baryonic decays will be an important check of the value of $V_{ub}$ obtained.

%\begin{figure}
%\begin{center}
%\includegraphics[width=.5\textwidth]{fpf0_compare_qsq_0_only.pdf}
%\caption{\label{fig:BstoK}}
%\end{center}
%\end{figure}

\subsection{$|V_{ub}|/|V_{cb}|$}
\label{sec:VubVcb}
\begin{figure}
\begin{center}
\includegraphics[width=.465\textwidth]{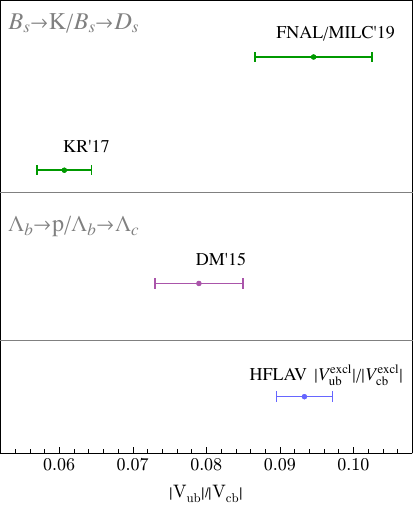}\hspace{.4cm}
\includegraphics[width=.487\textwidth]{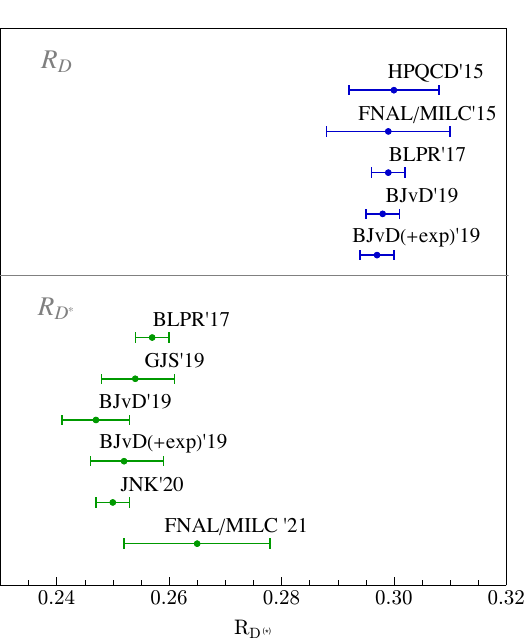}
\caption{We summarise the most recent determinations for the ratio of exclusive $V_{ub}/V_{cb}$ (left) measured at LHCb~\cite{LHCb:2020ist,LHCb:2015eia}. For the determination from $B_s$ decays, the form factor for the $B_s\to D_s$ transition is taken from the 2020 HPQCD result~\cite{McLean:2019qcx}, and for the $B_s\to K$ from FNAL/MILC'19~\cite{FermilabLattice:2019ikx} and LCSR KR'17~\cite{Khodjamirian:2017fxg} as indicated. For the determination from $\Lambda_b$ decays the form factors are from DM'15~\cite{Detmold:2015aaa}. These are compared to the ratio of the HFLAV experimental averages for $V_{ub}^{\rm excl}$ and $V_{cb}^{\rm excl}$~\cite{Amhis:2022mac}. We further present the most recent theory predictions for $R_D$ (right) (HPQCD'15~\cite{Na:2015kha}, FNAL/MILC'15~\cite{MILC:2015uhg}, BLPR'17~\cite{Bernlochner:2017jka}, BJvD'19~\cite{Bordone:2019vic}) and for $R_{D*}$(BLPR'17~\cite{Bernlochner:2017jka}, GJS'19~\cite{Gambino:2019sif}, BJvD'19~\cite{Bordone:2019vic}, JNK'20~\cite{Jaiswal:2020wer}, FNAL/MILC'21~\cite{FermilabLattice:2021cdg}).\label{fig:VubVcbRD}}
\end{center}
\end{figure}

While semileptonic measurements are difficult at LHCb due to the missing energy carried by the neutrino, and also the fact that LHCb measures relative branching fractions such that a normalization channel is required. $V_{ub}$ measurements are particularly challenging due to the large $b\to c$ background. One interesting possibility therefore is to measure a semileptonic decay sensitive to $V_{ub}$ making use of a normalization channel depending on $V_{cb}$, such that one obtains the ratio $|V_{ub}|/|V_{cb}|$. Due to the large number of $\Lambda_b$ baryons and $B_s$ mesons produced at LHCb compared to the B factories, LHCb have focussed on the ratios $\mathcal{B}(\Lambda_b\to p\ell\nu)/\mathcal{B}(\Lambda_b\to \Lambda_c\ell\nu)$~\cite{LHCb:2015eia} and $\mathcal{B}(B_s\to K\ell\nu)/\mathcal{B}(B_s\to D_s\ell\nu)$~\cite{LHCb:2020ist}. These measurements are interesting in light of the exclusive-inclusive discrepancies, as they provide an independent constraint in the $|V_{ub}|-|V_{cb}|$ plane (see plots by FLAG and HFLAV in \cite{Aoki:2021kgd,Amhis:2022mac}).
In order to extract the ratio of the CKM matrix elements, one requires the form factors for the four relevant channels. For the baryonic decays, combining the LHCb measurement with the form factors calculated in \cite{Detmold:2015aaa} results in an impressive result for $|V_{ub}|/|V_{cb}|$. For the $B_s$ decays, a precise calculation of the form factors by the HPQCD collaborationfor the $B_s\to D_s$ channel is available~\cite{Harrison:2021tol}. As for $B_s\to K$, several calculations exist, both in LCSR (KR 2017)~\cite{Khodjamirian:2017fxg} and LQCD from the HPQCD, RBC UKQCD and FNAP-MILC collaborations~\cite{Bouchard:2014ypa,Flynn:2015mha,FermilabLattice:2019ikx}.
The results are shown in Fig.~\ref{fig:VubVcbRD}.
 LHCb intend to achieve a 1\% uncertainty on the measurement of $|V_{ub}|/|V_{cb}|$ with the full upgrade II dataset of 300 fb$^{-1}$, assuming that LQCD will achieve a similar uncertainty.

\subsection{Semileptonic radiative decays}
\label{sec:semileprad}

The decay $B^+\to\gamma\ell^+\nu$ is sensitive to the structure of the $B$ meson, in particular to $\lambda_B$, the first moment of the $B$ meson light-cone distribution amplitude.~\cite{Beneke:2000wa,Grozin:1996pq,Bosch:2003fc}.
On calculating the BR for this decay at leading order in QCD factorization (in the region where the photon energy is of order $m_b$, it depends inversely no $\lambda_B$.
This parameter is of crucial important in calculating observables for $B$ meson decays in QCD factorization, and also when calculating form factors in light-cone sum rules in terms of the $B$ meson LCDA.
 The theoretical prediction is now at an advanced stage, with next-to-leading order corrections from \cite{Beneke:2011nf,Beneke:2018wjp,Wang:2018wfj,Wang:2016qii} and the soft corrections calculated in \cite{Braun:2012kp}.
  Limits on this decay have been obtained by BaBar~\cite{Aubert:2009ya} ($
	\mathcal{BR}(B^+\to \ell^+\nu\gamma)< 15.6 \times 10^{-6}$ at 90\% C.L.) and Belle~\cite{Heller:2015vvm} ($	\mathcal{BR}(B^+\to \ell^+\nu\gamma)s< 3.5 \times 10^{-6},$	both at $90\%$ C.L.),
the latter resulting in the most stringent lower limit of $\lambda_B>238$ MeV, using the expression for the partial branching fraction include state-of-the-art theoretical prediction.
 Belle II will improve this measurement, and the projected statistical uncertainty with 50 ab$^{-1}$ is at the level of a few percent (assuming a branching fraction of  $(5.0\pm 0.6) 10^{-6}$)~\cite{Gelb:2018end,Kou:2018nap}.

At the LHC however the measurement of $B^+\to \ell^+\nu\gamma$ is challenging as the low energy photon in combination with the neutrino in the final state provide complications for the trigger. LHCb have therefore  instead performed a search for the decay $B^+\to\mu^+\nu\mu^+\mu^-$  in the region where the lowest of the muon pair mass combination is below $980$ MeV~\cite{Aaij:2018pka},  obtaining the 95\% C.L.~upper limit $\mathcal{BR}(B^+\to \mu^+ \mu^- \mu^+ \nu)<1.6 \times 10^{-8}$.
First predictions for this mode can be found in \cite{Danilina:2018uzr,Danilina:2019dji}, based on the vector meson dominance approach. Later the QCD factorization approach was applied by several groups, see \cite{Beneke:2021rjf,Wang:2021yrr,Bharucha:2021zay}. It was found that at small $q^2$ (the dilepton momentum) the sensitivity to $\lambda_B$ is promising, and will provide an interesting comparison to the future Belle II result.
 It would also be interesting to study $B^+\to\ell^+\ell^-\ell'^+\nu$ for the case $\ell\neq\ell'$, as this allows one to avoid the ambiguity in identifying the two charged leptons~\cite{Bharucha:2021zay}.
Similar radiative decays in the neutral current channels are also of interest as these remove the helicity suppression of the fully leptonic decays. These have recently been studied in \cite{Guadagnoli:2017quo,Albrecht:2019zul,Janowski:2021yvz}. In \cite{Janowski:2021yvz} one finds the LCSR form factors for both the radiative charged current and neutral current decays.

\subsection{Determinations of $V_{cd}$ and  $V_{cs}$}
\label{sec:Vcdcs}
In the charm sector the determinations of the CKM matrix elements is extremely precise.  Previous results for $V_{cd}$ came from neutrino scattering data. More recently, progress has been made from the semileptonic modes, $D\to\pi\ell\nu$ and $D\to K\ell\nu$, for which the form factors are well known from LQCD, as well as from fully leptonic decays $D_{(s)}\to\ell\nu$, where the only hadronic information required is the decay constants, calculated to high precision on the Lattice. The latest $N_f=2+1+1$ calculations for the $D\to\pi$ and $D\to K$ form factors come from the HPQCD~\cite{Chakraborty:2021qav} and ETM~\cite{Lubicz:2017syv} collaborations. Latest results for the decay constants can be found in the FLAG review~\cite{Aoki:2021kgd}. 
Note that progress on the experimental side in the last ten years has been to a large extent due to BESIII (see e.g.~\cite{BESIII:2013iro,BESIII:2015tql,BESIII:2019vhn,BESIII:2021anh}), but also CLEO-c, BaBar and Belle (for a full list of references see \cite{HFLAV:2016hnz}).
The results for $V_{cd}$ and  $V_{cs}$ from semileptonic and leptonic decays using lattice $N_f=2+1+1$ results are summarised in Tab.~\ref{tab:VcdVcs},
\begin{table}
\centering
\begin{tabular}{c|ccc}
\hline
Channel & $V_{cd}$ & $V_{cs}$ &Ref.\\
\hline
Semileptonic & $0.2341(74)$ & $0.9714(69)$ & \cite{Lubicz:2017syv,Riggio:2017zwh,Chakraborty:2021qav}\\
Leptonic&  $0.2179(7)(57)$ & $0.983(2)(18)$ &\cite{Carrasco:2014poa,Bazavov:2017lyh}\\
Average & $0.2236(37)$ & $0.9741(65)$  &\cite{Aoki:2021kgd}\\
\hline
\end{tabular}
\caption{Determinations of $V_{cd}$ and $V_{cs}$ from leptonic and semileptonic $D$ decays and the averaged results~\cite{Aoki:2021kgd}. \label{tab:VcdVcs}}
\end{table}

\section{Lepton universality tests}
\label{sec:lu}

Out of the various observables exhibiting anomalies in $B$ decays, those that are theoretically cleanest are the lepton universality ratios $R_{K^{(\ast)}}$ and $R_{D^{(\ast)}}$, where any deviation from one in the SM stems from lepton mass effects. The former set of ratios involve neutral current $b\to s\ell\ell$ transitions, and the latter charged current $b\to c \ell\nu$ transitions, and are defined as follows:
\begin{equation}
R_{\mathcal{M}}=
\frac{\int_{q_{\mathrm{min}}^2}^{q_{\mathrm{max}}^2}\,dq^2d\mathcal{B}(\mathcal{B}\to \mathcal{M}\ell\ell/\nu)/dq^2}
{\int_{q_{\mathrm{min}}^2}^{q_{\mathrm{max}}^2}\,dq^2\,d\mathcal{B}(\mathcal{B}\to \mathcal{M}\mu\mu/\nu)/dq^2},
\end{equation}
where in the final state contains two leptons or a lepton neutrino pair depending on whether the decay involves a neutral or charged current respectively. For the neutral current ratios, as first proposed in Ref.~\cite{Hiller:2003js}, the lepton in the numerator is the electron, i.e.~$\ell=e$, and for the charged current ratios:  $\ell=\tau$, and for both the lepton in the numerator is the muon. The limits of the integration in $q^2$ are given by $q_{\mathrm{min}}^2$ and $q_{\mathrm{max}}^2$.

For the neutral current $b\to s$ ratios, the SM prediction is close to one due to the small size of the electron and muon masses. The measurement of  $R_{K^{(\ast)}}$ has been performed both by BaBar~\cite{BaBar:2012mrf} and Belle~\cite{Belle:2019oag,BELLE:2019xld},  and measurements by LHCb are available for the $K^+$~\cite{LHCb:2019hip} and $K^{\ast0}$~\cite{LHCb:2017avl} as well as the more challenging $K_S^0$ and $K^{\ast+}$~\cite{LHCb:2021lvy}. The ratios are found to be 2.1-2.5\,$\sigma$ below the predictions in the SM~\cite{Jager:2014rwa,Straub:2015ica,Bordone:2016gaq,Serra:2016ivr,Capdevila:2016ivx,Altmannshofer:2017fio,Straub:2018kue,Isidori:2020acz}.\footnote{Note that recently $R_{K^{(*)}}$ was updated and with the updated analysis with improved background simulations the central value is now in agreement with the SM.}

Verifying this anomaly is a priority for LHCb and Belle II. 
The ratio for the $\Lambda_b\to p\,K\,\ell\ell$ decay has also been measured~\cite{LHCb:2019efc}, and agrees with the SM at the 1\,$\sigma$ level. At Belle II, for example the ratio for the $b\to s$ inclusive channel will be analysed~\cite{Kou:2018nap}. Due to Belle II's excellent sensitivity to electrons, it will provide precise access to $R_K^{(*)}$ and $R_X$ (where $R_X$ is the inclusive $b\to s \ell\ell$ lepton universality ratio,  see~\cite{Altmannshofer:2014rta,Hurth:2016fbr,Capdevila:2016ivx,Serra:2016ivr}), with statistically dominated uncertainties at the 
few percent level with 50 ab$^{-1}$~\cite{Kou:2018nap}.
Note that it has also been suggested to use the difference between angular observables for different lepton final states as tests of lepton flavour universality~\cite{Altmannshofer:2015mqa,Capdevila:2016ivx,Serra:2016ivr}. Belle has measured two of these observables, $Q_{4,5}\equiv P_{4,5}^{\prime,\mu}-P_{4,5}^{\prime,e}$~\cite{Belle:2016fev} (these will be statistically limited at Belle II~\cite{Kou:2018nap}), but further measurements are needed at LHCb and Belle II in order to draw any conclusions.

Possible explanations for the deviation from the SM in terms of QED corrections have been ruled out~\cite{Bordone:2016gaq,Isidori:2020acz}, and if with increased statistics the anomaly persists (particularly if confirmed by Belle II) the most plausible explanation is physics beyond the SM. Therefore updates on these observables with the LHCb upgrade and from Belle II and much awaited.

For ratios involving the charged current, $b\to c\ell\nu$, the SM prediction deviates significantly from unity due to the large $\tau$ mass.  Latest theory predictions can be found in \cite{MILC:2015uhg,Na:2015kha,Bernlochner:2017jka,Bordone:2019vic,Gambino:2019sif,Jaiswal:2020wer,FermilabLattice:2021cdg}, as summarised in Fig.~\ref{fig:VubVcbRD}. A detailed discussion of the form factors adopted in these predictions can be found in Sect.~\ref{sec:semilep}.
The first experiment to measure these observables, and find a large deviation from the SM was BaBar~\cite{BaBar:2012obs,BaBar:2013mob}, claiming a significance of 3.4\,$\sigma$ on combining $R_D$ and $R_{D^*}$.
$R_{D^*}$  was subsequently measured by LHCb via the tau semileptonic decay~\cite{LHCb:2015gmp} and via the three prong tau decay ~\cite{LHCb:2017smo,LHCb:2017rln}. Belle has performed several measurements both of 
$R_D$ and $R_{D^*}$, via the $\tau$ fully hadronic decay~\cite{Belle:2015qfa},
the semileptonic decay~\cite{Belle:2016dyj} and the fully leptonic decay~\cite{Belle:2019rba}.\footnote{Note there are two recent analyses from LHCb of $R_{D^*}$~\cite{LHCb:2023zxo,LHCb:2023cjr}.}

The HFLAV average of these experimental measurements for $R_{D^*}$ is $0.295\pm0.010\pm0.010$ and for $R_D$ is $0.339\pm0.026\pm0.014$~\cite{Amhis:2022mac}. Note that LHCb has also measured  $R_{J/\Psi}$~\cite{LHCb:2017vlu}.

Theoretical predictions for $R_D$ agree well, two compatible Lattice QCD results for the form factors at non-zero recoil have been available since 2015 ~\cite{Na:2015kha,MILC:2015uhg} and the current deviation of the experimental average from the SM is at the 1.4\,$\sigma$ level. As for $R_{D^*}$, results from Fermilabl-MILC collaboration for the form factors at non-zero recoil have only recently become available, and a combination with experiment and a comparison from another Lattice collaboration would be desirable~\cite{FermilabLattice:2021cdg}. In the meantime, 
several groups calculated the form factors obtained on fitting the BGL parameterization to the unfolded spectrum from Belle for both tagged~\cite{Belle:2017rcc} and untagged~\cite{Belle:2018ezy}, a well as from BaBar~\cite{BaBar:2019vpl}, see \cite{Bernlochner:2017jka,Gambino:2019sif,Jaiswal:2020wer}. Further, the form factors were fitted, incorporating the HQET corrections at $\mathcal{O}(\alpha_s, 1/m_c^2)$, and $R_{D^{(*)}}$,  to theory (Lattice, QCDSR and LCSR) with and without Belle data from 2017 and 2018~\cite{Bordone:2019vic} (see also \cite{Bernlochner:2017jka} where HQET corrections at order $1/m_c$ were included).

HFLAV finds the combined significance of these discrepancies to be 3.3\,$\sigma$~\cite{Amhis:2022mac}. It will be of great interest to see how this evolves with increased precision on the form factors from Lattice QCD and updated measurements of the unfolded spectrum from Belle II.

%\subsection{Mixing of $B$ and $D$ mesons}
%\label{sec:cpv}
%\paragraph{B mixing}
%\paragraph{D mixing}

%\begin{figure}
%% Use the relevant command to insert your figure file.
%% For example, with the graphicx package use
%  \includegraphics{example.eps}
%\end{figure}
%%
%% For two-column wide figures use
%\begin{figure*}
%% Use the relevant command to insert your figure file.
%% For example, with the graphicx package use
%  \includegraphics[width=0.75\textwidth]{example.eps}
%\end{figure*}
%\begin{table}
%\begin{tabular}{lll}
%\hline\noalign{\smallskip}
%\end{tabular}
%\end{table}

\section{Summary}
\label{sec:summary}
We have reviewed the progress in the last ten years in heavy flavour physics, with an emphasis on neutral and charged current radiative and semileptonic decays. This progress has been greatly driven by the $B$ anomalies, as well as by the exclusive-inclusive discrepancies in the determinations of $V_{ub}$ and $V_{cb}$. 
In light of the anomalies, we summarised the progress in theoretical predictions related to searches for BSM via flavour changing neutral current processes, in particular rare radiative and semileptonic $b\to s$ decays, both at LHCb and the $B$-factories, and mention the prospects at Belle II.
Note that we have not yet discussed global analyses of decays involving $b\to s$ transitions.
Several groups have performed such analyses, with the aim to pinpoint which of the Wilson coefficients involved are most likely to be affected by BSM physics (for a recent review see \cite{Blake:2016olu}).
Often a single or pair of WCs  $C_7^{(\prime)}$ and $C_{9,10}^{(\prime)}$ are considered, or since the 2019 experimental updates  $C_{9,10}^{(\prime),e}$ and $C_{9,10}^{(\prime),\mu}$ are differentiated between.
  The conclusions that can currently be drawn are the following: excellent fits to the data are found for non-zero, negative BSM contributions to $C_9^\mu$, or BSM contributions to  certain combinations of two WCs out of $C_{9,10}^{(\prime),e},C_{9,10}^{(\prime),\mu}$; the sign of the WCs is probably SM-like and that there is not yet any compelling evidence for CP violation (for some recent analyses see~\cite{Arbey:2019duh,Altmannshofer:2021qrr,Alguero:2021anc} and references therein).
Note that  sub-percent accuracy on the angular observables for $B\to K^*\ell^+\ell^-$ is expected with the Upgrade II dataset at LHCb will provide decisive information for future fits.

Coming to semileptonic decays, while tremendous progress has been made in $V_{ub}$ and $V_{cb}$, the inclusive-exclusive discrepancies are not yet resolved, while for the former the results are within 2$\sigma$ of each other, for the latter the tension remains at the 2.4$\sigma$ level~\cite{Workman:2022ynf}.
The current status of inclusive and exclusive determinations of $V_{cb}$ and $V_{ub}$  are shown in Figs.~\ref{fig:Vcb} and \ref{fig:Vub}.
Note that hadronic uncertainties for inclusive determinations are under better theoretical control, in particular at low $q^2$, compared to their exclusive counterparts.
The analyses from Belle II of $B\to D^{(*)}\ell\nu$, $B\to\pi\ell\nu$ and $B\to X_u\ell\nu$ should finally resolve the discrepancy between inclusive and exclusive  $|V_{cb}|$ and  $V_{ub}$.
For the lepton universality ratios, LHCb and Belle II will achieve percent level precision on $R_{D^*}$ with 300 fb$^{-1}$ and 50 ab$^{-1}$ respectively,  but with just  5 ab$^{-1}$ of data from Belle these anomalies should be confirmed or refuted  II~\cite{Kou:2018nap}.
For the case of $R_K$, Belle II has excellent sensitivity, and with 20 ab$^{-1}$ will be able to claim a discovery at 5 $\sigma$ if the central value remains unchanged. At LHCb, the situation is similar, and sub-percent measurements of $R_{K}$ and $R_{K^*}$ with 300 fb$^{-1}$ will be achieved.
  
We are therefore now at a very exciting time in flavour physics, with data-taking well under way at Belle II, and just starting for Run 3 at the LHC, both having ambitious physics programs. We look forward to examining the results and hope to resolve some of the puzzling discrepancies.

%
%\begin{acknowledgements}
%We thanks Martin Jung for helpful comments.

%\end{acknowledgements}


\bibliographystyle{spmpsci}      % mathematics and physical sciences
\bibliography{refs-flavour-review}   % name your BibTeX data base

\begin{thebibliography}{100}
\providecommand{\url}[1]{{#1}}
\providecommand{\urlprefix}{URL }
\expandafter\ifx\csname urlstyle\endcsname\relax
  \providecommand{\doi}[1]{DOI~\discretionary{}{}{}#1}\else
  \providecommand{\doi}{DOI~\discretionary{}{}{}\begingroup
  \urlstyle{rm}\Url}\fi

\bibitem{CMS:2022dbz}
{Measurement of ${\rm B^0_s}\to\mu^+\mu^-$ decay properties and search for the
  ${\rm B}^0\to\mu\mu$ decay in proton-proton collisions at
  $\sqrt{s}=13~\rm{TeV}$}  (2022)

\bibitem{ATLAS:2018gqc}
Aaboud, M., et~al.: {Angular analysis of $B^0_d \rightarrow K^{*}\mu^+\mu^-$
  decays in $pp$ collisions at $\sqrt{s}= 8$ TeV with the ATLAS detector}.
\newblock JHEP \textbf{10}, 047 (2018).
\newblock \doi{10.1007/JHEP10(2018)047}

\bibitem{LHCb:2013zuf}
Aaij, R., et~al.: {Differential branching fraction and angular analysis of the
  decay $B^{0} \to K^{*0} \mu^{+}\mu^{-}$}.
\newblock JHEP \textbf{08}, 131 (2013).
\newblock \doi{10.1007/JHEP08(2013)131}

\bibitem{LHCb:2013jyo}
Aaij, R., et~al.: {Search for the rare decay $D^0 \to \mu^+ \mu^-$}.
\newblock Phys. Lett. B \textbf{725}, 15--24 (2013).
\newblock \doi{10.1016/j.physletb.2013.06.037}

\bibitem{LHCb:2014auh}
Aaij, R., et~al.: {Angular analysis of charged and neutral $B \to K \mu^+\mu^-$
  decays}.
\newblock JHEP \textbf{05}, 082 (2014).
\newblock \doi{10.1007/JHEP05(2014)082}

\bibitem{LHCb:2014cxe}
Aaij, R., et~al.: {Differential branching fractions and isospin asymmetries of
  $B \to K^{(*)} \mu^+ \mu^-$ decays}.
\newblock JHEP \textbf{06}, 133 (2014).
\newblock \doi{10.1007/JHEP06(2014)133}

\bibitem{LHCb:2014mit}
Aaij, R., et~al.: {Measurement of $C\!P$ asymmetries in the decays $B^0
  \rightarrow K^{*0} \mu^+ \mu^-$ and $B^+ \rightarrow K^{+} \mu^+ \mu^-$}.
\newblock JHEP \textbf{09}, 177 (2014).
\newblock \doi{10.1007/JHEP09(2014)177}

\bibitem{LHCb:2014vnw}
Aaij, R., et~al.: {Observation of Photon Polarization in the $b\to s\gamma$
  Transition}.
\newblock Phys. Rev. Lett. \textbf{112}(16), 161801 (2014).
\newblock \doi{10.1103/PhysRevLett.112.161801}

\bibitem{LHCb:2015wdu}
Aaij, R., et~al.: {Angular analysis and differential branching fraction of the
  decay $B^0_s\to\phi\mu^+\mu^-$}.
\newblock JHEP \textbf{09}, 179 (2015).
\newblock \doi{10.1007/JHEP09(2015)179}

\bibitem{LHCb:2015eia}
Aaij, R., et~al.: {Determination of the quark coupling strength $|V_{ub}|$
  using baryonic decays}.
\newblock Nature Phys. \textbf{11}, 743--747 (2015).
\newblock \doi{10.1038/nphys3415}

\bibitem{LHCb:2015tgy}
Aaij, R., et~al.: {Differential branching fraction and angular analysis of
  $\Lambda^{0}_{b} \rightarrow \Lambda \mu^+\mu^-$ decays}.
\newblock JHEP \textbf{06}, 115 (2015).
\newblock \doi{10.1007/JHEP06(2015)115}.
\newblock [Erratum: JHEP 09, 145 (2018)]

\bibitem{LHCb:2015hsa}
Aaij, R., et~al.: {First measurement of the differential branching fraction and
  $C\!P$ asymmetry of the $B^\pm\to\pi^\pm\mu^+\mu^-$ decay}.
\newblock JHEP \textbf{10}, 034 (2015).
\newblock \doi{10.1007/JHEP10(2015)034}

\bibitem{LHCb:2015gmp}
Aaij, R., et~al.: {Measurement of the ratio of branching fractions
  $\mathcal{B}(\bar{B}^0 \to
  D^{*+}\tau^{-}\bar{\nu}_{\tau})/\mathcal{B}(\bar{B}^0 \to
  D^{*+}\mu^{-}\bar{\nu}_{\mu})$}.
\newblock Phys. Rev. Lett. \textbf{115}(11), 111803 (2015).
\newblock \doi{10.1103/PhysRevLett.115.111803}.
\newblock [Erratum: Phys.Rev.Lett. 115, 159901 (2015)]

\bibitem{LHCb:2014yov}
Aaij, R., et~al.: {Study of the rare $B_s^0$ and $B^0$ decays into the
  $\pi^+\pi^-\mu^+\mu^-$ final state}.
\newblock Phys. Lett. B \textbf{743}, 46--55 (2015).
\newblock \doi{10.1016/j.physletb.2015.02.010}

\bibitem{LHCb:2015svh}
Aaij, R., et~al.: {Angular analysis of the $B^{0} \to K^{*0} \mu^{+} \mu^{-}$
  decay using 3 fb$^{-1}$ of integrated luminosity}.
\newblock JHEP \textbf{02}, 104 (2016).
\newblock \doi{10.1007/JHEP02(2016)104}

\bibitem{LHCb:2016eyu}
Aaij, R., et~al.: {Differential branching fraction and angular moments analysis
  of the decay $B^0 \to K^+ \pi^- \mu^+ \mu^-$ in the $K^*_{0,2}(1430)^0$
  region}.
\newblock JHEP \textbf{12}, 065 (2016).
\newblock \doi{10.1007/JHEP12(2016)065}

\bibitem{LHCb:2017rmj}
Aaij, R., et~al.: {Measurement of the $B^0_s\to\mu^+\mu^-$ branching fraction
  and effective lifetime and search for $B^0\to\mu^+\mu^-$ decays}.
\newblock Phys. Rev. Lett. \textbf{118}(19), 191801 (2017).
\newblock \doi{10.1103/PhysRevLett.118.191801}

\bibitem{LHCb:2016due}
Aaij, R., et~al.: {Measurement of the phase difference between short- and
  long-distance amplitudes in the $B^{+}\to K^{+}\mu^{+}\mu^{-}$ decay}.
\newblock Eur. Phys. J. C \textbf{77}(3), 161 (2017).
\newblock \doi{10.1140/epjc/s10052-017-4703-2}

\bibitem{LHCb:2017lpt}
Aaij, R., et~al.: {Observation of the suppressed decay $\Lambda^{0}_{b}\to
  p\pi^{-}\mu^{+}\mu^{-}$}.
\newblock JHEP \textbf{04}, 029 (2017).
\newblock \doi{10.1007/JHEP04(2017)029}

\bibitem{LHCb:2017myy}
Aaij, R., et~al.: {Search for the decays $B_s^0\to\tau^+\tau^-$ and
  $B^0\to\tau^+\tau^-$}.
\newblock Phys. Rev. Lett. \textbf{118}(25), 251802 (2017).
\newblock \doi{10.1103/PhysRevLett.118.251802}

\bibitem{LHCb:2017avl}
Aaij, R., et~al.: {Test of lepton universality with $B^{0} \rightarrow
  K^{*0}\ell^{+}\ell^{-}$ decays}.
\newblock JHEP \textbf{08}, 055 (2017).
\newblock \doi{10.1007/JHEP08(2017)055}

\bibitem{LHCb:2018jna}
Aaij, R., et~al.: {Angular moments of the decay $\Lambda_b^0 \rightarrow
  \Lambda \mu^{+} \mu^{-}$ at low hadronic recoil}.
\newblock JHEP \textbf{09}, 146 (2018).
\newblock \doi{10.1007/JHEP09(2018)146}

\bibitem{LHCb:2018rym}
Aaij, R., et~al.: {Evidence for the decay $ {B}_S^0\to {\overline{K}}^{\ast
  0}{\mu}^{+}{\mu}^{-} $}.
\newblock JHEP \textbf{07}, 020 (2018).
\newblock \doi{10.1007/JHEP07(2018)020}

\bibitem{LHCb:2017vlu}
Aaij, R., et~al.: {Measurement of the ratio of branching fractions
  $\mathcal{B}(B_c^+\,\to\,J/\psi\tau^+\nu_\tau)$/$\mathcal{B}(B_c^+\,\to\,J/\psi\mu^+\nu_\mu)$}.
\newblock Phys. Rev. Lett. \textbf{120}(12), 121801 (2018).
\newblock \doi{10.1103/PhysRevLett.120.121801}

\bibitem{LHCb:2017smo}
Aaij, R., et~al.: {Measurement of the ratio of the $B^0 \to D^{*-} \tau^+
  \nu_{\tau}$ and $B^0 \to D^{*-} \mu^+ \nu_{\mu}$ branching fractions using
  three-prong $\tau$-lepton decays}.
\newblock Phys. Rev. Lett. \textbf{120}(17), 171802 (2018).
\newblock \doi{10.1103/PhysRevLett.120.171802}

\bibitem{LHCb:2018roe}
Aaij, R., et~al.: {Physics case for an LHCb Upgrade II - Opportunities in
  flavour physics, and beyond, in the HL-LHC era}  (2018)

\bibitem{Aaij:2018pka}
Aaij, R., et~al.: {Search for the rare decay $B^{+} \rightarrow
  {\mu}^{+}{\mu}^{-}{\mu}^{+}{\nu}_{{\mu}}$}.
\newblock Submitted to: Eur. Phys. J.  (2018)

\bibitem{LHCb:2017rln}
Aaij, R., et~al.: {Test of Lepton Flavor Universality by the measurement of the
  $B^0 \to D^{*-} \tau^+ \nu_{\tau}$ branching fraction using three-prong
  $\tau$ decays}.
\newblock Phys. Rev. D \textbf{97}(7), 072013 (2018).
\newblock \doi{10.1103/PhysRevD.97.072013}

\bibitem{LHCb:2019hip}
Aaij, R., et~al.: {Search for lepton-universality violation in $B^+\to
  K^+\ell^+\ell^-$ decays}.
\newblock Phys. Rev. Lett. \textbf{122}(19), 191801 (2019).
\newblock \doi{10.1103/PhysRevLett.122.191801}

\bibitem{LHCb:2020lmf}
Aaij, R., et~al.: {Measurement of $CP$-Averaged Observables in the
  $B^{0}\rightarrow K^{*0}\mu^{+}\mu^{-}$ Decay}.
\newblock Phys. Rev. Lett. \textbf{125}(1), 011802 (2020).
\newblock \doi{10.1103/PhysRevLett.125.011802}

\bibitem{LHCb:2020cyw}
Aaij, R., et~al.: {Measurement of $|V_{cb}|$ with $B_s^0 \to D_s^{(*)-} \mu^+
  \nu_{\mu}$ decays}.
\newblock Phys. Rev. D \textbf{101}(7), 072004 (2020).
\newblock \doi{10.1103/PhysRevD.101.072004}

\bibitem{LHCb:2020dof}
Aaij, R., et~al.: {Strong constraints on the $b \to s\gamma$ photon
  polarisation from $B^0 \to K^{*0} e^+ e^-$ decays}.
\newblock JHEP \textbf{12}, 081 (2020).
\newblock \doi{10.1007/JHEP12(2020)081}

\bibitem{LHCb:2019efc}
Aaij, R., et~al.: {Test of lepton universality with $ {\Lambda}_b^0\to
  {pK}^{-}{\mathrm{\ell}}^{+}{\mathrm{\ell}}^{-} $ decays}.
\newblock JHEP \textbf{05}, 040 (2020).
\newblock \doi{10.1007/JHEP05(2020)040}

\bibitem{LHCb:2020gog}
Aaij, R., et~al.: {Angular Analysis of the $B^{+}\rightarrow
  K^{\ast+}\mu^{+}\mu^{-}$ Decay}.
\newblock Phys. Rev. Lett. \textbf{126}(16), 161802 (2021).
\newblock \doi{10.1103/PhysRevLett.126.161802}

\bibitem{LHCb:2021xxq}
Aaij, R., et~al.: {Angular analysis of the rare decay $B_s^0
  \to\phi\mu^+\mu^-$}.
\newblock JHEP \textbf{11}, 043 (2021).
\newblock \doi{10.1007/JHEP11(2021)043}

\bibitem{LHCb:2021zwz}
Aaij, R., et~al.: {Branching Fraction Measurements of the Rare
  $B^0_s\rightarrow\phi\mu^+\mu^-$ and $B^0_s\rightarrow
  f_2^\prime(1525)\mu^+\mu^-$- Decays}.
\newblock Phys. Rev. Lett. \textbf{127}(15), 151801 (2021).
\newblock \doi{10.1103/PhysRevLett.127.151801}

\bibitem{LHCb:2020ist}
Aaij, R., et~al.: {First observation of the decay $B_s^0 \to K^-\mu^+\nu_\mu$
  and Measurement of $|V_{ub}|/|V_{cb}|$}.
\newblock Phys. Rev. Lett. \textbf{126}(8), 081804 (2021).
\newblock \doi{10.1103/PhysRevLett.126.081804}

\bibitem{LHCb:2021vsc}
Aaij, R., et~al.: {Analysis of Neutral B-Meson Decays into Two Muons}.
\newblock Phys. Rev. Lett. \textbf{128}(4), 041801 (2022).
\newblock \doi{10.1103/PhysRevLett.128.041801}

\bibitem{LHCb:2021trn}
Aaij, R., et~al.: {Test of lepton universality in beauty-quark decays}.
\newblock Nature Phys. \textbf{18}(3), 277--282 (2022).
\newblock \doi{10.1038/s41567-021-01478-8}

\bibitem{LHCb:2021lvy}
Aaij, R., et~al.: {Tests of lepton universality using $B^0\to K^0_S \ell^+
  \ell^-$ and $B^+\to K^{*+} \ell^+ \ell^-$ decays}.
\newblock Phys. Rev. Lett. \textbf{128}(19), 191802 (2022).
\newblock \doi{10.1103/PhysRevLett.128.191802}

\bibitem{CDF:2011buy}
Aaltonen, T., et~al.: {Observation of the Baryonic Flavor-Changing Neutral
  Current Decay $\Lambda_{b} \to \Lambda \mu^{+} \mu^{-}$}.
\newblock Phys. Rev. Lett. \textbf{107}, 201802 (2011).
\newblock \doi{10.1103/PhysRevLett.107.201802}

\bibitem{CDF:2011tds}
Aaltonen, T., et~al.: {Measurements of the Angular Distributions in the Decays
  $B \to K^{(*)} \mu^+ \mu^-$ at CDF}.
\newblock Phys. Rev. Lett. \textbf{108}, 081807 (2012).
\newblock \doi{10.1103/PhysRevLett.108.081807}

\bibitem{Belle:2016mtj}
Abdesselam, A., et~al.: {Observation of $D^0\to \rho^0\gamma$ and search for
  $CP$ violation in radiative charm decays}.
\newblock Phys. Rev. Lett. \textbf{118}(5), 051801 (2017).
\newblock \doi{10.1103/PhysRevLett.118.051801}

\bibitem{Belle:2017rcc}
Abdesselam, A., et~al.: {Precise determination of the CKM matrix element
  $\left| V_{cb}\right|$ with $\bar B^0 \to D^{*\,+} \, \ell^- \, \bar
  \nu_\ell$ decays with hadronic tagging at Belle}  (2017)

\bibitem{Belle:2019oag}
Abdesselam, A., et~al.: {Test of Lepton-Flavor Universality in $B\to
  K^\ast\ell^+\ell^-$ Decays at Belle}.
\newblock Phys. Rev. Lett. \textbf{126}(16), 161801 (2021).
\newblock \doi{10.1103/PhysRevLett.126.161801}

\bibitem{BESIII:2013iro}
Ablikim, M., et~al.: {Precision measurements of $B(D^+ \rightarrow \mu^+
  \nu_{\mu})$, the pseudoscalar decay constant $f_{D^+}$, and the quark mixing
  matrix element $|V_{\rm cd}|$}.
\newblock Phys. Rev. D \textbf{89}(5), 051104 (2014).
\newblock \doi{10.1103/PhysRevD.89.051104}

\bibitem{BESIII:2015tql}
Ablikim, M., et~al.: {Study of Dynamics of $D^0 \to K^- e^+ \nu_{e}$ and
  $D^0\to\pi^- e^+ \nu_{e}$ Decays}.
\newblock Phys. Rev. D \textbf{92}(7), 072012 (2015).
\newblock \doi{10.1103/PhysRevD.92.072012}

\bibitem{BESIII:2019vhn}
Ablikim, M., et~al.: {Observation of the leptonic decay $D^+ \to \tau^+
  \nu_\tau$}.
\newblock Phys. Rev. Lett. \textbf{123}(21), 211802 (2019).
\newblock \doi{10.1103/PhysRevLett.123.211802}

\bibitem{BESIII:2021anh}
Ablikim, M., et~al.: {Measurement of the absolute branching fractions for
  purely leptonic $D_s^+$ decays}.
\newblock Phys. Rev. D \textbf{104}(5), 052009 (2021).
\newblock \doi{10.1103/PhysRevD.104.052009}

\bibitem{Belle-II:2021rof}
Abudin\'en, F., et~al.: {Search for $B^+\to K^+\nu\overline{\nu}$ Decays Using
  an Inclusive Tagging Method at Belle II}.
\newblock Phys. Rev. Lett. \textbf{127}(18), 181802 (2021).
\newblock \doi{10.1103/PhysRevLett.127.181802}

\bibitem{Belle-II:2022fky}
Abudin\'en, F., et~al.: {Measurement of the branching fraction for the decay $B
  \to K^{\ast}(892)\ell^+\ell^-$ at Belle II}  (2022)

\bibitem{Aglietti:2007ik}
Aglietti, U., Di~Lodovico, F., Ferrera, G., Ricciardi, G.: {Inclusive measure
  of |V(ub)| with the analytic coupling model}.
\newblock Eur. Phys. J. C \textbf{59}, 831--840 (2009).
\newblock \doi{10.1140/epjc/s10052-008-0817-x}

\bibitem{Alberti:2012dn}
Alberti, A., Ewerth, T., Gambino, P., Nandi, S.: {Kinetic operator effects in
  $\bar{B}\to X_c l \nu$ at O($\alpha_s$)}.
\newblock Nucl. Phys. B \textbf{870}, 16--29 (2013).
\newblock \doi{10.1016/j.nuclphysb.2013.01.005}

\bibitem{Alberti:2014yda}
Alberti, A., Gambino, P., Healey, K.J., Nandi, S.: {Precision Determination of
  the Cabibbo-Kobayashi-Maskawa Element $V_{cb}$}.
\newblock Phys. Rev. Lett. \textbf{114}(6), 061802 (2015).
\newblock \doi{10.1103/PhysRevLett.114.061802}

\bibitem{Alberti:2016fba}
Alberti, A., Gambino, P., Healey, K.J., Nandi, S.: {The Inclusive Determination
  of |V$_{cb}$|}.
\newblock Nucl. Part. Phys. Proc. \textbf{273-275}, 1325--1329 (2016).
\newblock \doi{10.1016/j.nuclphysbps.2015.09.212}

\bibitem{Alberti:2013kxa}
Alberti, A., Gambino, P., Nandi, S.: {Perturbative corrections to power
  suppressed effects in semileptonic B decays}.
\newblock JHEP \textbf{01}, 147 (2014).
\newblock \doi{10.1007/JHEP01(2014)147}

\bibitem{Albrecht:2019zul}
Albrecht, J., Stamou, E., Ziegler, R., Zwicky, R.: {Flavoured axions in the
  tail of $B_{q} \to\mu^{+}\mu^-$ and $B\to\gamma^*$ form factors}.
\newblock JHEP \textbf{21}, 139 (2020).
\newblock \doi{10.1007/JHEP09(2021)139}

\bibitem{Alguero:2021anc}
Alguer\'o, M., Capdevila, B., Descotes-Genon, S., Matias, J., Novoa-Brunet, M.:
  {$b\rightarrow s\ell ^+\ell ^-$ global fits after $R_{K_S}$ and
  $R_{K^{*+}}$}.
\newblock Eur. Phys. J. C \textbf{82}(4), 326 (2022).
\newblock \doi{10.1140/epjc/s10052-022-10231-1}

\bibitem{Altmannshofer:2008dz}
Altmannshofer, W., Ball, P., Bharucha, A., Buras, A.J., Straub, D.M., Wick, M.:
  {Symmetries and Asymmetries of $B \to K^{*} \mu^{+} \mu^{-}$ Decays in the
  Standard Model and Beyond}.
\newblock JHEP \textbf{01}, 019 (2009).
\newblock \doi{10.1088/1126-6708/2009/01/019}

\bibitem{Altmannshofer:2009ma}
Altmannshofer, W., Buras, A.J., Straub, D.M., Wick, M.: {New strategies for New
  Physics search in $B \to K^{*} \nu \bar{\nu}$, $B \to K \nu \bar{\nu}$ and $B
  \to X_{s} \nu \bar{\nu}$ decays}.
\newblock JHEP \textbf{04}, 022 (2009).
\newblock \doi{10.1088/1126-6708/2009/04/022}

\bibitem{Altmannshofer:2017fio}
Altmannshofer, W., Niehoff, C., Stangl, P., Straub, D.M.: {Status of the
  $B\rightarrow K^*\mu ^+\mu ^-$ anomaly after Moriond 2017}.
\newblock Eur. Phys. J. C \textbf{77}(6), 377 (2017).
\newblock \doi{10.1140/epjc/s10052-017-4952-0}

\bibitem{Altmannshofer:2021qrr}
Altmannshofer, W., Stangl, P.: {New physics in rare B decays after Moriond
  2021}.
\newblock Eur. Phys. J. C \textbf{81}(10), 952 (2021).
\newblock \doi{10.1140/epjc/s10052-021-09725-1}

\bibitem{Altmannshofer:2014rta}
Altmannshofer, W., Straub, D.M.: {New physics in $b\rightarrow s$ transitions
  after LHC run 1}.
\newblock Eur. Phys. J. C \textbf{75}(8), 382 (2015).
\newblock \doi{10.1140/epjc/s10052-015-3602-7}

\bibitem{Altmannshofer:2015mqa}
Altmannshofer, W., Yavin, I.: {Predictions for lepton flavor universality
  violation in rare B decays in models with gauged $L_\mu - L_\tau$}.
\newblock Phys. Rev. D \textbf{92}(7), 075022 (2015).
\newblock \doi{10.1103/PhysRevD.92.075022}

\bibitem{Kou:2018nap}
Altmannshofer, W., et~al.: {The Belle II Physics Book}.
\newblock PTEP \textbf{2019}(12), 123C01 (2019).
\newblock \doi{10.1093/ptep/ptz106}.
\newblock [Erratum: PTEP 2020, 029201 (2020)]

\bibitem{Belle-II:2018jsg}
Altmannshofer, W., et~al.: {The Belle II Physics Book}.
\newblock PTEP \textbf{2019}(12), 123C01 (2019).
\newblock \doi{10.1093/ptep/ptz106}.
\newblock [Erratum: PTEP 2020, 029201 (2020)]

\bibitem{HFLAV:2016hnz}
Amhis, Y., et~al.: {Averages of $b$-hadron, $c$-hadron, and $\tau$-lepton
  properties as of summer 2016}.
\newblock Eur. Phys. J. C \textbf{77}(12), 895 (2017).
\newblock \doi{10.1140/epjc/s10052-017-5058-4}

\bibitem{Amhis:2022mac}
Amhis, Y., et~al.: {Averages of $b$-hadron, $c$-hadron, and $\tau$-lepton
  properties as of 2021}  (2022)

\bibitem{HFLAV:2019otj}
Amhis, Y.S., et~al.: {Averages of b-hadron, c-hadron, and $\tau $-lepton
  properties as of 2018}.
\newblock Eur. Phys. J. C \textbf{81}(3), 226 (2021).
\newblock \doi{10.1140/epjc/s10052-020-8156-7}

\bibitem{BaBar:2010oqg}
del Amo~Sanchez, P., et~al.: {Search for the Rare Decay $B \to K \nu
  \bar{\nu}$}.
\newblock Phys. Rev. D \textbf{82}, 112002 (2010).
\newblock \doi{10.1103/PhysRevD.82.112002}

\bibitem{BaBar:2015chw}
del Amo~Sanchez, P., et~al.: {Time-dependent analysis of $B^0 \to {{K^0_{S}}}
  \pi^- \pi^+ \gamma$ decays and studies of the $K^+\pi^-\pi^+$ system in $B^+
  \to K^+ \pi^- \pi^+ \gamma$ decays}.
\newblock Phys. Rev. D \textbf{93}(5), 052013 (2016).
\newblock \doi{10.1103/PhysRevD.93.052013}

\bibitem{Andersen:2005mj}
Andersen, J.R., Gardi, E.: {Inclusive spectra in charmless semileptonic B
  decays by dressed gluon exponentiation}.
\newblock JHEP \textbf{01}, 097 (2006).
\newblock \doi{10.1088/1126-6708/2006/01/097}

\bibitem{Aoki:2021kgd}
Aoki, Y., et~al.: {FLAG Review 2021}  (2021)

\bibitem{Arbey:2018ics}
Arbey, A., Hurth, T., Mahmoudi, F., Neshatpour, S.: {Hadronic and New Physics
  Contributions to $b \to s$ Transitions}.
\newblock Phys. Rev. D \textbf{98}(9), 095027 (2018).
\newblock \doi{10.1103/PhysRevD.98.095027}

\bibitem{Arbey:2019duh}
Arbey, A., Hurth, T., Mahmoudi, F., Santos, D.M., Neshatpour, S.: {Update on
  the $b\to s$ anomalies}.
\newblock Phys. Rev. D \textbf{100}(1), 015045 (2019).
\newblock \doi{10.1103/PhysRevD.100.015045}

\bibitem{BaBar:2006tnv}
Aubert, B., et~al.: {Measurements of branching fractions, rate asymmetries, and
  angular distributions in the rare decays $B \to K \ell^{+} \ell^{-}$ and $B
  \to K^{*} \ell^{+} \ell^{-}$}.
\newblock Phys. Rev. D \textbf{73}, 092001 (2006).
\newblock \doi{10.1103/PhysRevD.73.092001}

\bibitem{Aubert:2009ya}
Aubert, B., et~al.: {A Model-independent search for the decay B+ ---> l+ nu(l)
  gamma}.
\newblock Phys. Rev. \textbf{D80}, 111105 (2009).
\newblock \doi{10.1103/PhysRevD.80.111105}

\bibitem{FermilabLattice:2014ysv}
Bailey, J.A., et~al.: {Update of $|V_{cb}|$ from the $\bar{B}\to
  D^*\ell\bar{\nu}$ form factor at zero recoil with three-flavor lattice QCD}.
\newblock Phys. Rev. D \textbf{89}(11), 114504 (2014).
\newblock \doi{10.1103/PhysRevD.89.114504}

\bibitem{MILC:2015uhg}
Bailey, J.A., et~al.: {$B\to D\ell\nu$ form factors at nonzero recoil and
  |V$_{cb}$| from 2+1-flavor lattice QCD}.
\newblock Phys. Rev. D \textbf{92}(3), 034506 (2015).
\newblock \doi{10.1103/PhysRevD.92.034506}

\bibitem{FermilabLattice:2015cdh}
Bailey, J.A., et~al.: {$B\to\pi\ell\ell$ form factors for new-physics searches
  from lattice QCD}.
\newblock Phys. Rev. Lett. \textbf{115}(15), 152002 (2015).
\newblock \doi{10.1103/PhysRevLett.115.152002}

\bibitem{FermilabLattice:2015mwy}
Bailey, J.A., et~al.: {$|V_{ub}|$ from $B\to\pi\ell\nu$ decays and (2+1)-flavor
  lattice QCD}.
\newblock Phys. Rev. D \textbf{92}(1), 014024 (2015).
\newblock \doi{10.1103/PhysRevD.92.014024}

\bibitem{Bailey:2015dka}
Bailey, J.A., et~al.: {$B\to Kl^+l^-$ Decay Form Factors from Three-Flavor
  Lattice QCD}.
\newblock Phys. Rev. D \textbf{93}(2), 025026 (2016).
\newblock \doi{10.1103/PhysRevD.93.025026}

\bibitem{Ball:2006eu}
Ball, P., Jones, G.W., Zwicky, R.: {$B \to V \gamma$ beyond QCD factorisation}.
\newblock Phys. Rev. D \textbf{75}, 054004 (2007).
\newblock \doi{10.1103/PhysRevD.75.054004}

\bibitem{Ball:2004rg}
Ball, P., Zwicky, R.: {$B_{d,s} \to \rho, \omega, K^*, \phi$ decay form-factors
  from light-cone sum rules revisited}.
\newblock Phys. Rev. D \textbf{71}, 014029 (2005).
\newblock \doi{10.1103/PhysRevD.71.014029}

\bibitem{Bause:2021cna}
Bause, R., Gisbert, H., Golz, M., Hiller, G.: {Interplay of dineutrino modes
  with semileptonic rare B-decays}.
\newblock JHEP \textbf{12}, 061 (2021).
\newblock \doi{10.1007/JHEP12(2021)061}

\bibitem{Bazavov:2017lyh}
Bazavov, A., et~al.: {$B$- and $D$-meson leptonic decay constants from
  four-flavor lattice QCD}.
\newblock Phys. Rev. D \textbf{98}(7), 074512 (2018).
\newblock \doi{10.1103/PhysRevD.98.074512}

\bibitem{FermilabLattice:2019ikx}
Bazavov, A., et~al.: {$B_s\to K\ell\nu$ decay from lattice QCD}.
\newblock Phys. Rev. D \textbf{100}(3), 034501 (2019).
\newblock \doi{10.1103/PhysRevD.100.034501}

\bibitem{FermilabLattice:2021cdg}
Bazavov, A., et~al.: {Semileptonic form factors for $B \to D^\ast\ell\nu$ at
  nonzero recoil from 2 + 1-flavor lattice QCD}  (2021)

\bibitem{Becher:2007tk}
Becher, T., Boos, H., Lunghi, E.: {Kinetic corrections to $B \to X_{c} \ell
  \bar{\nu}$ at one loop}.
\newblock JHEP \textbf{12}, 062 (2007).
\newblock \doi{10.1088/1126-6708/2007/12/062}

\bibitem{Belle:2020xgu}
Bele\~no, C., et~al.: {Measurement of the branching fraction of the decay
  $\boldsymbol{B^{+}\to\pi^{+}\pi^{-}\ell^{+}\nu_\ell}$ in fully reconstructed
  events at Belle}.
\newblock Phys. Rev. D \textbf{103}(11), 112001 (2021).
\newblock \doi{10.1103/PhysRevD.103.112001}

\bibitem{Belle-II:2022fug}
Belle-II: {Measurement of Lepton Mass Squared Moments in $B \to X_c \ell \bar
  \nu_{\ell}$ Decays with the Belle II Experiment}  (2022)

\bibitem{Beneke:2019slt}
Beneke, M., Bobeth, C., Szafron, R.: {Power-enhanced leading-logarithmic QED
  corrections to $B_q \to \mu^+\mu^-$}.
\newblock JHEP \textbf{10}, 232 (2019).
\newblock \doi{10.1007/JHEP10(2019)232}

\bibitem{Beneke:2021rjf}
Beneke, M., B\"oer, P., Rigatos, P., Vos, K.K.: {QCD factorization of the
  four-lepton decay $B^-\rightarrow \ell \bar{\nu }_\ell \ell ^{(\prime )}
  \bar{\ell }^{(\prime )}$}.
\newblock Eur. Phys. J. C \textbf{81}(7), 638 (2021).
\newblock \doi{10.1140/epjc/s10052-021-09388-y}

\bibitem{Beneke:2018wjp}
Beneke, M., Braun, V., Ji, Y., Wei, Y.B.: {Radiative leptonic decay $B\to
  \gamma \ell \nu_\ell$ with subleading power corrections}.
\newblock JHEP \textbf{07}, 154 (2018).
\newblock \doi{10.1007/JHEP07(2018)154}

\bibitem{Beneke:2000wa}
Beneke, M., Feldmann, T.: {Symmetry breaking corrections to heavy to light B
  meson form-factors at large recoil}.
\newblock Nucl. Phys. \textbf{B592}, 3--34 (2001).
\newblock \doi{10.1016/S0550-3213(00)00585-X}

\bibitem{Beneke:2001at}
Beneke, M., Feldmann, T., Seidel, D.: {Systematic approach to exclusive $B \to
  V l^+ l^-$, $V \gamma$ decays}.
\newblock Nucl. Phys. B \textbf{612}, 25--58 (2001).
\newblock \doi{10.1016/S0550-3213(01)00366-2}

\bibitem{Beneke:2004dp}
Beneke, M., Feldmann, T., Seidel, D.: {Exclusive radiative and electroweak $b
  \to d$ and $b \to s$ penguin decays at NLO}.
\newblock Eur. Phys. J. C \textbf{41}, 173--188 (2005).
\newblock \doi{10.1140/epjc/s2005-02181-5}

\bibitem{Beneke:2011nf}
Beneke, M., Rohrwild, J.: {B meson distribution amplitude from $B \to \gamma l
  \nu$}.
\newblock Eur. Phys. J. C \textbf{71}, 1818 (2011).
\newblock \doi{10.1140/epjc/s10052-011-1818-8}

\bibitem{Benson:2004sg}
Benson, D., Bigi, I.I., Uraltsev, N.: {On the photon energy moments and their
  `bias' corrections in B ---\ensuremath{>} X(s) + gamma}.
\newblock Nucl. Phys. B \textbf{710}, 371--401 (2005).
\newblock \doi{10.1016/j.nuclphysb.2004.12.035}

\bibitem{Benzke:2010js}
Benzke, M., Lee, S.J., Neubert, M., Paz, G.: {Factorization at Subleading Power
  and Irreducible Uncertainties in $\bar B\to X_s\gamma$ Decay}.
\newblock JHEP \textbf{08}, 099 (2010).
\newblock \doi{10.1007/JHEP08(2010)099}

\bibitem{Bernlochner:2022ucr}
Bernlochner, F., Fael, M., Olschewsky, K., Persson, E., van Tonder, R., Vos,
  K.K., Welsch, M.: {First extraction of inclusive $V_{cb}$ from $q^2$ moments}
   (2022)

\bibitem{Bernlochner:2020jlt}
Bernlochner, F.U., Lacker, H., Ligeti, Z., Stewart, I.W., Tackmann, F.J.,
  Tackmann, K.: {Precision Global Determination of the $B\rightarrow X_s\gamma$
  Decay Rate}.
\newblock Phys. Rev. Lett. \textbf{127}(10), 102001 (2021).
\newblock \doi{10.1103/PhysRevLett.127.102001}

\bibitem{Bernlochner:2017jka}
Bernlochner, F.U., Ligeti, Z., Papucci, M., Robinson, D.J.: {Combined analysis
  of semileptonic $B$ decays to $D$ and $D^*$: $R(D^{(*)})$, $|V_{cb}|$, and
  new physics}.
\newblock Phys. Rev. D \textbf{95}(11), 115008 (2017).
\newblock \doi{10.1103/PhysRevD.95.115008}.
\newblock [Erratum: Phys.Rev.D 97, 059902 (2018)]

\bibitem{Bernlochner:2021rel}
Bernlochner, F.U., Prim, M.T., Robinson, D.J.: {$B\to\rho\ell\nu$ and
  $B\to\omega\ell\nu$ in and beyond the Standard Model: Improved predictions
  and |Vub|}.
\newblock Phys. Rev. D \textbf{104}(3), 034032 (2021).
\newblock \doi{10.1103/PhysRevD.104.034032}

\bibitem{Bharucha:2012wy}
Bharucha, A.: {Two-loop Corrections to the $B to \pi$ Form Factor from QCD Sum
  Rules on the Light-Cone and $|V_{ub}|$}.
\newblock JHEP \textbf{05}, 092 (2012).
\newblock \doi{10.1007/JHEP05(2012)092}

\bibitem{Bharucha:2020eup}
Bharucha, A., Boito, D., M\'eaux, C.: {Disentangling QCD and new physics in
  $D^+\to\pi^+\ell^+\ell^-$}.
\newblock JHEP \textbf{04}, 158 (2021).
\newblock \doi{10.1007/JHEP04(2021)158}

\bibitem{Bharucha:2021zay}
Bharucha, A., Kindra, B., Mahajan, N.: {Probing the structure of the $B$ meson
  with $B\to\ell\ell\ell'\nu$}  (2021)

\bibitem{Bharucha:2010bb}
Bharucha, A., Reece, W.: {Constraining new physics with $B \to K^* \mu^+ \mu^-$
  in the early LHC era}.
\newblock Eur. Phys. J. C \textbf{69}, 623--640 (2010).
\newblock \doi{10.1140/epjc/s10052-010-1433-0}

\bibitem{Straub:2015ica}
Bharucha, A., Straub, D.M., Zwicky, R.: {$B\to V\ell^+\ell^-$ in the Standard
  Model from light-cone sum rules}.
\newblock JHEP \textbf{08}, 098 (2016).
\newblock \doi{10.1007/JHEP08(2016)098}

\bibitem{Bigi:2017njr}
Bigi, D., Gambino, P., Schacht, S.: {A fresh look at the determination of
  $|V_{cb}|$ from $B\to D^{*} \ell \nu$}.
\newblock Phys. Lett. B \textbf{769}, 441--445 (2017).
\newblock \doi{10.1016/j.physletb.2017.04.022}

\bibitem{Bigi:2017jbd}
Bigi, D., Gambino, P., Schacht, S.: {$R(D^*)$, $|V_{cb}|$, and the Heavy Quark
  Symmetry relations between form factors}.
\newblock JHEP \textbf{11}, 061 (2017).
\newblock \doi{10.1007/JHEP11(2017)061}

\bibitem{Bigi:2011em}
Bigi, I.I., Paul, A.: {On CP Asymmetries in Two-, Three- and Four-Body D
  Decays}.
\newblock JHEP \textbf{03}, 021 (2012).
\newblock \doi{10.1007/JHEP03(2012)021}

\bibitem{Bigi:2005bh}
Bigi, I.I., Uraltsev, N., Zwicky, R.: {On the nonperturbative charm effects in
  inclusive B ---\ensuremath{>} X(c) l nu decays}.
\newblock Eur. Phys. J. C \textbf{50}, 539--556 (2007).
\newblock \doi{10.1140/epjc/s10052-007-0216-8}

\bibitem{Blake:2016olu}
Blake, T., Lanfranchi, G., Straub, D.M.: {Rare $B$ Decays as Tests of the
  Standard Model}.
\newblock Prog. Part. Nucl. Phys. \textbf{92}, 50--91 (2017).
\newblock \doi{10.1016/j.ppnp.2016.10.001}

\bibitem{Bobeth:2021cxm}
Bobeth, C., Buras, A.J.: {Searching for New Physics with \(\bar {\mathcal
  {B}}(B_{s,d}\to \mu \bar \mu )/\Delta M_{s,d}\)}.
\newblock Acta Phys. Polon. B \textbf{52}(10), 1189 (2021).
\newblock \doi{10.5506/APhysPolB.52.1189}

\bibitem{Bobeth:2001jm}
Bobeth, C., Buras, A.J., Kruger, F., Urban, J.: {QCD corrections to $\bar{B}
  \to X_{d,s} \nu \bar{\nu}$, $\bar{B}_{d,s} \to \ell^{+} \ell^{-}$, $K \to \pi
  \nu \bar{\nu}$ and $K_{L} \to \mu^{+} \mu^{-}$ in the MSSM}.
\newblock Nucl. Phys. B \textbf{630}, 87--131 (2002).
\newblock \doi{10.1016/S0550-3213(02)00141-4}

\bibitem{Bobeth:2017vxj}
Bobeth, C., Chrzaszcz, M., van Dyk, D., Virto, J.: {Long-distance effects in
  $B\rightarrow K^*\ell \ell $ from analyticity}.
\newblock Eur. Phys. J. C \textbf{78}(6), 451 (2018).
\newblock \doi{10.1140/epjc/s10052-018-5918-6}

\bibitem{Bobeth:2003at}
Bobeth, C., Gambino, P., Gorbahn, M., Haisch, U.: {Complete NNLO QCD analysis
  of anti-B ---\ensuremath{>} X(s) l+ l- and higher order electroweak effects}.
\newblock JHEP \textbf{04}, 071 (2004).
\newblock \doi{10.1088/1126-6708/2004/04/071}

\bibitem{Bobeth:2013tba}
Bobeth, C., Gorbahn, M., Stamou, E.: {Electroweak Corrections to $B_{s,d} \to
  \ell^+ \ell^-$}.
\newblock Phys. Rev. D \textbf{89}(3), 034023 (2014).
\newblock \doi{10.1103/PhysRevD.89.034023}

\bibitem{Bobeth:2008ij}
Bobeth, C., Hiller, G., Piranishvili, G.: {CP Asymmetries in bar $B \to
  \bar{K}^* (\to \bar{K} \pi) \bar{\ell} \ell$ and Untagged $\bar{B}_s$, $B_s
  \to \phi (\to K^{+} K^-) \bar{\ell} \ell$ Decays at NLO}.
\newblock JHEP \textbf{07}, 106 (2008).
\newblock \doi{10.1088/1126-6708/2008/07/106}

\bibitem{Bobeth:1999mk}
Bobeth, C., Misiak, M., Urban, J.: {Photonic penguins at two loops and $m_t$
  dependence of $BR[B \to X_s l^+ l^-]$}.
\newblock Nucl. Phys. B \textbf{574}, 291--330 (2000).
\newblock \doi{10.1016/S0550-3213(00)00007-9}

\bibitem{deBoer:2015boa}
de~Boer, S., Hiller, G.: {Flavor and new physics opportunities with rare charm
  decays into leptons}.
\newblock Phys. Rev. D \textbf{93}(7), 074001 (2016).
\newblock \doi{10.1103/PhysRevD.93.074001}

\bibitem{deBoer:2017que}
de~Boer, S., Hiller, G.: {Rare radiative charm decays within the standard model
  and beyond}.
\newblock JHEP \textbf{08}, 091 (2017).
\newblock \doi{10.1007/JHEP08(2017)091}

\bibitem{Boer_WC}
de~Boer, S., Müller, B., Seidel, D.: {Higher-order Wilson coefficients for $c
  \to u$ transitions in the standard model}.
\newblock JHEP \textbf{08}, 091 (2016).
\newblock \doi{10.1007/JHEP08(2016)091}

\bibitem{Bordone:2021oof}
Bordone, M., Capdevila, B., Gambino, P.: {Three loop calculations and inclusive
  Vcb}.
\newblock Phys. Lett. B \textbf{822}, 136679 (2021).
\newblock \doi{10.1016/j.physletb.2021.136679}

\bibitem{Bordone:2016gaq}
Bordone, M., Isidori, G., Pattori, A.: {On the Standard Model predictions for
  $R_K$ and $R_{K^*}$}.
\newblock Eur. Phys. J. C \textbf{76}(8), 440 (2016).
\newblock \doi{10.1140/epjc/s10052-016-4274-7}

\bibitem{Bordone:2019vic}
Bordone, M., Jung, M., van Dyk, D.: {Theory determination of $\bar{B}\to
  D^{(*)}\ell^-\bar\nu$ form factors at $\mathcal{O}(1/m_c^2)$}.
\newblock Eur. Phys. J. C \textbf{80}(2), 74 (2020).
\newblock \doi{10.1140/epjc/s10052-020-7616-4}

\bibitem{Bosch:2003fc}
Bosch, S.W., Hill, R.J., Lange, B.O., Neubert, M.: {Factorization and Sudakov
  resummation in leptonic radiative B decay}.
\newblock Phys. Rev. \textbf{D67}, 094014 (2003).
\newblock \doi{10.1103/PhysRevD.67.094014}

\bibitem{Bouchard:2013pna}
Bouchard, C., Lepage, G.P., Monahan, C., Na, H., Shigemitsu, J.: {Rare decay $B
  \to K \ell^+ \ell^-$ form factors from lattice QCD}.
\newblock Phys. Rev. \textbf{D88}(5), 054509 (2013).
\newblock \doi{10.1103/PhysRevD.88.079901, 10.1103/PhysRevD.88.054509}.
\newblock [Erratum: Phys. Rev.D88,no.7,079901(2013)]

\bibitem{Bouchard:2014ypa}
Bouchard, C.M., Lepage, G.P., Monahan, C., Na, H., Shigemitsu, J.: {$B_s \to K
  \ell \nu$ form factors from lattice QCD}.
\newblock Phys. Rev. D \textbf{90}, 054506 (2014).
\newblock \doi{10.1103/PhysRevD.90.054506}

\bibitem{Bourrely:2008za}
Bourrely, C., Caprini, I., Lellouch, L.: {Model-independent description of B
  ---\ensuremath{>} pi l nu decays and a determination of |V(ub)|}.
\newblock Phys. Rev. D \textbf{79}, 013008 (2009).
\newblock \doi{10.1103/PhysRevD.82.099902}.
\newblock [Erratum: Phys.Rev.D 82, 099902 (2010)]

\bibitem{Boyd:1994tt}
Boyd, C.G., Grinstein, B., Lebed, R.F.: {Constraints on form-factors for
  exclusive semileptonic heavy to light meson decays}.
\newblock Phys. Rev. Lett. \textbf{74}, 4603--4606 (1995).
\newblock \doi{10.1103/PhysRevLett.74.4603}

\bibitem{Boyd:1995sq}
Boyd, C.G., Grinstein, B., Lebed, R.F.: {Model independent determinations of
  anti-B ---\ensuremath{>} D (lepton), D* (lepton) anti-neutrino form-factors}.
\newblock Nucl. Phys. B \textbf{461}, 493--511 (1996).
\newblock \doi{10.1016/0550-3213(95)00653-2}

\bibitem{Boyd:1997kz}
Boyd, C.G., Grinstein, B., Lebed, R.F.: {Precision corrections to dispersive
  bounds on form-factors}.
\newblock Phys. Rev. D \textbf{56}, 6895--6911 (1997).
\newblock \doi{10.1103/PhysRevD.56.6895}

\bibitem{Braun:2012kp}
Braun, V.M., Khodjamirian, A.: {Soft contribution to $B\to \gamma \ell
  \nu_\ell$ and the $B$-meson distribution amplitude}.
\newblock Phys. Lett. \textbf{B718}, 1014--1019 (2013).
\newblock \doi{10.1016/j.physletb.2012.11.047}

\bibitem{Browder:2021hbl}
Browder, T.E., Deshpande, N.G., Mandal, R., Sinha, R.: {Impact of $B\to
  K\nu\bar{\nu}$ measurements on beyond the Standard Model theories}.
\newblock Phys. Rev. D \textbf{104}(5), 053007 (2021).
\newblock \doi{10.1103/PhysRevD.104.053007}

\bibitem{Buras:2014fpa}
Buras, A.J., Girrbach-Noe, J., Niehoff, C., Straub, D.M.: {$ B\to
  {K}^{\left(\ast \right)}\nu \overline{\nu} $ decays in the Standard Model and
  beyond}.
\newblock JHEP \textbf{02}, 184 (2015).
\newblock \doi{10.1007/JHEP02(2015)184}

\bibitem{Calibbi:2015kma}
Calibbi, L., Crivellin, A., Ota, T.: {Effective Field Theory Approach to $b\to
  s\ell\ell^{(')}$, $B\to K^{(*)}\nu\overline{\nu}$ and $B\to D^{(*)}\tau\nu$
  with Third Generation Couplings}.
\newblock Phys. Rev. Lett. \textbf{115}, 181801 (2015).
\newblock \doi{10.1103/PhysRevLett.115.181801}

\bibitem{Belle:2021ymg}
Cao, L., et~al.: {Measurement of Differential Branching Fractions of Inclusive
  $B \to X_u \, \ell^+\, \nu_{\ell}$ Decays}.
\newblock Phys. Rev. Lett. \textbf{127}(26), 261801 (2021).
\newblock \doi{10.1103/PhysRevLett.127.261801}

\bibitem{Belle:2021eni}
Cao, L., et~al.: {Measurements of Partial Branching Fractions of Inclusive $B
  \to X_u \, \ell^+\, \nu_{\ell}$ Decays with Hadronic Tagging}.
\newblock Phys. Rev. D \textbf{104}(1), 012008 (2021).
\newblock \doi{10.1103/PhysRevD.104.012008}

\bibitem{Capdevila:2016ivx}
Capdevila, B., Descotes-Genon, S., Matias, J., Virto, J.: {Assessing
  lepton-flavour non-universality from $B\to K^*\ell\ell$ angular analyses}.
\newblock JHEP \textbf{10}, 075 (2016).
\newblock \doi{10.1007/JHEP10(2016)075}

\bibitem{Capdevila:2021vkf}
Capdevila, B., Gambino, P., Nandi, S.: {Perturbative corrections to power
  suppressed effects in $\bar B\to X_u\ell\nu$}.
\newblock JHEP \textbf{04}, 137 (2021).
\newblock \doi{10.1007/JHEP04(2021)137}

\bibitem{Cappiello:2012vg}
Cappiello, L., Cata, O., D'Ambrosio, G.: {Standard Model prediction and new
  physics tests for $D^0 \to h^+ h^- \ell^+ \ell^- (h=\pi,K: \ell=e,\mu)$}.
\newblock JHEP \textbf{04}, 135 (2013).
\newblock \doi{10.1007/JHEP04(2013)135}

\bibitem{Caprini:1997mu}
Caprini, I., Lellouch, L., Neubert, M.: {Dispersive bounds on the shape of
  anti-B ---\ensuremath{>} D(*) lepton anti-neutrino form-factors}.
\newblock Nucl. Phys. B \textbf{530}, 153--181 (1998).
\newblock \doi{10.1016/S0550-3213(98)00350-2}

\bibitem{Belle:2019rba}
Caria, G., et~al.: {Measurement of $\mathcal{R}(D)$ and $\mathcal{R}(D^*)$ with
  a semileptonic tagging method}.
\newblock Phys. Rev. Lett. \textbf{124}(16), 161803 (2020).
\newblock \doi{10.1103/PhysRevLett.124.161803}

\bibitem{Carrasco:2014poa}
Carrasco, N., et~al.: {Leptonic decay constants $f_{K}$, $f_{D}$, and
  $f_{{D}_{s}}$ with $N_{f} = 2+1+1$ twisted-mass lattice QCD}.
\newblock Phys. Rev. \textbf{D91}(5), 054507 (2015).
\newblock \doi{10.1103/PhysRevD.91.054507}

\bibitem{Chakraborty:2021qav}
Chakraborty, B., Parrott, W.G., Bouchard, C., Davies, C.T.H., Koponen, J.,
  Lepage, G.P.: {Improved Vcs determination using precise lattice QCD form
  factors for $D\to K\ell\nu$}.
\newblock Phys. Rev. D \textbf{104}(3), 034505 (2021).
\newblock \doi{10.1103/PhysRevD.104.034505}

\bibitem{Cheng:2017smj}
Cheng, S., Khodjamirian, A., Virto, J.: {$B\to\pi\pi$ Form Factors from
  Light-Cone Sum Rules with $B$-meson Distribution Amplitudes}.
\newblock JHEP \textbf{05}, 157 (2017).
\newblock \doi{10.1007/JHEP05(2017)157}

\bibitem{Cheng:2017sfk}
Cheng, S., Khodjamirian, A., Virto, J.: {Timelike-helicity $B\to \pi\pi$ form
  factor from light-cone sum rules with dipion distribution amplitudes}.
\newblock Phys. Rev. D \textbf{96}(5), 051901 (2017).
\newblock \doi{10.1103/PhysRevD.96.051901}

\bibitem{BELLE:2019xld}
Choudhury, S., et~al.: {Test of lepton flavor universality and search for
  lepton flavor violation in $B \rightarrow K\ell \ell$ decays}.
\newblock JHEP \textbf{03}, 105 (2021).
\newblock \doi{10.1007/JHEP03(2021)105}

\bibitem{Ciuchini:2015qxb}
Ciuchini, M., Fedele, M., Franco, E., Mishima, S., Paul, A., Silvestrini, L.,
  Valli, M.: {$B\to K^* \ell^+ \ell^-$ decays at large recoil in the Standard
  Model: a theoretical reappraisal}.
\newblock JHEP \textbf{06}, 116 (2016).
\newblock \doi{10.1007/JHEP06(2016)116}

\bibitem{Colquhoun:2022atw}
Colquhoun, B., et~al.: {Form factors of $B\to\pi\ell\nu$ and a determination of
  $|V_{ub}|$ with M\"obius domain-wall-fermions}  (2022)

\bibitem{Dalgic:2006dt}
Dalgic, E., Gray, A., Wingate, M., Davies, C.T.H., Lepage, G.P., Shigemitsu,
  J.: {B meson semileptonic form-factors from unquenched lattice QCD}.
\newblock Phys. Rev. D \textbf{73}, 074502 (2006).
\newblock \doi{10.1103/PhysRevD.75.119906}.
\newblock [Erratum: Phys.Rev.D 75, 119906 (2007)]

\bibitem{Danilina:2019dji}
Danilina, A., Nikitin, N., Toms, K.: {Decays of charged $B$-mesons into three
  charged leptons and a neutrino}.
\newblock Phys. Rev. D \textbf{101}(9), 096007 (2020).
\newblock \doi{10.1103/PhysRevD.101.096007}

\bibitem{Danilina:2018uzr}
Danilina, A.V., Nikitin, N.V.: {Four-Leptonic Decays of Charged and Neutral $B$
  Mesons within the Standard Model}.
\newblock Phys. Atom. Nucl. \textbf{81}(3), 347--359 (2018).
\newblock \doi{10.1134/S1063778818030092}.
\newblock [Yad. Fiz.81,no.3,331(2018)]

\bibitem{Dassinger:2006md}
Dassinger, B.M., Mannel, T., Turczyk, S.: {Inclusive semi-leptonic B decays to
  order 1 / m(b)**4}.
\newblock JHEP \textbf{03}, 087 (2007).
\newblock \doi{10.1088/1126-6708/2007/03/087}

\bibitem{DeBruyn:2012wk}
De~Bruyn, K., Fleischer, R., Knegjens, R., Koppenburg, P., Merk, M.,
  Pellegrino, A., Tuning, N.: {Probing New Physics via the $B^0_s\to
  \mu^+\mu^-$ Effective Lifetime}.
\newblock Phys. Rev. Lett. \textbf{109}, 041801 (2012).
\newblock \doi{10.1103/PhysRevLett.109.041801}

\bibitem{Descotes-Genon:2020buf}
Descotes-Genon, S., Fajfer, S., Kamenik, J.F., Novoa-Brunet, M.: {Implications
  of $b\to s\mu\mu$ anomalies for future measurements of $B \to K^{(*)} \nu
  \bar \nu$ and $K\to \pi \nu \bar \nu$}.
\newblock Phys. Lett. B \textbf{809}, 135769 (2020).
\newblock \doi{10.1016/j.physletb.2020.135769}

\bibitem{Descotes-Genon:2013vna}
Descotes-Genon, S., Hurth, T., Matias, J., Virto, J.: {Optimizing the basis of
  $B\to K^*ll$ observables in the full kinematic range}.
\newblock JHEP \textbf{05}, 137 (2013).
\newblock \doi{10.1007/JHEP05(2013)137}

\bibitem{Descotes-Genon:2019bud}
Descotes-Genon, S., Khodjamirian, A., Virto, J.: {Light-cone sum rules for
  $B\to K\pi$ form factors and applications to rare decays}.
\newblock JHEP \textbf{12}, 083 (2019).
\newblock \doi{10.1007/JHEP12(2019)083}

\bibitem{Descotes-Genon:2012isb}
Descotes-Genon, S., Matias, J., Ramon, M., Virto, J.: {Implications from clean
  observables for the binned analysis of $B -> K*\mu^+\mu^-$ at large recoil}.
\newblock JHEP \textbf{01}, 048 (2013).
\newblock \doi{10.1007/JHEP01(2013)048}

\bibitem{Detmold:2015aaa}
Detmold, W., Lehner, C., Meinel, S.: {$\Lambda_b \to p \ell^- \bar{\nu}_\ell$
  and $\Lambda_b \to \Lambda_c \ell^- \bar{\nu}_\ell$ form factors from lattice
  QCD with relativistic heavy quarks}.
\newblock Phys. Rev. D \textbf{92}(3), 034503 (2015).
\newblock \doi{10.1103/PhysRevD.92.034503}

\bibitem{Detmold:2016pkz}
Detmold, W., Meinel, S.: {$\Lambda_b \to \Lambda \ell^+ \ell^-$ form factors,
  differential branching fraction, and angular observables from lattice QCD
  with relativistic $b$ quarks}.
\newblock Phys. Rev. D \textbf{93}(7), 074501 (2016).
\newblock \doi{10.1103/PhysRevD.93.074501}

\bibitem{Dimou:2012un}
Dimou, M., Lyon, J., Zwicky, R.: {Exclusive Chromomagnetism in heavy-to-light
  FCNCs}.
\newblock Phys. Rev. D \textbf{87}(7), 074008 (2013).
\newblock \doi{10.1103/PhysRevD.87.074008}

\bibitem{Du:2015tda}
Du, D., El-Khadra, A.X., Gottlieb, S., Kronfeld, A.S., Laiho, J., Lunghi, E.,
  Van~de Water, R.S., Zhou, R.: {Phenomenology of semileptonic B-meson decays
  with form factors from lattice QCD}.
\newblock Phys. Rev. D \textbf{93}(3), 034005 (2016).
\newblock \doi{10.1103/PhysRevD.93.034005}

\bibitem{Egede:2008uy}
Egede, U., Hurth, T., Matias, J., Ramon, M., Reece, W.: {New observables in the
  decay mode $\bar B_d \to \bar K^{*0} l^+ l^-$}.
\newblock JHEP \textbf{11}, 032 (2008).
\newblock \doi{10.1088/1126-6708/2008/11/032}

\bibitem{Fael:2018vsp}
Fael, M., Mannel, T., Keri~Vos, K.: {$V_{cb}$ determination from inclusive $b
  \to c$ decays: an alternative method}.
\newblock JHEP \textbf{02}, 177 (2019).
\newblock \doi{10.1007/JHEP02(2019)177}

\bibitem{Fael:2020tow}
Fael, M., Sch\"onwald, K., Steinhauser, M.: {Third order corrections to the
  semileptonic $b\to c$ and the muon decays}.
\newblock Phys. Rev. D \textbf{104}(1), 016003 (2021).
\newblock \doi{10.1103/PhysRevD.104.016003}

\bibitem{Fajfer:2015mia}
Fajfer, S., Ko\v{s}nik, N.: {Prospects of discovering new physics in rare charm
  decays}.
\newblock Eur. Phys. J. C \textbf{75}(12), 567 (2015).
\newblock \doi{10.1140/epjc/s10052-015-3801-2}

\bibitem{FK2015}
Fajfer, S., Košnik, N.: {Prospects of discovering new physics in rare charm
  decays}.
\newblock Eur. Phys. J. \textbf{C75}(12), 567 (2015).
\newblock \doi{10.1140/epjc/s10052-015-3801-2}

\bibitem{Faller:2013dwa}
Faller, S., Feldmann, T., Khodjamirian, A., Mannel, T., van Dyk, D.:
  {Disentangling the Decay Observables in $B^- \to
  \pi^+\pi^-\ell^-\bar\nu_\ell$}.
\newblock Phys. Rev. D \textbf{89}(1), 014015 (2014).
\newblock \doi{10.1103/PhysRevD.89.014015}

\bibitem{Feldmann:2017izn}
Feldmann, T., M\"uller, B., Seidel, D.: {$D \to \rho \,\ell^+\ell^-$ decays in
  the QCD factorization approach}.
\newblock JHEP \textbf{08}, 105 (2017).
\newblock \doi{10.1007/JHEP08(2017)105}

\bibitem{Feldmann:2018kqr}
Feldmann, T., Van~Dyk, D., Vos, K.K.: {Revisiting $B \to \pi\pi \ell \nu$ at
  large dipion masses}.
\newblock JHEP \textbf{10}, 030 (2018).
\newblock \doi{10.1007/JHEP10(2018)030}

\bibitem{Felkl:2021uxi}
Felkl, T., Li, S.L., Schmidt, M.A.: {A tale of invisibility: constraints on new
  physics in $b \to s\nu\overline{\nu}$}.
\newblock JHEP \textbf{12}, 118 (2021).
\newblock \doi{10.1007/JHEP12(2021)118}

\bibitem{Flynn:2015mha}
Flynn, J.M., Izubuchi, T., Kawanai, T., Lehner, C., Soni, A., Van~de Water,
  R.S., Witzel, O.: {$B \to \pi \ell \nu$ and $B_s \to K \ell \nu$ form factors
  and $|V_{ub}|$ from 2+1-flavor lattice QCD with domain-wall light quarks and
  relativistic heavy quarks}.
\newblock Phys. Rev. D \textbf{91}(7), 074510 (2015).
\newblock \doi{10.1103/PhysRevD.91.074510}

\bibitem{Gambino:2015ima}
Gambino, P.: {Inclusive semileptonic B decays and |V$_{cb}$|: In memoriam Kolya
  Uraltsev}.
\newblock Int. J. Mod. Phys. A \textbf{30}(10), 1543002 (2015).
\newblock \doi{10.1142/S0217751X15430022}

\bibitem{Gambino:2007rp}
Gambino, P., Giordano, P., Ossola, G., Uraltsev, N.: {Inclusive semileptonic B
  decays and the determination of |V(ub)|}.
\newblock JHEP \textbf{10}, 058 (2007).
\newblock \doi{10.1088/1126-6708/2007/10/058}

\bibitem{Gambino:2003zm}
Gambino, P., Gorbahn, M., Haisch, U.: {Anomalous dimension matrix for radiative
  and rare semileptonic B decays up to three loops}.
\newblock Nucl. Phys. B \textbf{673}, 238--262 (2003).
\newblock \doi{10.1016/j.nuclphysb.2003.09.024}

\bibitem{Gambino:2020crt}
Gambino, P., Hashimoto, S.: {Inclusive Semileptonic Decays from Lattice QCD}.
\newblock Phys. Rev. Lett. \textbf{125}(3), 032001 (2020).
\newblock \doi{10.1103/PhysRevLett.125.032001}

\bibitem{Gambino:2022dvu}
Gambino, P., Hashimoto, S., M\"achler, S., Panero, M., Sanfilippo, F., Simula,
  S., Smecca, A., Tantalo, N.: {Lattice QCD study of inclusive semileptonic
  decays of heavy mesons}  (2022)

\bibitem{Gambino:2016fdy}
Gambino, P., Healey, K.J., Mondino, C.: {Neural network approach to $B\to X_u
  \ell \nu$}.
\newblock Phys. Rev. D \textbf{94}(1), 014031 (2016).
\newblock \doi{10.1103/PhysRevD.94.014031}

\bibitem{Gambino:2016jkc}
Gambino, P., Healey, K.J., Turczyk, S.: {Taming the higher power corrections in
  semileptonic B decays}.
\newblock Phys. Lett. B \textbf{763}, 60--65 (2016).
\newblock \doi{10.1016/j.physletb.2016.10.023}

\bibitem{Gambino:2019sif}
Gambino, P., Jung, M., Schacht, S.: {The $V_{cb}$ puzzle: An update}.
\newblock Phys. Lett. B \textbf{795}, 386--390 (2019).
\newblock \doi{10.1016/j.physletb.2019.06.039}

\bibitem{Gambino:2013rza}
Gambino, P., Schwanda, C.: {Inclusive semileptonic fits, heavy quark masses,
  and $V_{cb}$}.
\newblock Phys. Rev. D \textbf{89}(1), 014022 (2014).
\newblock \doi{10.1103/PhysRevD.89.014022}

\bibitem{Gambino:2020jvv}
Gambino, P., et~al.: {Challenges in semileptonic $B$ decays}.
\newblock Eur. Phys. J. C \textbf{80}(10), 966 (2020).
\newblock \doi{10.1140/epjc/s10052-020-08490-x}

\bibitem{Gelb:2018end}
Gelb, M., et~al.: {Search for the rare decay of $B^+ \to \ell^{+} \nu_{\ell}
  \gamma$ with improved hadronic tagging}.
\newblock Phys. Rev. D \textbf{98}(11), 112016 (2018).
\newblock \doi{10.1103/PhysRevD.98.112016}

\bibitem{Gershon:2016fda}
Gershon, T., Gligorov, V.V.: {$CP$ violation in the $B$ system}.
\newblock Rept. Prog. Phys. \textbf{80}(4), 046201 (2017).
\newblock \doi{10.1088/1361-6633/aa5514}

\bibitem{Gorbahn:2004my}
Gorbahn, M., Haisch, U.: {Effective Hamiltonian for non-leptonic $|\Delta F| =
  1$ decays at NNLO in QCD}.
\newblock Nucl. Phys. B \textbf{713}, 291--332 (2005).
\newblock \doi{10.1016/j.nuclphysb.2005.01.047}

\bibitem{Gorbahn:2005sa}
Gorbahn, M., Haisch, U., Misiak, M.: {Three-loop mixing of dipole operators}.
\newblock Phys. Rev. Lett. \textbf{95}, 102004 (2005).
\newblock \doi{10.1103/PhysRevLett.95.102004}

\bibitem{Gratrex:2015hna}
Gratrex, J., Hopfer, M., Zwicky, R.: {Generalised helicity formalism, higher
  moments and the $B \to K_{J_K}(\to K \pi) \bar{\ell}_1 \ell_2$ angular
  distributions}.
\newblock Phys. Rev. D \textbf{93}(5), 054008 (2016).
\newblock \doi{10.1103/PhysRevD.93.054008}

\bibitem{Gremm:1996df}
Gremm, M., Kapustin, A.: {Order $1/m_b^3$ corrections to $B \to X_c
  \ell\bar\nu$ decay and their implication for the measurement of Lambda-bar
  and lambda(1)}.
\newblock Phys. Rev. D \textbf{55}, 6924--6932 (1997).
\newblock \doi{10.1103/PhysRevD.55.6924}

\bibitem{Grinstein:2017nlq}
Grinstein, B., Kobach, A.: {Model-Independent Extraction of $|V_{cb}|$ from
  $\bar{B}\rightarrow D^* \ell \overline{\nu}$}.
\newblock Phys. Lett. B \textbf{771}, 359--364 (2017).
\newblock \doi{10.1016/j.physletb.2017.05.078}

\bibitem{Gronau:2001ng}
Gronau, M., Grossman, Y., Pirjol, D., Ryd, A.: {Measuring the photon
  polarization in B ---\ensuremath{>} K pi pi gamma}.
\newblock Phys. Rev. Lett. \textbf{88}, 051802 (2002).
\newblock \doi{10.1103/PhysRevLett.88.051802}

\bibitem{Gronau:2002rz}
Gronau, M., Pirjol, D.: {Photon polarization in radiative B decays}.
\newblock Phys. Rev. D \textbf{66}, 054008 (2002).
\newblock \doi{10.1103/PhysRevD.66.054008}

\bibitem{Grozin:1996pq}
Grozin, A.G., Neubert, M.: {Asymptotics of heavy meson form-factors}.
\newblock Phys. Rev. \textbf{D55}, 272--290 (1997).
\newblock \doi{10.1103/PhysRevD.55.272}

\bibitem{Belle:2017oht}
Grygier, J., et~al.: {Search for $\boldsymbol{B\to h\nu\bar{\nu}}$ decays with
  semileptonic tagging at Belle}.
\newblock Phys. Rev. D \textbf{96}(9), 091101 (2017).
\newblock \doi{10.1103/PhysRevD.96.091101}.
\newblock [Addendum: Phys.Rev.D 97, 099902 (2018)]

\bibitem{Guadagnoli:2017quo}
Guadagnoli, D., Reboud, M., Zwicky, R.: {$B_s^0\to\ell^+\ell^-\gamma$ as a test
  of lepton flavor universality}.
\newblock JHEP \textbf{11}, 184 (2017).
\newblock \doi{10.1007/JHEP11(2017)184}

\bibitem{Gubernari:2020eft}
Gubernari, N., van Dyk, D., Virto, J.: {Non-local matrix elements in
  $B_{(s)}\to K^{(*)},\phi\ell^+\ell^-$}.
\newblock JHEP \textbf{02}, 088 (2021).
\newblock \doi{10.1007/JHEP02(2021)088}

\bibitem{Gubernari:2022hrq}
Gubernari, N., Khodjamirian, A., Mandal, R., Mannel, T.: {$B\to D_1(2420)$ and
  $B\to D_1'(2430)$ form factors from QCD light-cone sum rules}  (2022)

\bibitem{Gubernari:2018wyi}
Gubernari, N., Kokulu, A., van Dyk, D.: {$B\to P$ and $B\to V$ Form Factors
  from $B$-Meson Light-Cone Sum Rules beyond Leading Twist}.
\newblock JHEP \textbf{01}, 150 (2019).
\newblock \doi{10.1007/JHEP01(2019)150}

\bibitem{Gunawardana:2017zix}
Gunawardana, A., Paz, G.: {On HQET and NRQCD Operators of Dimension 8 and
  Above}.
\newblock JHEP \textbf{07}, 137 (2017).
\newblock \doi{10.1007/JHEP07(2017)137}

\bibitem{Hambrock:2015aor}
Hambrock, C., Khodjamirian, A.: {Form factors in $\bar B^0 \to
  \pi\pi\ell\bar\nu_\ell$ from QCD light-cone sum rules}.
\newblock Nucl. Phys. B \textbf{905}, 373--390 (2016).
\newblock \doi{10.1016/j.nuclphysb.2016.02.035}

\bibitem{Harrison:2017fmw}
Harrison, J., Davies, C., Wingate, M.:

\bibitem{Harrison:2021tol}
Harrison, J., Davies, C.T.H.: {$B_s \rightarrow D_s^*$ Form Factors for the
  full $q^2$ range from Lattice QCD}  (2021)

\bibitem{Harrison:2020gvo}
Harrison, J., Davies, C.T.H., Lytle, A.: {$B_c \rightarrow J/\psi$ form factors
  for the full $q^2$ range from lattice QCD}.
\newblock Phys. Rev. D \textbf{102}(9), 094518 (2020).
\newblock \doi{10.1103/PhysRevD.102.094518}

\bibitem{Hayashi:2022hjk}
Hayashi, Y., Sumino, Y., Takaura, H.: {Determination of $|V_{cb}|$ using N3LO
  perturbative corrections to $\Gamma(B\to X_c\ell\nu)$ and 1S masses}.
\newblock Phys. Lett. B \textbf{829}, 137068 (2022).
\newblock \doi{10.1016/j.physletb.2022.137068}

\bibitem{Heinonen:2016cwm}
Heinonen, J., Mannel, T.: {Revisiting Uraltsev's BPS limit for Heavy Quarks}
  (2016)

\bibitem{Heller:2015vvm}
Heller, A., et~al.: {Search for $B^+ \to \ell^+ \nu_\ell \gamma$ decays with
  hadronic tagging using the full Belle data sample}.
\newblock Phys. Rev. \textbf{D91}(11), 112009 (2015).
\newblock \doi{10.1103/PhysRevD.91.112009}

\bibitem{Hermann:2013kca}
Hermann, T., Misiak, M., Steinhauser, M.: {Three-loop QCD corrections to $B_s
  \to \mu^+ \mu^-$}.
\newblock JHEP \textbf{12}, 097 (2013).
\newblock \doi{10.1007/JHEP12(2013)097}

\bibitem{Hiller:2001zj}
Hiller, G., Kagan, A.: {Probing for new physics in polarized $\Lambda_b$ decays
  at the $Z$}.
\newblock Phys. Rev. D \textbf{65}, 074038 (2002).
\newblock \doi{10.1103/PhysRevD.65.074038}

\bibitem{Hiller:2003js}
Hiller, G., Kruger, F.: {More model-independent analysis of $b \to s$
  processes}.
\newblock Phys. Rev. D \textbf{69}, 074020 (2004).
\newblock \doi{10.1103/PhysRevD.69.074020}

\bibitem{Belle:2016dyj}
Hirose, S., et~al.: {Measurement of the $\tau$ lepton polarization and $R(D^*)$
  in the decay $\bar{B} \to D^* \tau^- \bar{\nu}_\tau$}.
\newblock Phys. Rev. Lett. \textbf{118}(21), 211801 (2017).
\newblock \doi{10.1103/PhysRevLett.118.211801}

\bibitem{Horgan:2013pva}
Horgan, R.R., Liu, Z., Meinel, S., Wingate, M.: {Calculation of $B^0 \to K^{*0}
  \mu^+ \mu^-$ and $B_s^0 \to \phi \mu^+ \mu^-$ observables using form factors
  from lattice QCD}.
\newblock Phys. Rev. Lett. \textbf{112}, 212003 (2014).
\newblock \doi{10.1103/PhysRevLett.112.212003}

\bibitem{Horgan:2013hoa}
Horgan, R.R., Liu, Z., Meinel, S., Wingate, M.: {Lattice QCD calculation of
  form factors describing the rare decays $B \to K^* \ell^+ \ell^-$ and $B_s
  \to \phi \ell^+ \ell^-$}.
\newblock Phys. Rev. D \textbf{89}(9), 094501 (2014).
\newblock \doi{10.1103/PhysRevD.89.094501}

\bibitem{Hostert:2020gou}
Hostert, M., Kaneta, K., Pospelov, M.: {Pair production of dark particles in
  meson decays}.
\newblock Phys. Rev. D \textbf{102}(5), 055016 (2020).
\newblock \doi{10.1103/PhysRevD.102.055016}

\bibitem{Huber:2014nna}
Huber, T., Poradzi\'nski, M., Virto, J.: {Four-body contributions to $
  \overline{B}\to {X}_s\gamma $ at NLO}.
\newblock JHEP \textbf{01}, 115 (2015).
\newblock \doi{10.1007/JHEP01(2015)115}

\bibitem{Hurth:2016fbr}
Hurth, T., Mahmoudi, F., Neshatpour, S.: {On the anomalies in the latest LHCb
  data}.
\newblock Nucl. Phys. B \textbf{909}, 737--777 (2016).
\newblock \doi{10.1016/j.nuclphysb.2016.05.022}

\bibitem{Hurth:2020rzx}
Hurth, T., Mahmoudi, F., Neshatpour, S.: {Implications of the new LHCb angular
  analysis of $B \to K^* \mu^+ \mu^-$ : Hadronic effects or new physics?}
\newblock Phys. Rev. D \textbf{102}(5), 055001 (2020).
\newblock \doi{10.1103/PhysRevD.102.055001}

\bibitem{Belle:2015qfa}
Huschle, M., et~al.: {Measurement of the branching ratio of $\bar{B} \to
  D^{(\ast)} \tau^- \bar{\nu}_\tau$ relative to $\bar{B} \to D^{(\ast)} \ell^-
  \bar{\nu}_\ell$ decays with hadronic tagging at Belle}.
\newblock Phys. Rev. D \textbf{92}(7), 072014 (2015).
\newblock \doi{10.1103/PhysRevD.92.072014}

\bibitem{Isidori:2020acz}
Isidori, G., Nabeebaccus, S., Zwicky, R.: {QED corrections in $ \overline{B}\to
  \overline{K}{\mathrm{\ell}}^{+}{\mathrm{\ell}}^{-} $ at the
  double-differential level}.
\newblock JHEP \textbf{12}, 104 (2020).
\newblock \doi{10.1007/JHEP12(2020)104}

\bibitem{Jager:2012uw}
J\"ager, S., Martin~Camalich, J.: {On $B \to V \ell \ell$ at small dilepton
  invariant mass, power corrections, and new physics}.
\newblock JHEP \textbf{05}, 043 (2013).
\newblock \doi{10.1007/JHEP05(2013)043}

\bibitem{Jager:2014rwa}
J\"ager, S., Martin~Camalich, J.: {Reassessing the discovery potential of the
  $B \to K^{*} \ell^+\ell^-$ decays in the large-recoil region: SM challenges
  and BSM opportunities}.
\newblock Phys. Rev. D \textbf{93}(1), 014028 (2016).
\newblock \doi{10.1103/PhysRevD.93.014028}

\bibitem{Jaiswal:2017rve}
Jaiswal, S., Nandi, S., Patra, S.K.: {Extraction of $|V_{cb}|$ from $B\to
  D^{(*)}\ell\nu_\ell$ and the Standard Model predictions of $R(D^{(*)})$}.
\newblock JHEP \textbf{12}, 060 (2017).
\newblock \doi{10.1007/JHEP12(2017)060}

\bibitem{Jaiswal:2020wer}
Jaiswal, S., Nandi, S., Patra, S.K.: {Updates on extraction of |V$_{cb}$| and
  SM prediction of R(D*) in $B\to D^{*}\ell\nu_\ell$ decays}.
\newblock JHEP \textbf{06}, 165 (2020).
\newblock \doi{10.1007/JHEP06(2020)165}

\bibitem{Janowski:2021yvz}
Janowski, T., Pullin, B., Zwicky, R.: {Charged and neutral $
  {\overline{B}}_{u,d,s} $ $\to \gamma$ form factors from light cone sum rules
  at NLO}.
\newblock JHEP \textbf{12}, 008 (2021).
\newblock \doi{10.1007/JHEP12(2021)008}

\bibitem{CMS:2014xfa}
Khachatryan, V., et~al.: {Observation of the rare $B^0_s\to\mu^+\mu^-$ decay
  from the combined analysis of CMS and LHCb data}.
\newblock Nature \textbf{522}, 68--72 (2015).
\newblock \doi{10.1038/nature14474}

\bibitem{CMS:2015bcy}
Khachatryan, V., et~al.: {Angular analysis of the decay $B^0 \to K^{*0} \mu^+
  \mu^-$ from pp collisions at $\sqrt s = 8$ TeV}.
\newblock Phys. Lett. B \textbf{753}, 424--448 (2016).
\newblock \doi{10.1016/j.physletb.2015.12.020}

\bibitem{Khodjamirian:2011jp}
Khodjamirian, A., Klein, C., Mannel, T., Wang, Y.M.: {Form Factors and Strong
  Couplings of Heavy Baryons from QCD Light-Cone Sum Rules}.
\newblock JHEP \textbf{09}, 106 (2011).
\newblock \doi{10.1007/JHEP09(2011)106}

\bibitem{Khodjamirian:2010vf}
Khodjamirian, A., Mannel, T., Pivovarov, A.A., Wang, Y.M.: {Charm-loop effect
  in $B \to K^{(*)} \ell^{+} \ell^{-}$ and $B\to K^*\gamma$}.
\newblock JHEP \textbf{09}, 089 (2010).
\newblock \doi{10.1007/JHEP09(2010)089}

\bibitem{Khodjamirian:2012rm}
Khodjamirian, A., Mannel, T., Wang, Y.M.: {$B \to K \ell^{+}\ell^{-}$ decay at
  large hadronic recoil}.
\newblock JHEP \textbf{02}, 010 (2013).
\newblock \doi{10.1007/JHEP02(2013)010}

\bibitem{Khodjamirian:2017fxg}
Khodjamirian, A., Rusov, A.V.: {$B_{s}\to K \ell \nu_\ell$ and $B_{(s)} \to \pi
  (K) \ell^+\ell^-$ decays at large recoil and CKM matrix elements}.
\newblock JHEP \textbf{08}, 112 (2017).
\newblock \doi{10.1007/JHEP08(2017)112}

\bibitem{Kim:2000dq}
Kim, C.S., Kim, Y.G., Lu, C.D., Morozumi, T.: {Azimuthal angle distribution in
  B ---\ensuremath{>} K* (---\ensuremath{>} K pi) lepton+ lepton- at low
  invariant m(lepton+ lepton-) region}.
\newblock Phys. Rev. D \textbf{62}, 034013 (2000).
\newblock \doi{10.1103/PhysRevD.62.034013}

\bibitem{Kou:2010kn}
Kou, E., Le~Yaouanc, A., Tayduganov, A.: {Determining the photon polarization
  of the b --\ensuremath{>} s gamma using the B --\ensuremath{>} K1(1270) gamma
  --\ensuremath{>} (K pi pi) gamma decay}.
\newblock Phys. Rev. D \textbf{83}, 094007 (2011).
\newblock \doi{10.1103/PhysRevD.83.094007}

\bibitem{Kou:2016iau}
Kou, E., Le~Yaouanc, A., Tayduganov, A.: {Angular analysis of B -\ensuremath{>}
  J/psi K1 : towards a model independent determination of the photon
  polarization with B-\ensuremath{>} K1 gamma}.
\newblock Phys. Lett. B \textbf{763}, 66--71 (2016).
\newblock \doi{10.1016/j.physletb.2016.10.013}

\bibitem{Kruger:2005ep}
Kruger, F., Matias, J.: {Probing new physics via the transverse amplitudes of
  $B^0\to K^{*0} (\to K^- \pi^+) l^+l^-$ at large recoil}.
\newblock Phys. Rev. D \textbf{71}, 094009 (2005).
\newblock \doi{10.1103/PhysRevD.71.094009}

\bibitem{Kruger:1999xa}
Kruger, F., Sehgal, L.M., Sinha, N., Sinha, R.: {Angular distribution and CP
  asymmetries in the decays $\bar B \to K^- \pi^+ e^- e^+$ and $\bar B \to
  \pi^- \pi^+ e^- e^+$}.
\newblock Phys. Rev. D \textbf{61}, 114028 (2000).
\newblock \doi{10.1103/PhysRevD.61.114028}.
\newblock [Erratum: Phys.Rev.D 63, 019901 (2001)]

\bibitem{Lange:2005yw}
Lange, B.O., Neubert, M., Paz, G.: {Theory of charmless inclusive B decays and
  the extraction of V(ub)}.
\newblock Phys. Rev. D \textbf{72}, 073006 (2005).
\newblock \doi{10.1103/PhysRevD.72.073006}

\bibitem{BaBar:2012obs}
Lees, J.P., et~al.: {Evidence for an excess of $\bar{B} \to D^{(*)}
  \tau^-\bar{\nu}_\tau$ decays}.
\newblock Phys. Rev. Lett. \textbf{109}, 101802 (2012).
\newblock \doi{10.1103/PhysRevLett.109.101802}

\bibitem{BaBar:2012mrf}
Lees, J.P., et~al.: {Measurement of Branching Fractions and Rate Asymmetries in
  the Rare Decays $B \to K^{(*)} l^+ l^-$}.
\newblock Phys. Rev. D \textbf{86}, 032012 (2012).
\newblock \doi{10.1103/PhysRevD.86.032012}

\bibitem{BaBar:2013mob}
Lees, J.P., et~al.: {Measurement of an Excess of $\bar{B} \to D^{(*)}\tau^-
  \bar{\nu}_\tau$ Decays and Implications for Charged Higgs Bosons}.
\newblock Phys. Rev. D \textbf{88}(7), 072012 (2013).
\newblock \doi{10.1103/PhysRevD.88.072012}

\bibitem{BaBar:2013npw}
Lees, J.P., et~al.: {Search for $B \to K^{(*)} \nu \overline \nu$ and invisible
  quarkonium decays}.
\newblock Phys. Rev. D \textbf{87}(11), 112005 (2013).
\newblock \doi{10.1103/PhysRevD.87.112005}

\bibitem{BaBar:2016rxh}
Lees, J.P., et~al.: {Measurement of the inclusive electron spectrum from B
  meson decays and determination of $|V_{ub}|$}.
\newblock Phys. Rev. D \textbf{95}(7), 072001 (2017).
\newblock \doi{10.1103/PhysRevD.95.072001}

\bibitem{BaBar:2019vpl}
Lees, J.P., et~al.: {Extraction of form Factors from a Four-Dimensional Angular
  Analysis of $\overline{B} \rightarrow D^\ast \ell^- \overline{\nu}_\ell$}.
\newblock Phys. Rev. Lett. \textbf{123}(9), 091801 (2019).
\newblock \doi{10.1103/PhysRevLett.123.091801}

\bibitem{Legger:2006cq}
Legger, F., Schietinger, T.: {Photon helicity in $\Lambda_b \to p K \gamma$
  decays}.
\newblock Phys. Lett. B \textbf{645}, 204--212 (2007).
\newblock \doi{10.1016/j.physletb.2006.12.011}.
\newblock [Erratum: Phys.Lett.B 647, 527--528 (2007)]

\bibitem{Leljak:2021vte}
Leljak, D., Meli\'c, B., van Dyk, D.: {The $ \overline{B} \to\pi$ form factors
  from QCD and their impact on |V$_{ub}$|}.
\newblock JHEP \textbf{07}, 036 (2021).
\newblock \doi{10.1007/JHEP07(2021)036}

\bibitem{Ligeti:2008ac}
Ligeti, Z., Stewart, I.W., Tackmann, F.J.: {Treating the b quark distribution
  function with reliable uncertainties}.
\newblock Phys. Rev. D \textbf{78}, 114014 (2008).
\newblock \doi{10.1103/PhysRevD.78.114014}

\bibitem{Lubicz:2017syv}
Lubicz, V., Riggio, L., Salerno, G., Simula, S., Tarantino, C.: {Scalar and
  vector form factors of $D \to \pi(K) \ell \nu$ decays with $N_f=2+1+1$
  twisted fermions}.
\newblock Phys. Rev. D \textbf{96}(5), 054514 (2017).
\newblock \doi{10.1103/PhysRevD.96.054514}.
\newblock [Erratum: Phys.Rev.D 99, 099902 (2019), Erratum: Phys.Rev.D 100,
  079901 (2019)]

\bibitem{Belle:2013tnz}
Lutz, O., et~al.: {Search for $B \to h^{(*)} \nu \bar{\nu}$ with the full Belle
  $\Upsilon(4S)$ data sample}.
\newblock Phys. Rev. D \textbf{87}(11), 111103 (2013).
\newblock \doi{10.1103/PhysRevD.87.111103}

\bibitem{Lyon:2013gba}
Lyon, J., Zwicky, R.: {Isospin asymmetries in $B\to(K^*,\rho)\gamma/l^+l^-$ and
  $B\to Kl^+l^-$ in and beyond the standard model}.
\newblock Phys. Rev. D \textbf{88}(9), 094004 (2013).
\newblock \doi{10.1103/PhysRevD.88.094004}

\bibitem{Mannel:2015osa}
Mannel, T., van Dyk, D.: {Zero-recoil sum rules for $\Lambda_b \to \Lambda_c$
  form factors}.
\newblock Phys. Lett. B \textbf{751}, 48--53 (2015).
\newblock \doi{10.1016/j.physletb.2015.10.016}

\bibitem{Mannel:2019qel}
Mannel, T., Pivovarov, A.A.: {QCD corrections to inclusive heavy hadron weak
  decays at $\Lambda_{\rm QCD}^3 /m_Q^3$}.
\newblock Phys. Rev. D \textbf{100}(9), 093001 (2019).
\newblock \doi{10.1103/PhysRevD.100.093001}

\bibitem{Mannel:2014xza}
Mannel, T., Pivovarov, A.A., Rosenthal, D.: {Inclusive semileptonic B decays
  from QCD with NLO accuracy for power suppressed terms}.
\newblock Phys. Lett. B \textbf{741}, 290--294 (2015).
\newblock \doi{10.1016/j.physletb.2014.12.058}

\bibitem{Mannel:2015jka}
Mannel, T., Pivovarov, A.A., Rosenthal, D.: {Inclusive weak decays of heavy
  hadrons with power suppressed terms at NLO}.
\newblock Phys. Rev. D \textbf{92}(5), 054025 (2015).
\newblock \doi{10.1103/PhysRevD.92.054025}

\bibitem{Mannel:2010wj}
Mannel, T., Turczyk, S., Uraltsev, N.: {Higher Order Power Corrections in
  Inclusive B Decays}.
\newblock JHEP \textbf{11}, 109 (2010).
\newblock \doi{10.1007/JHEP11(2010)109}

\bibitem{Mannel:2011xg}
Mannel, T., Wang, Y.M.: {Heavy-to-light baryonic form factors at large recoil}.
\newblock JHEP \textbf{12}, 067 (2011).
\newblock \doi{10.1007/JHEP12(2011)067}

\bibitem{Martinelli:2022tte}
Martinelli, G., Simula, S., Vittorio, L.: {Exclusive semileptonic $B \to \pi
  \ell \nu_\ell$ and $B_s \to K \ell \nu_\ell$ decays through unitarity and
  lattice QCD}  (2022)

\bibitem{McLean:2019qcx}
McLean, E., Davies, C.T.H., Koponen, J., Lytle, A.T.: {$B_s\to D_s \ell\nu$
  Form Factors for the full $q^2$ range from Lattice QCD with
  non-perturbatively normalized currents}.
\newblock Phys. Rev. D \textbf{101}(7), 074513 (2020).
\newblock \doi{10.1103/PhysRevD.101.074513}

\bibitem{Meinel:2016dqj}
Meinel, S.: {$\Lambda_c \to \Lambda l^+ \nu_l$ form factors and decay rates
  from lattice QCD with physical quark masses}.
\newblock Phys. Rev. Lett. \textbf{118}(8), 082001 (2017).
\newblock \doi{10.1103/PhysRevLett.118.082001}

\bibitem{Meinel:2017ggx}
Meinel, S.: {$\Lambda_c \to N$ form factors from lattice QCD and phenomenology
  of $\Lambda_c \to n \ell^+ \nu_\ell$ and $\Lambda_c \to p \mu^+ \mu^-$
  decays}.
\newblock Phys. Rev. D \textbf{97}(3), 034511 (2018).
\newblock \doi{10.1103/PhysRevD.97.034511}

\bibitem{Meinel:2021mdj}
Meinel, S., Rendon, G.:

\bibitem{Meinel:2020owd}
Meinel, S., Rendon, G.: {$\Lambda_b \to \Lambda^*(1520)\ell^+\ell^-$ form
  factors from lattice QCD}.
\newblock Phys. Rev. D \textbf{103}(7), 074505 (2021).
\newblock \doi{10.1103/PhysRevD.103.074505}

\bibitem{Misiak:2017woa}
Misiak, M., Rehman, A., Steinhauser, M.: {NNLO QCD counterterm contributions to
  $\bar B \to X_{s\gamma}$ for the physical value of $m_c$}.
\newblock Phys. Lett. B \textbf{770}, 431--439 (2017).
\newblock \doi{10.1016/j.physletb.2017.05.008}

\bibitem{Misiak:2020vlo}
Misiak, M., Rehman, A., Steinhauser, M.: {Towards $ \overline{B}\to {X}_s\gamma
  $ at the NNLO in QCD without interpolation in m$_{c}$}.
\newblock JHEP \textbf{06}, 175 (2020).
\newblock \doi{10.1007/JHEP06(2020)175}

\bibitem{Monahan:2018lzv}
Monahan, C.J., Bouchard, C.M., Lepage, G.P., Na, H., Shigemitsu, J.: {Form
  factor ratios for $B_s \rightarrow K \, \ell \, \nu$ and $B_s \rightarrow D_s
  \, \ell \, \nu$ semileptonic decays and $|V_{ub}/V_{cb}|$}.
\newblock Phys. Rev. D \textbf{98}(11), 114509 (2018).
\newblock \doi{10.1103/PhysRevD.98.114509}

\bibitem{Na:2015kha}
Na, H., Bouchard, C.M., Lepage, G.P., Monahan, C., Shigemitsu, J.: {$B
  \rightarrow D l \nu$ form factors at nonzero recoil and extraction of
  $|V_{cb}|$}.
\newblock Phys. Rev. D \textbf{92}(5), 054510 (2015).
\newblock \doi{10.1103/PhysRevD.93.119906}.
\newblock [Erratum: Phys.Rev.D 93, 119906 (2016)]

\bibitem{Nir:2020mgy}
Nir, Y., Vagnoni, V.: {$CP$ violation in $B$ decays}.
\newblock Comptes Rendus Physique \textbf{21}(1), 61--74 (2020).
\newblock \doi{10.5802/crphys.11}

\bibitem{Petrov:2017nwo}
Petrov, A.A.: {Theory of rare charm decays into leptons}.
\newblock PoS \textbf{CKM2016}, 059 (2017).
\newblock \doi{10.22323/1.291.0059}

\bibitem{Riggio:2017zwh}
Riggio, L., Salerno, G., Simula, S.: {Extraction of $|V_{cd}|$ and $|V_{cs}|$
  from experimental decay rates using lattice QCD $D \to \pi(K) \ell \nu$ form
  factors}.
\newblock Eur. Phys. J. C \textbf{78}(6), 501 (2018).
\newblock \doi{10.1140/epjc/s10052-018-5943-5}

\bibitem{Belle:2016ure}
Sato, Y., et~al.: {Measurement of the branching ratio of $\bar{B}^0 \rightarrow
  D^{*+} \tau^- \bar{\nu}_{\tau}$ relative to $\bar{B}^0 \rightarrow D^{*+}
  \ell^- \bar{\nu}_{\ell}$ decays with a semileptonic tagging method}.
\newblock Phys. Rev. D \textbf{94}(7), 072007 (2016).
\newblock \doi{10.1103/PhysRevD.94.072007}

\bibitem{Imsong:2014oqa}
Sentitemsu~Imsong, I., Khodjamirian, A., Mannel, T., van Dyk, D.:
  {Extrapolation and unitarity bounds for the $B\to\pi$ form factor}.
\newblock JHEP \textbf{02}, 126 (2015).
\newblock \doi{10.1007/JHEP02(2015)126}

\bibitem{Serra:2016ivr}
Serra, N., Silva~Coutinho, R., van Dyk, D.: {Measuring the breaking of lepton
  flavor universality in $B\to K^*\ell^+\ell^-$}.
\newblock Phys. Rev. D \textbf{95}(3), 035029 (2017).
\newblock \doi{10.1103/PhysRevD.95.035029}

\bibitem{CMS:2018qih}
Sirunyan, A.M., et~al.: {Angular analysis of the decay B$^+\to$ K$^+\mu^+\mu^-$
  in proton-proton collisions at $\sqrt{s} =$ 8 TeV}.
\newblock Phys. Rev. D \textbf{98}(11), 112011 (2018).
\newblock \doi{10.1103/PhysRevD.98.112011}

\bibitem{CMS:2017rzx}
Sirunyan, A.M., et~al.: {Measurement of angular parameters from the decay
  $\mathrm{B}^0 \to \mathrm{K}^{*0} \mu^+ \mu^-$ in proton-proton collisions at
  $\sqrt{s} = $ 8 TeV}.
\newblock Phys. Lett. B \textbf{781}, 517--541 (2018).
\newblock \doi{10.1016/j.physletb.2018.04.030}

\bibitem{CMS:2020oqb}
Sirunyan, A.M., et~al.: {Angular analysis of the decay B$^+$ $\to$
  K$^*$(892)$^+\mu^+\mu^-$ in proton-proton collisions at $\sqrt{s} =$ 8 TeV}.
\newblock JHEP \textbf{04}, 124 (2021).
\newblock \doi{10.1007/JHEP04(2021)124}

\bibitem{Straub:2018kue}
Straub, D.M.: {flavio: a Python package for flavour and precision phenomenology
  in the Standard Model and beyond}  (2018)

\bibitem{Belle:2021idw}
van Tonder, R., et~al.: {Measurements of $q^2$ Moments of Inclusive $B
  \rightarrow X_c \ell^+ \nu_{\ell}$ Decays with Hadronic Tagging}.
\newblock Phys. Rev. D \textbf{104}(11), 112011 (2021).
\newblock \doi{10.1103/PhysRevD.104.112011}

\bibitem{Belle:2018ezy}
Waheed, E., et~al.: {Measurement of the CKM matrix element $|V_{cb}|$ from
  $B^0\to D^{*-}\ell^{+} \nu_\ell$ at Belle}.
\newblock Phys. Rev. D \textbf{100}(5), 052007 (2019).
\newblock \doi{10.1103/PhysRevD.100.052007}.
\newblock [Erratum: Phys.Rev.D 103, 079901 (2021)]

\bibitem{Wang:2021yrr}
Wang, C., Wang, Y.M., Wei, Y.B.: {QCD factorization for the four-body leptonic
  B-meson decays}.
\newblock JHEP \textbf{02}, 141 (2022).
\newblock \doi{10.1007/JHEP02(2022)141}

\bibitem{Wang:2016qii}
Wang, Y.M.: {Factorization and dispersion relations for radiative leptonic $B$
  decay}.
\newblock JHEP \textbf{09}, 159 (2016).
\newblock \doi{10.1007/JHEP09(2016)159}

\bibitem{Wang:2018wfj}
Wang, Y.M., Shen, Y.L.: {Subleading-power corrections to the radiative leptonic
  $B \to \gamma \ell \nu$ decay in QCD}.
\newblock JHEP \textbf{05}, 184 (2018).
\newblock \doi{10.1007/JHEP05(2018)184}

\bibitem{Belle:2016fev}
Wehle, S., et~al.: {Lepton-Flavor-Dependent Angular Analysis of $B\to K^\ast
  \ell^+\ell^-$}.
\newblock Phys. Rev. Lett. \textbf{118}(11), 111801 (2017).
\newblock \doi{10.1103/PhysRevLett.118.111801}

\bibitem{Belle:2009zue}
Wei, J.T., et~al.: {Measurement of the Differential Branching Fraction and
  Forward-Backward Asymmetry for $B \to K^{(*)}\ell^+\ell^-$}.
\newblock Phys. Rev. Lett. \textbf{103}, 171801 (2009).
\newblock \doi{10.1103/PhysRevLett.103.171801}

\bibitem{Workman:2022ynf}
Workman, R.L.: {Review of Particle Physics}.
\newblock PTEP \textbf{2022}, 083C01 (2022).
\newblock \doi{10.1093/ptep/ptac097}

\end{thebibliography}

%\begin{thebibliography}{}
%%
%% and use \bibitem to create references. Consult the Instructions
%% for authors for reference list style.
%\end{thebibliography}

\end{document}